\documentclass[aps,prd,10pt,superscriptaddress,twocolumn,nofootinbib,dvipsnames,longbibliography]{revtex4-2}

\usepackage[pdftex]{color}
\usepackage[sort&compress]{natbib}
\usepackage[colorlinks=true,linkcolor=purple,filecolor=orange,urlcolor=blue,citecolor=purple,pdftex,plainpages=false]{hyperref}
\usepackage{url}
\pdfoutput=1
\usepackage{ifpdf}
\usepackage[utf8]{inputenc}
\usepackage{color,hhline,psfrag,rotating}
\usepackage{dcolumn}
\usepackage{verbatim}
\usepackage{lipsum}
\usepackage{bm}
\usepackage{slashed}
\usepackage{braket}
\usepackage{rotating}
\usepackage{multirow,tabularx}
\usepackage{graphicx}
\usepackage[dvipsnames]{xcolor}
\usepackage{amsfonts,amssymb,amsmath}
\usepackage{booktabs}  
\newcommand\brachat[1]{\mathord{\mathop{#1}\limits^{\scriptscriptstyle(\wedge)}}}  

\mathchardef\mhyphen="2D

\begin{document}

\global\let\newpage\relax

\title{$B\to K$ and $D\to K$ form factors from fully relativistic lattice QCD}
\author{W.~G.~Parrott}
\email[]{w.parrott.1@research.gla.ac.uk}
\affiliation{SUPA, School of Physics and Astronomy, University of Glasgow, Glasgow, G12 8QQ, UK}
\author{C.~Bouchard}
\email[]{chris.bouchard@glasgow.ac.uk}
\affiliation{SUPA, School of Physics and Astronomy, University of Glasgow, Glasgow, G12 8QQ, UK}
\author{C.~T.~H.~Davies}
\email[]{christine.davies@glasgow.ac.uk}
\affiliation{SUPA, School of Physics and Astronomy, University of Glasgow, Glasgow, G12 8QQ, UK}

\collaboration{HPQCD collaboration}
\homepage{http://www.physics.gla.ac.uk/HPQCD}
\noaffiliation

\date{\today}

\begin{abstract}
We present the result of lattice QCD calculation of the scalar, vector and tensor form factors for the $B\to K\ell^+\ell^-$ decay, across the full physical range of momentum transfer.  We use the highly improved staggered quark (HISQ) formalism for all valence quarks on eight ensembles of gluon field configurations generated by the MILC collaboration. These include four flavours of HISQ quarks in the sea, with three ensembles having the light $u/d$ quarks at physical masses. In the first fully relativistic calculation of these form factors, we use the heavy-HISQ method. This allows us to determine the form factors as a function of heavy quark mass from the $c$ to the $b$, and so we also obtain new results for the $D\to K$ tensor form factor. The advantage of the relativistic formalism is that we can match the lattice weak currents to their continuum counterparts much more accurately than in previous calculations; our scalar and vector currents are renormalised fully nonperturbatively and we use a well-matched intermediate momentum-subtraction scheme for our tensor current. Our scalar and vector $B\to K$ form factors have uncertainties of less than 4\% across the entire physical $q^2$ range and the uncertainty in our tensor form factor is less than 7\%. Our heavy-HISQ method allows us to map out the dependence on heavy-quark mass of the form factors and we can also see the impact of changing spectator quark mass by comparing to earlier HPQCD results for the same quark weak transition but for heavier mesons. 
\end{abstract}

\maketitle
\section{Introduction} \label{sec:motiv}
Here we study the $B\to K\ell\bar{\ell}$ decay, where $\ell$ can be a charged lepton or a neutrino. The decay involves the $b\to s$ flavour changing neutral current (FCNC) and is highly suppressed in the Standard Model (SM) since it must proceed through loop diagrams with at least one off-diagonal (and hence small) element of the  Cabbibo-Kobayashi-Maskawa (CKM) matrix~\cite{Cabibbo:1963yz,Kobayashi:1973fv}. This means that the process is highly sensitive to the existence of `new' particles which may appear in the loops. 

The increasing quantity of experimental data being collected~\cite{Aubert:2008ps,Lees:2012tva,TheBaBar:2016xwe,Wei:2009zv,BELLE:2019xld,Aaltonen:2011qs,Aaij:2012cq,Aaij:2012vr,Aaij:2014pli,Aaij:2014tfa,Aaij:2014ora,LHCb:2016due,Aaij:2021vac} allows for much stronger bounds to be placed on rare decays such as this one, which often rely on huge numbers of collisions to be observed to register just a handful of events. In order to take advantage of this improved precision in our search for new physics beyond the SM~\cite{Altmannshofer:2012az,Bobeth:2007dw,Bobeth:2011nj,Bobeth:2012vn,Du:2015tda,Bouchard:2013mia,Khodjamirian:2012rm,Wang:2012ab}, we must meet these results with improved theoretical uncertainty. At present, lattice Quantum Chromodynamics (QCD) is the only model independent method for calculating hadronic form factors for such decays. The form factors can be used to construct the dominant contribution to the differential branching fraction, for comparison to experiment, in regions of $q^2$ away from $c\overline{c}$ and $u\overline{u}$ resonances. 

Previous full lattice QCD calculations used gluon field configurations generated by the MILC collaboration that include the effect of 3 flavours of sea quarks in the asqtad formalism~\cite{Bernard:2001av}. Ref.~\cite{Bouchard:2013pna} used the non-relativistic (NRQCD)~\cite{Lepage:1992tx} formalism for the $b$ quarks and highly improved staggered quarks (HISQ)~\cite{Follana:2006rc} for other valence flavours. Similarly,~\cite{Bailey:2015dka} used Fermilab $b$ quarks~\cite{ElKhadra:1996mp} and the asqtad formalism for other flavours. Both of these calculations used $\mathcal{O}(\alpha_s)$ perturbation theory to match the lattice weak current operators to their continuum counterparts. Missing higher-order effects in the matching are then a significant source of uncertainty in the form factors. In addition the calculations were done at relatively low values of the $K$ meson spatial momentum in the $B$ meson rest-frame (i.e. close to zero-recoil). 

In this paper we present the first fully relativistic calculation, using HISQ formalism for all valence quarks and working on MILC $N_f=2+1+1$ gluon field ensembles that include HISQ quarks in the sea~\cite{MILC:2010pul,MILC:2012znn}. The calculation mirrors the heavy-HISQ approach used successfully in several other recent HPQCD calculations (e.g.~\cite{Parrott:2020vbe,Harrison:2020gvo,Cooper:2020wnj,McLean:2019qcx}). By using a relativistic treatment we eliminate the matching errors arising from effective theory treatment of the $b$ quark in previous methods. We are also able to cover the full physical $q^2$ range of the decay process directly. 

Our method involves calculating the form factors for a range of heavy quark masses from that of the $c$ quark up to that of the $b$. We thus obtain results for form factors for both $D\to K$ decay and for $B\to K$ decay and the functional form in heavy quark mass that connects them. The vector and scalar $D\to K$ form factors were recently used in an analysis of the $D\to K\ell \overline{\nu}$ weak semileptonic decay process and, combined with experimental results, gave a 1\%-accurate determination of the CKM element $V_{cs}$~\cite{Chakraborty:2021qav}.  That analysis showed very good agreement between the $q^2$-dependence of the vector form factor calculated in lattice QCD and that inferred from the experimental results for the differential decay rate. This provides a very solid test that our lattice QCD form factors at the $c$ quark mass end of our heavy quark mass range describe experimental results for a case ($D\to K\ell\overline{\nu}$) where no new physics is expected. Here we give the tensor form factor for $D\to K$ decay, not calculated in~\cite{Chakraborty:2021qav}.

In this work we will focus on the calculation of the form factors themselves, whilst an accompanying article~\cite{Parrott:2022dnu} will study the phenomenological implications from our results. Section~\ref{sec:latticecalc} sets out the calculational framework and then Section~\ref{sec:fits} describes the fits and data analysis. Section~\ref{sec:results} shows the results as a function of $q^2$, detailing the changes between $D\to K$ and $B\to K$, as well as comparing $B\to K$ form factors with those for $B_c\rightarrow D_s$~\cite{Cooper:2021ofu}, which differ in spectator quark mass. We also provide complete error budgets for the form factors and compare to expectations from Heavy Quark Effective Theory. Section~\ref{sec:conclusions} gives our conclusions.

\section{Lattice calculation}
\label{sec:latticecalc}
\subsection{Form factors}
\label{sec:formfactors}
The quantities of interest here are the scalar, vector and tensor form factors $f_0(q^2)$, $f_+(q^2)$ and $f_T(q^2)$, which are functions of $q^2=(p_{B}-p_{K})^2$. We can construct these form factors from hadronic matrix elements between the $B$ and $K$ which we calculate on the lattice. 

Our heavy-HISQ approach works by determining a set of matrix elements for mesons in which the $b$ quark is replaced by a heavy quark with mass $m_h < m_b$. The heaviest mass on the finest ensemble we use is close to the $b$ mass ($m_h/m_b\approx0.85$). We denote the resulting pseudoscalar heavy-light mesons generically by $H$. We compute these matrix elements for a variety of masses ranging from that of the charm quark upwards, across the range $0\leq{}q^2\leq{}q_{\text{max}}^2=(M_{H}-M_{K})^2$, which is the full physical range of $q^2$ for the decay of a heavy-light meson of mass $M_H$. As $m_h \rightarrow m_b$ this becomes the full range for the $B\to K$ decay.   

The connection between the matrix elements of the lattice scalar, vector and tensor currents and the form factors is,
\begin{equation}\label{Eq:vec}
\begin{split}
  Z_V\bra{K}V^{\mu}_{\mathrm{latt}}\ket{\hat{H}}&=\\
  &f_+(q^2)\Big(p_{H}^{\mu}+p_{K}^{\mu}-\frac{M_{H}^2-M_{K}^2}{q^2}q^{\mu}\Big)\\
    &+f_0(q^2)\frac{M_{H}^2-M_{K}^2}{q^2}q^{\mu},
\end{split}
\end{equation}
\begin{equation}\label{Eq:sca}
  \bra{K}S_{\mathrm{latt}}\ket{H}=\frac{M_{H}^2-M_{K}^2}{m_h-m_s}f_0(q^2),
\end{equation}
\begin{equation}\label{Eq:ten}
  Z_T(\mu)\bra{\hat{K}}T^{k0}_{\mathrm{latt}}\ket{\hat{H}}=\frac{2iM_{H}p_K^k}{M_H+M_K}f_T(q^2,\mu).
\end{equation}
Here $q^{\mu}$ is the 4-momentum transfer and $q^2$ its square. We work in the rest frame of the $H$ such that $p^0_H=M_H$. The $K$ meson is given spatial momentum $\vec{p}_K$ in the $(1,1,1)$ direction, making all spatial directions equivalent, and we take spatial component $k=1$ for the tensor form factor. $Z_V$ and $Z_T$ are renormalisation factors for the lattice vector and tensor currents that we discuss below. Note that the tensor form factor has a renormalisation scale $\mu$ associated with it. $m_h$ and $m_s$ in Eq.~\eqref{Eq:sca} are the lattice valence quark masses for the $h$ and $s$ quarks. 

Requiring that the matrix elements are finite as $q^2\to{}0$ gives the constraint
\begin{equation}\label{eq:f0=fp}
  f_+(0)=f_0(0).
\end{equation}
We will make use of this condition later. 

Bilinears constructed from staggered quarks have a `taste' degree of freedom, $\xi$, and we need to arrange the tastes of mesons and lattice currents appropriately so that tastes cancel in the correlation functions that we calculate. Here we follow the approach used in~\cite{Chakraborty:2021qav}, where the rationale is described in more detail. We aim to have local operators for all of the currents (with $\psi$ and $\bar{\psi}$ at the same point) because these are least noisy and have no tree-level discretisation errors. This means that we must use point-split operators for the meson creation and annihilation operators in some cases. In spin-taste notation~\cite{Follana:2006rc}, the lattice scalar, vector and tensor currents are $S = \bar{\psi}_s1\otimes{}1\psi_h$, $V^{\mu}=\bar{\psi}_s\gamma^{\mu}\otimes\xi^{\mu}\psi_h$ and $T^{k0}=\bar{\psi}_s\gamma^{k}\gamma^0\otimes\xi^k\xi^0\psi_h$. $H=\bar{\psi}_l\gamma^5\otimes{}\xi^5\psi_h$ and $\hat{H}=\bar{\psi}_l\gamma^5\gamma^0\otimes{}\xi^5\xi^0\psi_h$ denote Goldstone and local non-Goldstone heavy-light pseudoscalar mesons, respectively. Similarly for the kaons, $K=\bar{\psi}_s\gamma^5\otimes{}\xi^5\psi_l$ and $\hat{K}=\bar{\psi}_s\gamma^5\otimes{}\xi^5\xi^1\psi_l$ denote Goldstone and point-split non-Goldstone strange-light pseudoscalar mesons, respectively. 

We use the local temporal component of the vector current, $V^0$, for most results but we also include some additional spatial current data with the local vector current $V^k$ in the $k=1$ direction.   The reason for this is that the vector form factor determined from the temporal vector current has a numerical problem at large $q^2$ from the way that it is constructed~\cite{Cooper:2020wnj}. Rearranging Eq.~\eqref{Eq:vec} we have
\begin{equation}
  f_+(q^2)=\frac{1}{A^{\mu}-B^{\mu}}(Z_V\bra{K}V^{\mu}\ket{\hat{H}}-f_0(q^2)B^{\mu}),
\end{equation}
where $A^{\mu}=p^{\mu}_H+p^{\mu}_K$ and $B^{\mu}=\frac{M_{H}^2-M_{K}^2}{q^2}q^{\mu}$. Both numerator and denominator vanish as $q^2\rightarrow q^2_{\mathrm{max}}$ amplifying the uncertainties and leading to large statistical errors in $f_+$ at large $q^2$. One solution to this issue is to use a spatial component of the vector current, with spin-taste $\gamma^1\otimes\xi^1$. This component requires a point-split $\gamma^5\otimes\xi^5\xi^1$ kaon ($\hat{K}$), which we have already used in the tensor case, and the Goldstone heavy-light pseudoscalar ($H$). At low $q^2$ the point split kaon makes this component noisier than the temporal case which we predominantly use, but at large $q^2$ the resulting $f_+$ does not suffer from the same dramatic growth in uncertainties and so is to be preferred.
We include in our data set a handful of $V^1$ matrix elements, at large $q^2$ and mass, on our two finest ensembles to supplement our comprehensive $V^0$ data. We use the same $Z_V$ for both $V^0$ and $V^1$ since any difference between the two cases for our relativistic action is purely a discretisation effect. We will denote the form factors obtained in the two cases $f_+^{V^0}$ and $f_+^{V^1}$ where the distinction is relevant. We discuss a comparison of the two cases in Section~\ref{sec:fits}. 

\subsection{Current normalisation}
The fact that the partially conserved vector current (PCVC) relation holds for the HISQ action means that the scalar form factor $f_0(q^2)$ can be obtained from the matrix element of the local scalar current using Eq.~\eqref{Eq:sca} with absolute normalisation~\cite{Na:2010uf}. 
We would also need no renormalisation for the vector current if we used the conserved current~\cite{Hatton:2019gha}. Here, however, we use the much simpler local vector current and this requires renormalisation. 
The renormalisation factor, $Z_V$, can be calculated fully non-perturbatively using the PCVC relation~\cite{Na:2010uf, Koponen:2013tua}. We apply it in the temporal vector case at zero-recoil, where both the $H$ and $K$ mesons are at rest and it gives the most accurate results~\cite{Cooper:2020wnj},
\begin{equation}\label{Eq:Zv}
    Z_V=\frac{(m_h-m_s)\bra{K}S\ket{H}}{(M_{H}-M_{K})\bra{K}V^{0}\ket{\hat{H}}}\Bigg\rvert_{q^2=q^2_{\mathrm{max}}}.
\end{equation}

We also calculate the tensor form factor and the tensor current requires renormalisation. $Z_T$ in Eq.~\eqref{Eq:ten} takes the lattice local tensor current to that in the $\overline{\mathrm{MS}}$ scheme at a specific renormalisation scale $\mu$. $Z_T$ can be determined accurately for the HISQ action~\cite{Hatton:2020vzp} using an intermediate momentum-subtraction scheme, called RI-SMOM, that can be matched through $\mathcal{O}(\alpha_s^3)$ to $\overline{\mathrm{MS}}$~\cite{Kniehl:2020sgo}. This makes the renormalisation factor much more accurate than the $\mathcal{O}(\alpha_s)$ renormalisation factors used in previous calculations of the $B\to K$ tensor form factor~\cite{Bouchard:2013pna, Bailey:2015dka}.  Note that the intermediate momentum-subtraction scheme is implemented nonperturbatively on the lattice and so attention must be paid to nonperturbative artefacts (`condensates') that can appear as inverse powers of the intermediate renormalisation scale. These are analysed using fits to multiple intermediate scales in~\cite{Hatton:2020vzp}. We use corrected $Z_T$ values from Table VIII of~\cite{Hatton:2020vzp} in which these artefacts have been removed. We will give final results for $f_T$ for $B\to K$ at a scale $\mu=4.8~\mathrm{GeV}$ appropriate to $m_b$ (taken as the approximate value of the $b$ quark pole mass); for $D\to K$ we will give values at a lower scale ($\mu=2~\mathrm{GeV}$). $f_T$ values can be run between scales straightforwardly~\cite{Hatton:2020vzp}. 

\begin{table*}
  \caption{Gluon field ensembles used in this work, numbered in column 1, with gauge coupling values, $\beta$, in column 2. The Wilson flow parameter~\cite{Borsanyi:2012zs} is used to calculate the lattice spacing $a$ via values for $w_0/a$~\cite{McLean:2019qcx} in column 3. We use $w_0=0.1715(9)\text{fm}$,  determined from $f_{\pi}$ in~\cite{Dowdall:2013rya}. Column 4 gives the approximate value of $a$ for each set. Column 5 gives the spatial ($N_x$) and temporal ($N_t$) dimensions of each lattice in lattice units and column 6, the number of configurations and time sources used in each case. Columns 7-11 give the masses of the valence and sea quarks in lattice units, noting that $m_u=m_d=m_l$ and the valence and sea masses are the same in the case of $m_l$. The valence $s$ quark masses are tuned to give $M_{\eta_s}$ = 0.6885(22) GeV~\cite{Dowdall:2013rya, Chakraborty:2014aca}. We include the values of $am_c^{\mathrm{val}}$ in column 11, since this is always the lightest of the heavy valence quark masses that we work with. A complete list of the heavy quark masses used on each set is given in Table~\ref{tab:renorm}. Column 12 shows values for the tensor normalisation $Z_T$ at scale $m_b=4.8~\mathrm{GeV}$ (\cite{Hatton:2020vzp}, Table VIII). Sets 1 and 2 did not include calculation of the tensor 3-point functions, so this is omitted in those cases.}
  \begin{center} 
    \begin{tabular}{c c c c c c c c c c c c}
      \hline
      Set & $\beta$ & $w_0/a$ & $a$ (fm) & $N_x^3\times N_t$  &$n_{\mathrm{cfg}}\times n_{\mathrm{src}}$ &    $am_{l}^{\mathrm{sea/val}}$ & $am_{s}^{\mathrm{sea}}$ & $am_c^{\mathrm{sea}}$& $am_{s}^{\mathrm{val}}$ & $am_c^{\mathrm{val}}$& $Z_T(m_b)$\\
  \hline
  \hline
    1   & 5.8    & 1.1367(5)  & 0.15 &  $32^3\times 48$    & $998\times 16$ &     0.00235        &  0.0647        &  0.831&  0.0678 &  0.8605& -\\
    \hline
    2    & 6.0    & 1.4149(6)   & 0.12 &  $48^3\times 64$    & $985\times 16$&   0.00184       &   0.0507        &  0.628  &0.0527  & 0.643& -\\
    \hline
    3   &   6.3  & 1.9518(7)   & 0.088 & $64^3\times 96$    & $620\times 8$&  0.00120       &  0.0363         &  0.432&   0.036  &  0.433& 1.0029(43)\\
      \hline
      \hline
    4    & 5.8    & 1.1119(10) & 0.15 &  $16^3\times 48$    & $1020\times 16$&     0.013        &  0.065        &  0.838 &  0.0705 &  0.888 & 0.9493(42)\\
    \hline
    5    & 6.0    &  1.3826(11) &  0.12 & $24^3\times 64$   & $1053\times 16$ &     0.0102        &  0.0509        &  0.635  &  0.0545  & 0.664 & 0.9740(43)\\
    \hline
    6    & 6.3    &  1.9006(20)  & 0.09 & $32^3\times 96$  & $499\times 16$  &   0.0074       &   0.037        &  0.440&  0.0376  &  0.449& 1.0029(43)\\
    \hline
    7   & 6.72    &  2.896(6)  & 0.059 & $48^3\times 144$    & $413\times 8$&  0.0048       &  0.024         &  0.286& 0.0234 &  0.274 & 1.0342(43)\\
    \hline
    8   & 7.0    &   3.892(12) & 0.044 & $64^3\times 192$    & $375\times 4$&  0.00316       &  0.0158         &  0.188& 0.0165 &  0.194 & 1.0476(42) \\
    \hline
    \hline
    \end{tabular}
  \end{center}
  \label{tab:ensembles}
\end{table*}

\subsection{Simulation details}
\label{sec:sim}
The calculation was run on MILC gluon field ensembles~\cite{MILC:2010pul,MILC:2012znn} that include in the sea two degenerate light quarks, strange and charm quarks, with masses $m_l^{\text{sea}},m_s^{\text{sea}},m_c^{\text{sea}}$, using the HISQ action~\cite{Follana:2006rc}. The eight ensembles used have parameters listed in Table~\ref{tab:ensembles}. Sets 1, 2 and 3 have physical light quark masses, whilst sets 4-8 have $m_l^{\mathrm{sea/val}}=0.2m_s^{\mathrm{sea}}$. Note that the valence light quark masses are the same as those in the sea; the valence strange quark masses are tuned more accurately than the sea strange quark masses and so differ slightly from them. The valence strange quark masses are tuned~\cite{Chakraborty:2014aca} to give the physical value for the mass of the $s\overline{s}$ pseudoscalar meson known as the $\eta_s$ (which does not appear in the real world), whose mass is determined in terms of the pion and kaon masses in~\cite{Dowdall:2013rya}. The gluon action is Symanzik-improved to remove discretisation errors through $\mathcal{O}(\alpha_sa^2)$~\cite{Hart:2008sq}. 

A significant portion of the data used here overlaps with that used for $D\to K$ form factors in~\cite{Chakraborty:2021qav}. Sets 1 and 2 are identical, whilst other sets share the lowest mass (the charm), but include additional masses and the extra tensor current insertion. This means that, whilst the calculation produced a slightly different set of $D\to K$ scalar and vector form factors, these are correlated to those in~\cite{Chakraborty:2021qav} and as such should not be viewed as an independent calculation. The $D\to{} K$ tensor form factor, however, was not calculated in~\cite{Chakraborty:2021qav} and will be presented here. The valence heavy quark masses used on each ensemble are given in Table~\ref{tab:renorm}.

\begin{table}[t]
  \caption{Masses in lattice units used for the valence heavy quarks on each set from Table~\ref{tab:ensembles}. The lightest heavy mass in each case corresponds to a well-tuned value for the charm quark mass~\cite{Hatton:2020qhk}. $m_h/m_c$ then reaches 4.1 on our finest lattices, set 8. Column 3 gives normalisation constants for the vector current from our results. $Z_V$ is calculated using Eq.~\eqref{Eq:Zv}. $Z_{\text{disc}}$ in column 4 is a small tree-level discretisation correction, beginning at $(am_h)^4$, that we make to all the matrix elements, see Eq.~\eqref{Eq:fitnormalisation}. It is defined in~\cite{Monahan:2012dq}.}
  \begin{center} 
    \begin{tabular}{c c c c}
      \hline
      Set &  $am^{\text{val}}_{h}$& $Z_V$ &$Z_{\text{disc}}$\\ [0.5ex]
      \hline
      1 & 0.8605 &1.0440(87)& 0.99197\\ [1ex]
      \hline
      2 & 0.643 &1.0199(54)& 0.99718\\ [1ex]
      \hline
      3 & 0.433 &1.0016(81)& 0.99938\\ [1ex]
        & 0.683 &1.011(10)& 0.99648  \\ [1ex]
        & 0.8 &1.017(12)& 0.99377   \\ [1ex]
      \hline
      \hline
      4 & 0.888 &1.0376(52)& 0.99050\\ [1ex]
      \hline
      5 & 0.664 &1.0221(41)& 0.99683\\ [1ex]
        & 0.8 &1.0300(47)& 0.99377  \\ [1ex]
        & 0.9 &1.0365(51)& 0.99063   \\ [1ex]
      \hline
      6 & 0.449 &0.9977(67)& 0.99892\\ [1ex]
        & 0.566 &1.0033(80)&  0.99826  \\ [1ex]
        & 0.683 &1.0091(85)& 0.99648  \\ [1ex]
        & 0.8 &1.055(32)& 0.99377   \\ [1ex]

      \hline
      7 & 0.274 &0.9901(94)&0.99990\\ [1ex]
        & 0.45 &0.992(12)& 0.99928 \\ [1ex]
        & 0.6 &0.996(13)& 0.99783  \\ [1ex]
        & 0.8 &1.006(14)& 0.99377  \\ [1ex]

      \hline
      8 &  0.194 &0.984(10)&0.99997\\ [1ex]
       & 0.45 &0.993(12)& 0.99928  \\ [1ex]
       & 0.6 &0.998(13)& 0.99783\\ [1ex]
       & 0.8 &1.006(16)& 0.99377  \\ [1ex]

      \hline
    \end{tabular}
  \end{center}
  \label{tab:renorm}
\end{table}
\begin{figure}
\includegraphics[width=0.48\textwidth]{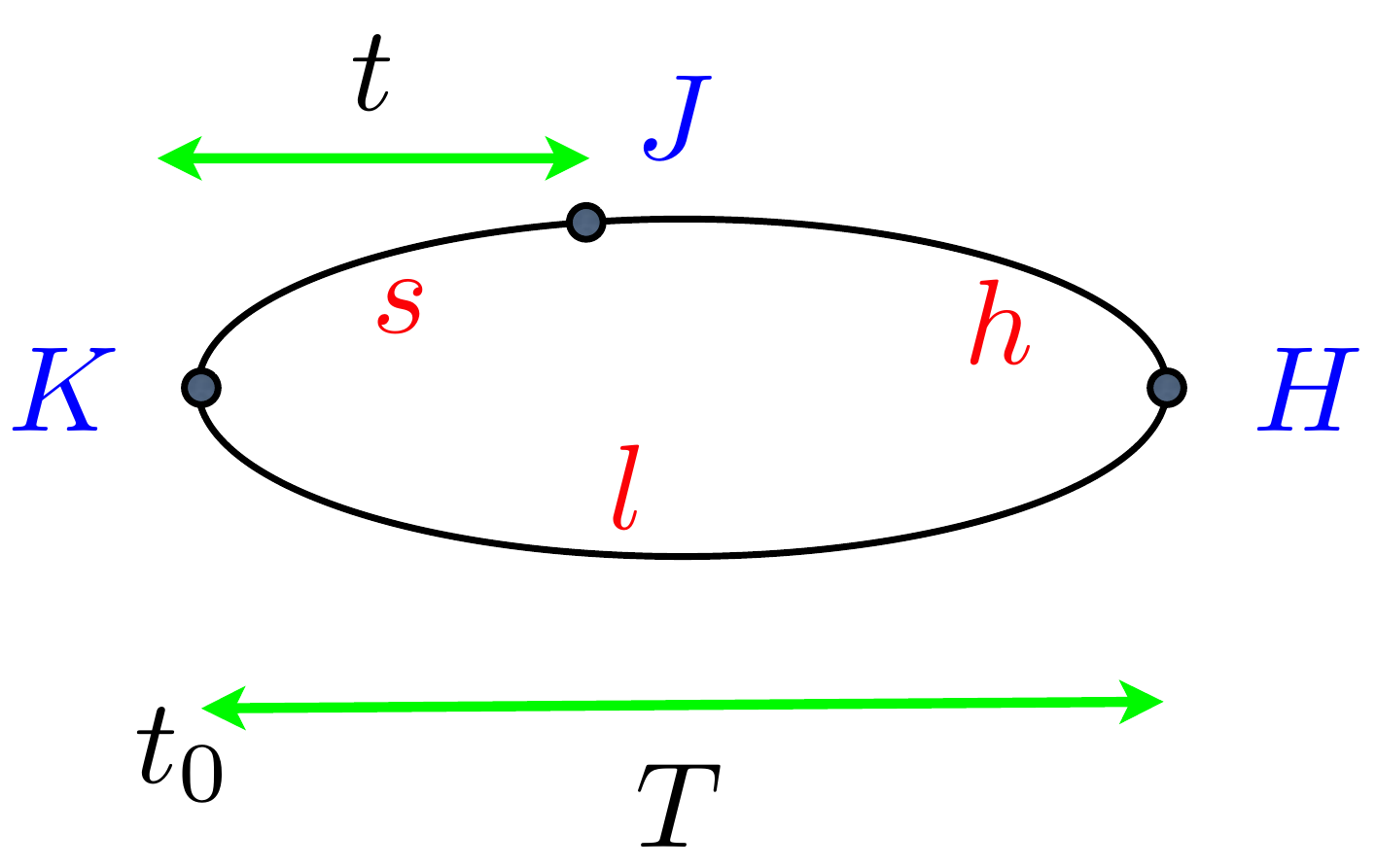}
\caption{Schematic of our three-point correlation function.}
\label{fig:3pt}
\end{figure}

In order to compute the matrix elements needed for our form factors, we must extract the amplitudes from three-point correlation functions built on the lattice. 
A schematic of our setup is shown in Figure~\ref{fig:3pt}.  An $h$ `parent' quark propagator is generated as an `extended' propagator from a source at timeslice $t_0+T$; the source is constructed from a light `spectator' quark propagator originating from timeslice $t_0$. The $h$ quark propagator is combined with an $s$ `daughter' quark propagator from $t_0$ at time $t_0+t$, where current $J$ is inserted. The propagators are combined with appropriate colour and spin (i.e. staggered spin-taste) for the quantum numbers of a pseudoscalar to pseudoscalar transition via current $J$. 
Our calculation is set up in this `backwards' arrangement for computational convenience, as the physics is unchanged by a time reversal. 

For each gluon field configuration multiple values of $t_0$, uniformly placed on the lattice, with the first being randomly selected to reduce autocorrelation, are used to increase our statistics. To improve the statistics on each ensemble further and to better fit the $T$ dependence, multiple values of the source-sink separation $T$ are also used for each $t_0$ value, with odd and even values included to capture oscillations in $t$. On most ensembles, we average the correlation functions for different $t_0$ values. On the finest ensemble (set 8), however, we do not do this. On this ensemble we have only four source $t_0$ values which are very widely spaced and tests confirm that correlations between them are negligible. It is then useful to keep the correlation functions for different $t_0$ as separate data to improve our determination of the covariance matrix. 
 
The $H$ meson is at rest on the lattice and momentum is given to the $K$ meson. This momentum, $\vec{p}_K$, is generated in the $(1,1,1)$ direction using twisted boundary conditions~\cite{Guadagnoli:2005be} for the $s$ quark propagator. The momentum in lattice units is related to the twist, $\theta$, by $|a\vec{p}_K|=\theta(\sqrt{3}\pi)/N_x$, where $N_x$ is the spatial extent of the lattice in lattice units. Different values of momentum are chosen so as to cover the full physical range of momentum transfer, $q$, on each lattice for the different heavy masses used. The corresponding twists are listed in Table~\ref{tab:twists}. Because we have a variety of heavy masses on each gluon field ensemble, the coverage of the momenta cannot be optimised for each mass - we settle for values which give the best coverage overall. This means that some masses can generate negative $q^2$ values at large twist. Whilst these points are unphysical, they are easily accommodated in our fit form, as we shall see below. 

As well as the aforementioned three-point functions,  we also generate two-point correlation functions in the standard way for each of the $\brachat{H}$ and $\brachat{K}$ masses and momenta, in order to extract energies and amplitudes for the mesons.

In addition to the $H\to K$ results discussed above we also include results for $H_s\to \eta_s$ correlation functions from~\cite{Parrott:2020vbe}. The $H_s \to \eta_s$ results are for sets 6 and 7 (called sets 1 and 2 in~\cite{Parrott:2020vbe}) and include scalar and temporal vector current insertions in the three-point functions only. We do not include results on set 3 (set 8 here) from~\cite{Parrott:2020vbe} as the statistics are much lower than for our $H\to K$ data, nor do we include the continuum $f_0(q^2_{\mathrm{max}})$ data point used in that paper. The heavy masses and twists used there are the same as those used here and given in Tables~\ref{tab:ensembles},~\ref{tab:twists} and~\ref{tab:renorm}. Instead of the spectator light quark that we have here, the earlier results have a spectator strange quark. For further details of the $H_s\to \eta_s$ data see~\cite{Parrott:2020vbe}. The $H_s\to\eta_s$ data was fitted simultaneously with the $H\to{}K$ data on each of the two sets 6 and 7 in order to preserve correlations between the two. This helps us to pin down the chiral extrapolation for the spectator quark to the physical light mass by giving a third light mass value: $m_l=m_s$, $m_l=m_s/5$ and $m_l\approx m_l^{\text{phys}}$. The effect of this extra light mass value on the overall results will be discussed in Section~\ref{sec:eval}.
\begin{table}
  \caption{Details of the twists used for the $K$ meson momenta on each gluon field ensemble. Momenta can be obtained from twist, $\theta$, via $|a\vec{p}_K|=\theta(\sqrt{3}\pi)/N_x$, where $N_x$ is the spatial dimension of the lattice in lattice units, given in Table~\ref{tab:ensembles}. $\vec{p}_K$ is in the (1,1,1) direction. Column 3 gives the $T$ values used for time extent, in lattice units, for the three-point correlation functions on each ensemble, see Fig.~\ref{fig:3pt}.}
  \begin{center} 
    \begin{tabular}{c c c}
      \hline
      Set & $\theta$ & $T$  \\
  \hline
  \hline
    1    & 0, 2.013, 3.050, 3.969 &9, 12, 15, 18   \\
    \hline
    2    & 0, 2.405, 3.641, 4.735 &12, 15, 18, 21\\
    \hline
    3   & 0, 0.8563, 2.998, 5.140 & 14, 17, 20 \\
      \hline
      \hline
    4    & 0, 0.3665, 1.097, 1.828 & 9, 12, 15, 18 \\
    \hline
    5    & 0,  0.441, 1.323, 2.205, 2.646 & 12, 15, 18, 21 \\
    \hline
    6    & 0, 0.4281, 1.282, 2.141, 2.570& 14, 17, 20 \\
    \hline
    7   & 0, 1.261, 2.108, 2.946, 3.624 & 20, 25, 30  \\
    \hline
    8   &0, 0.706, 1.529, 2.235, 4.705 & 24, 33, 40 \\
    \hline
    \hline
    \end{tabular}
  \end{center}
  \label{tab:twists}
\end{table}

\begin{figure}
\includegraphics[width=0.48\textwidth]{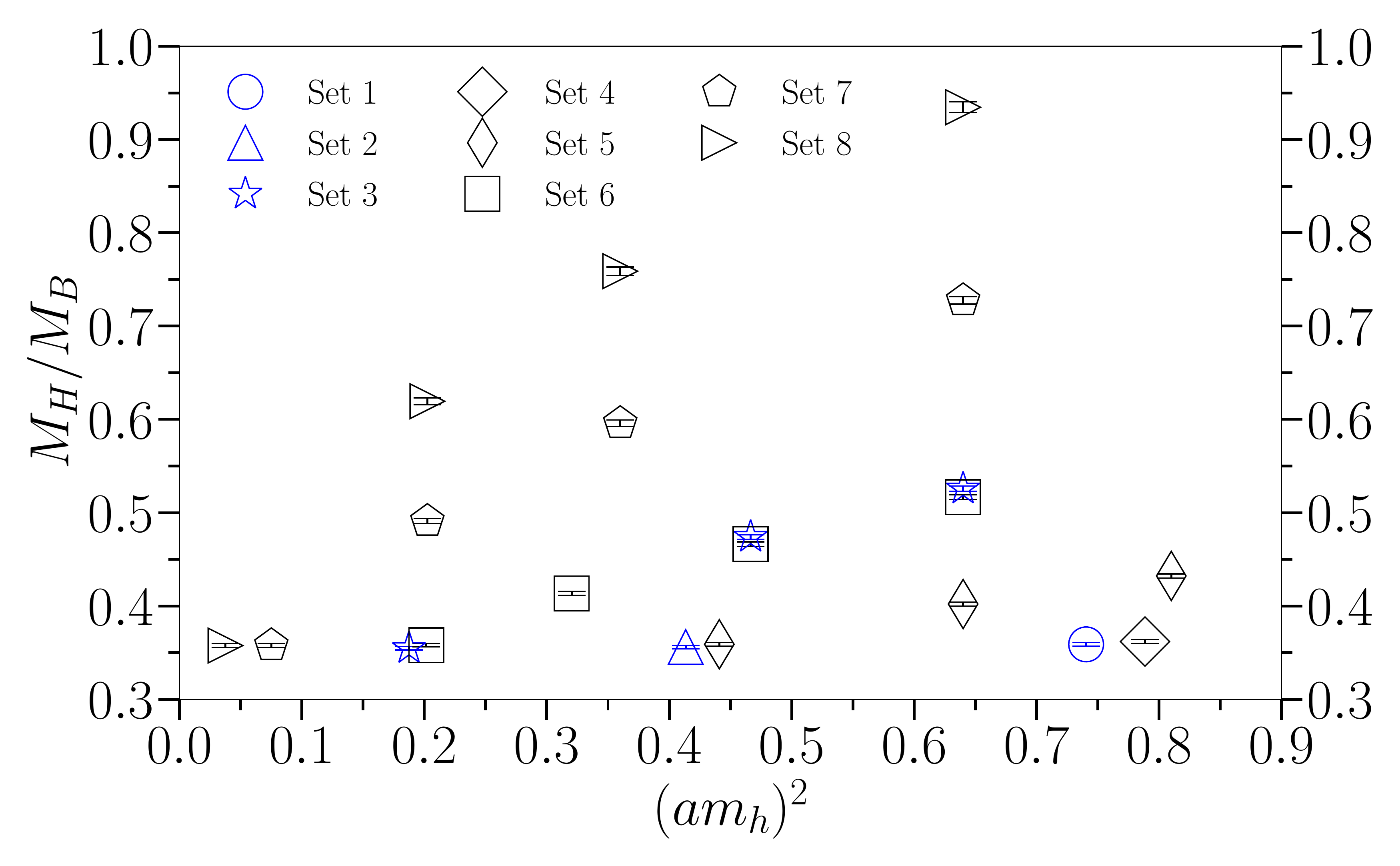}
\caption{The range of $am_h$ values used in this work (Table~\ref{tab:renorm}), and their corresponding $M_H/M_B$ values. Ensembles with physical light quarks are shown in blue.}
\label{fig:amhMH}
\end{figure}

Our ensembles contain a range of different $am_h$ values (see Table~\ref{tab:renorm}), as well as values for $M_H/M_B$ which are correlated between the masses on a given ensemble. This is demonstrated in Figure~\ref{fig:amhMH}, and makes it possible for our fit to distinguish between $am_h$ dependent discretisation effects and $M_H$ dependence. In particular, all ensembles have data points at the physical charm mass, which differ only in their $am_h$ values, and some $am_h$ values, such as $am_h=0.8$ are common to multiple ensembles with different $M_H$ values. Additionally, the included $H_s\to\eta_s$ data discussed above provide an increased lever arm in the $M_H$ dependence, via $M_{H_s}$, for a range of $am_h$ values.

\section{Fits and analysis}\label{sec:fits}
\begin{table}[t]
  \caption{Priors used in the fit on each set. Priors are based on previous experience and given large widths. Sometimes, initial priors are tightened or loosened in order to find a fit with an acceptable $\chi^2$. These changes are balanced against the resulting change in log(GBF) (see text). On rare occasions, the fit finds spurious states (with zero amplitude). This renders the fit very obviously wrong, and is easily remedied with an adjustment to the offending priors. The effect of doubling and halving the standard deviation on all priors on the final fit result is shown in Figure~\ref{fig:corrstab}. $d^{M}_{i\neq{}0}$ indicates the amplitudes for oscillating and non-oscillating $H$ mesons and for non-oscillating kaons. $d^{K,o}_i$ is the amplitude for oscillating kaons, which we expect to be smaller, particularly in the case of zero momentum. $P[S_{ij\neq{}00}^{kl}]=P[V^{0,kl}_{ij\neq{}00}]=0.0(5)$ and $P[V^{1,kl}_{ij\neq{}00}]=P[T_{ij\neq{}00}^{kl}]=0.0(1)$ in all cases, whilst $P[V^{1,kl\neq{}\mathrm{nn}}_{00}]=0.0(3)$.}
  \begin{center} 
    \begin{tabular}{c c c c c c c c c }
      \hline
      Set & $P[d^{M}_{i\neq{}0}]$ & $P[d^{K,\mathrm{o}}_i]$  &$P[S_{00}^{kl\neq{}\mathrm{nn}}]$&$P[V_{00}^{0,kl\neq{}\mathrm{nn}}]$& $P[T_{00}^{kl\neq{}\mathrm{nn}}]$   \\ [0.5ex]
      \hline
      1 & 0.15(20) &0.05(5)&0.0(1.0)&0.0(1.0)&- \\ [1ex]
      2 & 0.15(10) &0.05(5)&0.0(1.0)&0.0(1.0)&- \\ [1ex]
      3 & 0.10(10) &0.05(5)&0.0(1.5)&0.0(1.5)&0.0(3) \\ [1ex]
      4 & 0.20(20) &0.05(5)&0.0(1.0)&0.0(1.0)&0.0(3) \\ [1ex]
      5 & 0.20(20) &0.03(3)&0.0(1.0)&0.0(1.5)&0.0(3) \\ [1ex]
      6 & 0.10(10) &0.05(5)&0.0(1.5)&0.0(1.5)&0.0(3) \\ [1ex]
      7 & 0.05(5)  &0.02(2)&0.0(1.0)&0.0(2.0)&0.0(3) \\ [1ex]
      8 & 0.08(10) &0.01(2)&0.0(1.0)&0.0(2.0)&0.0(2) \\ [1ex]

      \hline
    \end{tabular}
  \end{center}
  \label{Tab:priorsforfit}
\end{table}

\begin{figure}

\includegraphics[width=0.48\textwidth]{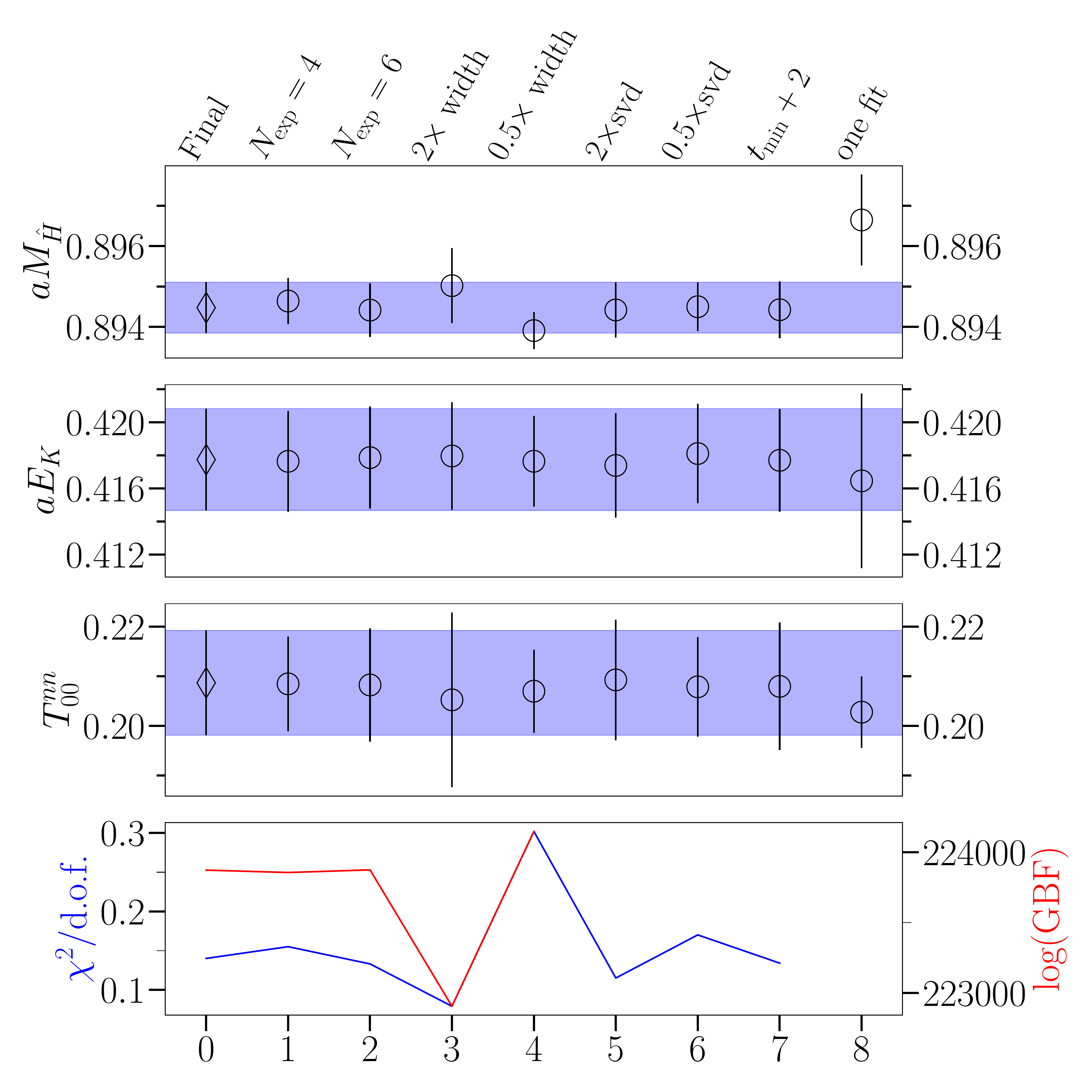}
\caption{Stability plot for different correlator fit choices on set 8, showing the mass of the ground-state non-goldstone $\hat{H}$ meson for $am_{h}=0.6$, the ground-state energy of the $K$ with twist $\theta=4.705$ and $T^{\mathrm{nn}}_{00}$ for $am_h=0.45$, $\theta=2.235$. Test 0 is the final result, corresponding to $N_{\mathrm{exp}}=5$ exponentials. Tests 1 and 2 use one fewer and one more exponential respectively. Tests 3-6 double and halve the prior widths and SVD cut. Test 7 increases $t_\text{min}$ by 2 across the whole fit. The final test, 8, is when the fit is done on its own, or in the case of the $T^{\mathrm{nn}}_{00}$, just with the $\hat{H}$ and $\hat{K}$ two-point correlators required, as opposed to being part of one big simultaneous fit. The $\chi^2$ per degree of freedom and log(GBF) value for each test are shown in the bottom pane in blue and red respectively. For the later tests (5--7), data is removed from the fit, resulting in a lower log(GBF) which is not comparable with the others and not displayed. As discussed in Section~\ref{sec:corrfits}, $\chi^2$ values are artificially lowered by our SVD cut and priors so are only meaningful relatively. $\chi^2$ values for tests 3-6, which change prior width and SVD, are thus not directly comparable with other tests. The final fit gives a $\chi^2$/d.o.f. close to 1 with SVD and prior noise.}
\label{fig:corrstab}
\end{figure}
\subsection{Correlator fits}\label{sec:corrfits}

Using a standard Bayesian approach, as outlined in~\cite{Lepage:2001ym}, we perform a simultaneous, multi-exponential fit to both the two and three point correlation functions. This allows us to extract the ground-state energies, ground-state amplitudes and ground-state to ground-state current matrix elements with uncertainties that allow for any unresolved excited-state contamination. 
Fit quality is judged using $\chi^2$ per degree of freedom (d.o.f.) values and the log of the Gaussian Bayes Factor, log(GBF). As discussed in~\cite{Parrott:2020vbe} and the appendix to~\cite{Dowdall:2019bea}, $\chi^2$ values are artificially reduced by Singular Value Decomposition (SVD) cuts and broad priors for the parameters. This means that $\chi^2$ values should not be taken at face value, but rather as a relative measure, comparable across fits where the SVD cut and priors are the same. The fitting packages we use~\cite{peter_lepage_2020_3707868,peter_lepage_2020_3715065,peter_lepage_2019_3563090} contain an inbuilt noise test~\cite{Dowdall:2019bea}, under which we check that our fits are stable and have an acceptable $\chi^2$/d.o.f. value close to unity when appropriately modified by the inclusion of prior and SVD noise. The log(GBF) value penalises overfitting, so by also using this measure, we are able to confirm that our fits describe the data without overfitting.  

We fit two point correlators for a meson $M$ to a set of exponentials representing a tower of possible states of energy $E^{M}_i$ and amplitude $d_i^{M}$,
\begin{equation}\label{eq:2ptcorr}
  \begin{split}
    C_2^M(t)&=\sum^{N_\mathrm{exp}}_{i=0}\big(|d_i^{M,\mathrm{n}}|^2(e^{-E_i^{M,\mathrm{n}}t}+e^{-E_i^{M,\mathrm{n}}(N_t-t)})\\
    &-(-1)^{t}|d_i^{M,\mathrm{o}}|^2(e^{-E_i^{M,\mathrm{o}}t}+e^{-E_i^{M,\mathrm{o}}(N_t-t)})\big).
  \end{split}
\end{equation}
The ground state is specified by $i=0$. Because of the nature of staggered quarks, states which oscillate in time (labelled `o' as opposed to `n' for non-oscillating states) are also present and are accounted for in the fit. Discarding the first $t_{\text{min}}$ data points allows us to fit to a finite number, $N_{\text{exp}}$, of exponentials, and $t_{\text{min}}$ takes values in the range 2 to 7 for different correlators and different lattice spacings. We estimate priors for the ground state energies and amplitudes using the effective mass and effective amplitudes, as in~\cite{Parrott:2020vbe,Chakraborty:2021qav}, and give each a broad uncertainty, ensuring that the final result of the fit is much more precisely determined than this prior. 
We use log-normal parameters throughout to enforce positive values on energy splittings and amplitudes. Amplitudes are guaranteed to be positive because we use the same interpolating operator at the source and sink. Priors for excited state non-oscillating and all oscillating amplitudes are based on previous experience of amplitude sizes in similar fits~\cite{Parrott:2020vbe,Chakraborty:2021qav}. Some priors are slightly adjusted by trial and error to maximise log(GBF), as well as to ensure that the fit does not find spurious states, which have amplitudes consistent with zero but interfere with the ground state determination. Priors for the oscillating ground state energy of the $\brachat{H}$ and $\brachat{K}$ are taken to be $0.4\,\text{GeV}$ and $0.25\,\text{GeV}$ larger than the non-oscillating ground states respectively, with prior widths on non-oscillating ground states typically in the range $2-10\%$ and oscillating ground states $5-20\%$. In both cases prior widths vary by ensemble, and the posteriors are much better determined than their priors. The energy splitting between excited states is taken as 0.50(25) GeV. Other priors are are listed in Table~\ref{Tab:priorsforfit}.

For the kaons with non-zero twist, as in~\cite{Chakraborty:2021qav}, we use the dispersion relation to inform our ground state priors, allowing for discretisation effects using the following ansatz,
\begin{equation}\label{Eq:disp}
  \begin{split}
    P[aE^{\brachat{K}}_{0,\vec{p}}]&=\sqrt{P[aE^{\brachat{K}}_{0,\vec{0}}]^2+(a\vec{p})^2}\Big(1+P[c_2]\Big(\frac{a\vec{p}}{\pi}\Big)^2\Big),\\
    P[d^{\brachat{K}}_{0,\vec{p}}]&=\frac{P[d^{\brachat{K}}_{0,\vec{0}}]}{[1+(a\vec{p}/P[aE^{\brachat{K}}_{0,\vec{0}}])^2]^{1/4}}\Big(1+P[d_2]\Big(\frac{a\vec{p}}{\pi}\Big)^2\Big).
  \end{split}
\end{equation}
Here $P[d]$ and $P[aE]$ represent the priors of the relevant amplitudes and energies.  
We take priors for $c_2$ and $d_2$ as $0\pm1$ based on observations of dispersion relations in similar fits~\cite{Parrott:2020vbe,Bouchard:2013pna,Chakraborty:2021qav}. We find their posteriors to fall comfortably within their priors in all fits, typically with a magnitude less than $0.5$. 

We perform three point fits (for mother and daughter mesons $M_2$ and $M_1$) with scalar, vector and tensor current insertions to the following form, 
\begin{equation}\label{eq:3ptcorr}
  \begin{split}
    &C^{M_1,M_2}_3(t,T)=\sum^{N_\mathrm{exp}}_{i,j=0}\big(d_i^{M_1,\mathrm{n}}J_{ij}^{\mathrm{nn}}d_j^{M_2,\mathrm{n}}e^{-E_i^{M_1,\mathrm{n}}t}e^{-E_j^{M_2,\mathrm{n}}(T-t)}\\
    &-(-1)^{(T-t)}d_i^{M_1,\mathrm{n}}J_{ij}^{\mathrm{no}}d_j^{M_2,\mathrm{o}}e^{-E_i^{M_1,\mathrm{n}}t}e^{-E_j^{M_2,\mathrm{o}}(T-t)}\\
    &-(-1)^{t}d_i^{M_1,\mathrm{o}}J_{ij}^{\mathrm{on}}d_j^{M_2,\mathrm{n}}e^{-E_i^{M_1,\mathrm{o}}t}e^{-E_j^{M_2,\mathrm{n}}(T-t)}\\
    &+(-1)^{T}d_i^{M_1,\mathrm{o}}J_{ij}^{\mathrm{oo}}d_j^{M_2,\mathrm{o}}e^{-E_i^{M_1,\mathrm{o}}t}e^{-E_j^{M_2,\mathrm{o}}(T-t)}\big).
  \end{split}
\end{equation}
Here $J_{ij}^{kl}$ ($i,j\in\{0,1,...,N_{\text{exp}}-1\}$, and $k,l\in\{\mathrm{n},\mathrm{o}\}$) are matrix elements of $J=S(V)[T]$ for the scalar (vector) [tensor] currents. For example, $J_{ij}^{\mathrm{no}}$, gives the matrix element for $J$ between the $i$th non-oscillating (`n') state of $M_1$ and the $j$th oscillating (`o') state of $M_2$.
$T$ and $t$ appear as in Figure~\ref{fig:3pt} (where we have taken $t_0=0$), and $T$ is not to be confused with the tensor current insertion. 
The key parameters that we want to determine are the $J_{00}^{\mathrm{nn}}$ for each current. 

Priors for $J_{00}^{\mathrm{nn}}$ are estimated by dividing the three-point correlation function by the relevant two-point correlators and multiplying by their effective amplitudes (as in~\cite{Parrott:2020vbe,Chakraborty:2021qav}). A broad uncertainty (typically 20-50\%) is then given to this effective amplitude. Other $J_{ij}^{kl}$ priors are listed in Table~\ref{Tab:priorsforfit}.

On each ensemble, using the corrfitter package,~\cite{peter_lepage_2020_3707868,peter_lepage_2020_3715065,peter_lepage_2019_3563090}, we perform a simultaneous fit to all of the two-point and three-point functions for all $a\vec{p}_K$ and $T$ values, selecting $N_{\text{exp}}$ for each lattice spacing such that it gives an acceptable $\chi^2$ and maximises the log(GBF). We use $N_{\text{exp}}=4$ for all ensembles except set 8 where we use $N_{\text{exp}}=5$. In the case of sets 6, 7 and 8, the fits are very large because of the number of heavy masses and twists, as well as the increasing number of timeslices. To handle this, we split them up, fitting each heavy mass sequentially and taking a correlated weighted average of any shared parameters at the end. This is especially necessary in the case of sets 6 and 7, where combining the $H_s\to\eta_s$ data with the $H\to K$ data as described in section~\ref{sec:sim} makes the fits even larger. Tests across the range of $J$, $\vec{p}_K$ and $m_h$ show that this method preserves correlations between $J_{00}^{\mathrm{nn}}$ very well. These correlations are small, typically less than 0.3. 

Since our fits involve a large number of different correlation functions with a finite number of samples there is a bias in the small eigenvalues of the  covariance matrix. We address this by applying an SVD cut to these eigenvalues; see Appendix D of~\cite{Dowdall:2019bea}.  This is a conservative move which increases errors. As discussed above, it also leads to an artificial reduction in $\chi^2$, something which we check for by introducing SVD noise, again using corrfitter (see documentation for further details~\cite{peter_lepage_2020_3707868,peter_lepage_2020_3715065,peter_lepage_2019_3563090}).

We check stability of our fitted results for the ground-state parameters to a variety of changes to the fit. An illustration of such tests is given in Figure~\ref{fig:corrstab} for set 8, showing the results for the ground-state to ground-state tensor current matrix element at one twist value at one heavy quark mass along with the ground-state $\hat{H}$ meson mass at a different heavy quark mass and the ground-state $K$ meson energy for a different twist (thus showing a broad range of results). A stability plot for a lower mass ($m_c$) on set 5 with the vector current matrix element is given in~\cite{Chakraborty:2021qav}. We check stability against changing the number of excited states included, doubling and halving all of the prior widths, doubling and halving the SVD cut (compared to the recommended cut given by the lsqfit package~\cite{peter_lepage_2020_3707868}) and changing $t_{\mathrm{min}}$. We also show the result of doing a  single fit, rather than a simultaneous fit to multiple correlators. This figure aims to give a representative range of examples on one ensemble; other ensembles were similarly well behaved, showing stable fits in all cases. We also check that the momentum dispersion relation for our $\brachat{K}$ fit results agrees with the twists specified in the lattice calculation. The two should differ by discretisation effects only and this is confirmed in~\cite{Chakraborty:2021qav} which uses the same kaon data on all ensembles as here. We can also infer this from the modest values we find for $c_2$ and $d_2$ from Equation~\eqref{Eq:disp} in all cases. 

Our fit parameters $J^{\mathrm{nn}}_{00}$ are converted into matrix elements for the corresponding lattice currents according to
\begin{equation}\label{Eq:fitnormalisation}
  \bra{K}J_{\mathrm{latt}}\ket{\brachat{H}}=2Z_{\mathrm{disc}}\sqrt{M_{H}E_{K}}J^{\mathrm{nn}}_{00}.
\end{equation}
These matrix elements can then be converted into values for the form factors using Eqs.~\eqref{Eq:vec},~\eqref{Eq:sca} and~\eqref{Eq:ten}. We have included a factor $Z_{\mathrm{disc}}$ to account for small ($\mathcal{O}(am_h)^4$) tree-level discretisation effects. Values for $Z_{\mathrm{disc}}$ are given in Table~\ref{tab:renorm}. 
We always use the mass of the Goldstone $H$ pseudoscalar for the conversion as the non-Goldstone mass is the same in the continuum limit. The difference is a small discretisation effect, less than 0.1\% in~\cite{Chakraborty:2021qav}, which is accounted for in our extrapolation to the physical point (Section~\ref{sec:eval}).

Numerical results for the left-hand-side of Eq.~\eqref{Eq:fitnormalisation} on each of our ensembles are summarised in Tables~\ref{tab:fitresults2},~\ref{tab:fitresults3} and~\ref{tab:fitresults4} in Appendix~\ref{sec:corr-results}. The vector current results must be multiplied by values of $Z_V$ from Table~\ref{tab:renorm} and the tensor current results by values of $Z_T$ from Table~\ref{tab:ensembles} before values for the form factors can be obtained. The form factors values are also given in the Tables in Appendix~\ref{sec:corr-results}.

\subsection{Extrapolating form factors using a modified $z$ expansion}\label{sec:eval}
Once we have our form factors over a range of $q^2$ values and on all ensembles, we perform a fit in $q^2$ space, heavy mass, light quark mass and lattice spacing. We can then evaluate our form factors at the physical quark masses, and zero lattice spacing, at any heavy-light meson mass from the physical $D$ mass to the physical $B$ mass. Following the method successfully employed in~\cite{McLean:2019qcx,Parrott:2020vbe,Chakraborty:2021qav}, we fit the form factors on the lattice using the Bourreley-Caprini-Lellouch (BCL) parameterisation~\cite{Bourrely:2008za},  
\begin{equation}\label{Eq:zexpansion}
  \begin{split}
    f_0(q^2)&=\frac{\mathcal{L}}{1-\frac{q^2}{M^2_{H_{s0}^{*}}}}\sum_{n=0}^{N-1}a_n^0z^n\\
    f_+(q^2)&=\frac{\mathcal{L}}{1-\frac{q^2}{M^2_{H_{s}^{*}}}}\sum_{n=0}^{N-1}a_n^+\Big(z^n-\frac{n}{N}(-1)^{n-N}z^N\Big)\\
    f_T(q^2)&=\frac{\mathcal{L}}{1-\frac{q^2}{M^2_{H_{s}^{*}}}}\sum_{n=0}^{N-1}a_n^T\Big(z^n-\frac{n}{N}(-1)^{n-N}z^N\Big).
  \end{split}
\end{equation}
This uses a mapping of $q^2$ to $z$, so that the physical $q^2$ range $0 \leq q^2 \leq (M_H-M_K)^2$ is mapped to a region within the unit circle in $z$:
\begin{equation}\label{eq:zofq}
  z(q^2,t_0)=\frac{\sqrt{t_+-q^2}-\sqrt{t_+-t_0}}{\sqrt{t_+-q^2}+\sqrt{t_+-t_0}}.
\end{equation}
$t_{+}=(M_{H} + M_{K})^2$ is the beginning of a branch cut in the complex $t=q^2$ plane corresponding to $HK$ production in the crossed channel. We choose to take $t_0=0$, which permits a simple enforcement of the kinematic constraint in Eq.~\eqref{eq:f0=fp} as it means $z(q^2=0)=0$. Fit results were compared for different values of $t_0$ in~\cite{Chakraborty:2021qav} and good agreement was found. 

The first term in the fit forms of Eq.~\eqref{Eq:zexpansion} removes poles in the form factor that appear from production of heavy-strange mesons with squared masses below $t_+$ (but above $t_-$).  These mesons are the scalar $H_{s0}^*$ and vector $H_s^*$ states. In our fits we need to use a mass for these mesons that is simply related to masses that we have measured in our calculation. We take $M_{H_{s0}^*}$ to be $M_{H}+\Delta$ with $\Delta=0.45\text{GeV}$. As discussed in~\cite{Parrott:2020vbe} the exact value used here is unimportant. The value of $\Delta$ is taken from experimental results for the $D$ system; there are no experimental results for the $B$ system but we expect the splitting to be largely independent of $m_h$. The vector mass $M_{H_s^*}$ can be estimated, as in~\cite{McLean:2019qcx,Parrott:2020vbe}, with the PDG~\cite{pdg} values $M^{\text{phys}}_{D^*_s}=2.1122(4)\text{GeV}$, $M^{\text{phys}}_{B^*_s}=5.4158(15)\text{GeV}$. We use 
\begin{equation}\label{eq:vectorpolemass}
  \begin{split}
    &M_{H^*_s}=M_{H}+\frac{M^{\text{phys}}_{D}}{M_{H}}\Delta(D) \\
    &+ \frac{M^{\text{phys}}_{B}}{M_{H}}\Big(\frac{M_{H}-M^{\text{phys}}_{D}}{M^{\text{phys}}_{B}-M^{\text{phys}}_{D}}\Big(\Delta(B)-\frac{M^{\text{phys}}_{D}}{M^{\text{phys}}_{B}}\Delta(D)\Big)\Big),
  \end{split}
\end{equation}
where $\Delta(H)=M^{\text{phys}}_{H^*_s}-M^{\text{phys}}_{H}$. The physical masses used are those for the isospin averages $(K^0+K^{\pm})/2$, $(B^0+B^{\pm})/2$ and $(D^0+D^{\pm})/2$ (all from~\cite{pdg}), corresponding to the fact that our lattice results have $m_u = m_d = m_l$.  We also need to consider isospin breaking effects and we will do this below.

\begin{figure*}

\includegraphics[width=0.90\textwidth]{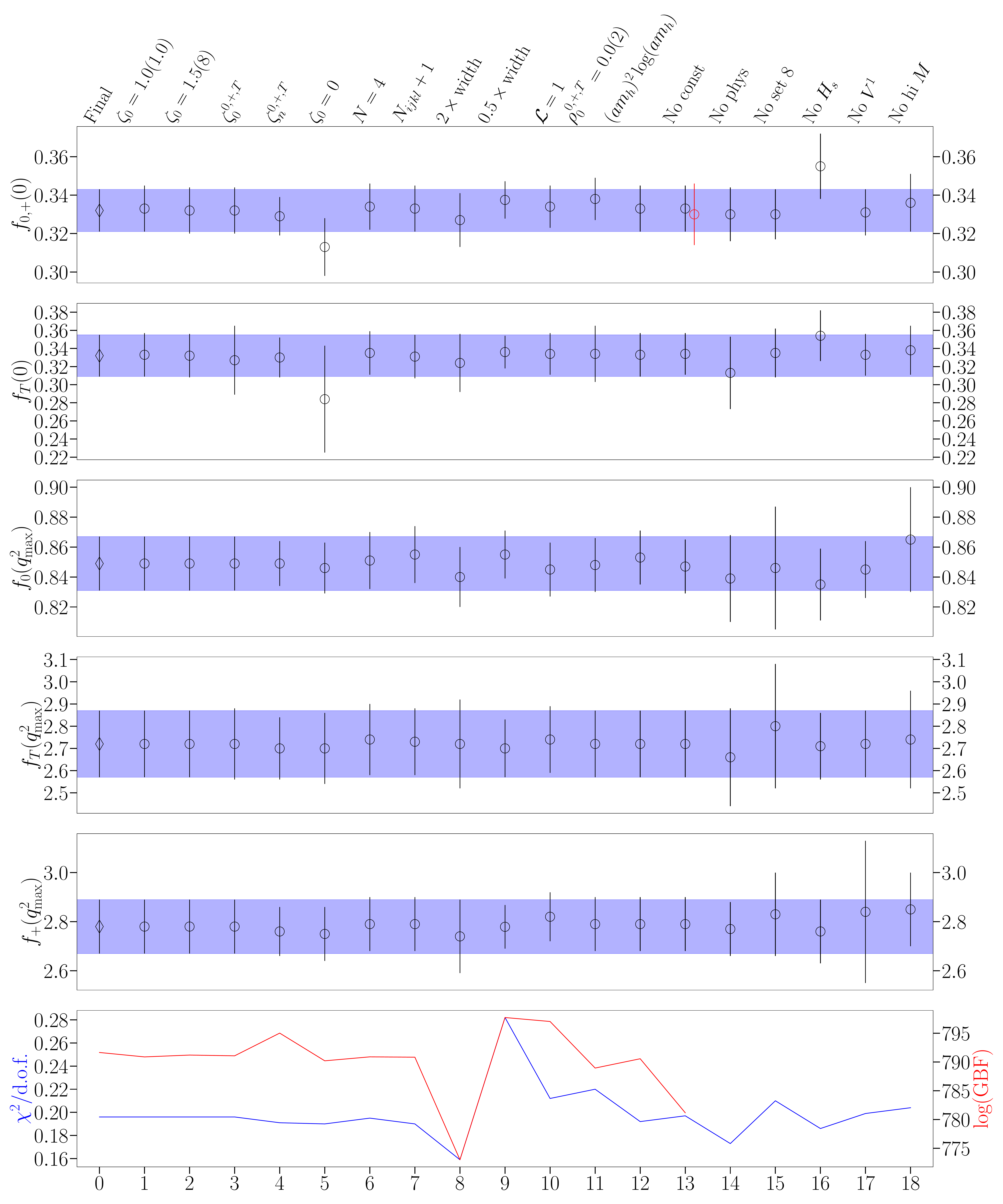}
\caption{Stability tests for the $z$ expansion fit evaluated at the physical $B$ mass. Test 0 is the final result, 1 and 2 take different priors for $\zeta_0$, test 3 allows $\zeta_0$ to vary between the form factors and test 4 (see text) allows for $\zeta_{n\neq0}\neq0$. Test 5 drops the term containing $\zeta$ entirely. Test 6 increases the number of the terms in the $z$-expansion, $N$, by 1 (to 4) and test 7 does the same for each component of $N_{ijkl}$ in each $a_n$ coefficient.  Test 8 doubles the width of $\zeta_n$ and all $d$ and $\rho$ priors, and 9 halves them. Test 10 removes the chiral logarithm term by setting $\mathcal{L}=1$, and 11 tightens the prior on the $\rho$ coefficients considerably. Test 12 allows for logarithmic heavy mass dependence $(am_h)^2\log(am_h)$ in the fit. Test 13 removes the $f_0(0)=f_+(0)$ constraint; in this case the black point is $f_0(0)$ and the red is $f_+(0)$. Tests 14, 15 and 16 remove all the lattices with physical light masses, all of set 8 data, and results with $m_l=m_s$ respectively. Test 17 removes the spatial vector data, and 18 removes the largest mass from all ensembles with multiple masses. The $\chi^2$ per degree of freedom and log(GBF) value for each test are shown in the bottom pane in blue and red respectively. For the latter tests, data is removed from the fit, resulting in a lower log(GBF) which is not comparable with others and so not displayed. As in our correlator fits, $\chi^2$ values are artificially lowered by our SVD cut and priors so are only meaningful relatively. $\chi^2$ values for tests 7 and 8, which change widths on many priors, are thus not directly comparable with other tests. No SVD cut is required, and our final fit has a $\chi^2$/d.o.f. of 0.3 when prior noise is included.}
\label{fig:extrapstab}
\end{figure*}
\begin{table*}
  \caption{Values of fit coefficients $a_n^{0,+,T}$, pole masses, and the $\mathcal{L}$ term with correlation matrix below, evaluated at the physical point and the $B$ mass. Note that $a_0^+=a_0^0$. Masses are in GeV. The pole masses and $\mathcal{L}$ are very slightly correlated due to the way the fit function is constructed. These correlations are too small to have any meaningful effect on the fit, but we include them for completeness. See Appendix~\ref{sec:reconstruct} for details of reconstructing our results.}
  \begin{center} 
    \begin{tabular}{c c c c c c c c c c c}
      \hline
      $a_0^{0/+}$&$a_1^0$&$a_2^0$&$a_1^+$&$a_2^+$&$a_0^T$&$a_1^T$&$a_2^T$&$M^{\text{phys}}_{B^*_{s0}}$& $M^{\text{phys}}_{B^*_s}$ & $\mathcal{L}$\\ [0.5ex]
      0.2545(90)&0.210(76)&0.02(17)&-0.71(14)&0.32(59)&0.255(18)&-0.66(23)&0.36(84)&5.729495(85)&5.4158(15)&1.304(10)\\ [1ex]
      \hline 

      1.00000&0.80619&0.56441&0.30543&0.04776&0.42939&0.19136&0.06240&-0.00032&-0.00197&-0.19815\\ [0.5ex]
      &1.00000&0.91180&0.35256&0.06186&0.31091&0.16899&0.05677&0.00006&-0.00250&0.02839\\ [0.5ex]
      &&1.00000&0.28531&0.08655&0.18297&0.09938&0.04827&0.00005&-0.00181&0.03245\\ [0.5ex]
      &&&1.00000&0.84649&0.06813&0.09633&0.05829&0.00074&-0.01316&0.09126\\ [0.5ex]
      &&&&1.00000&-0.02470&0.02366&0.04442&-0.00054&0.00963&0.00353\\ [0.5ex]
      &&&&&1.00000&0.59841&0.32316&-0.00030&0.00167&-0.11487\\ [0.5ex]
      &&&&&&1.00000&0.85349&0.00032&-0.00574&0.04788\\ [0.5ex]
      &&&&&&&1.00000&-0.00046&0.00825&0.00184\\ [0.5ex]
      &&&&&&&&1.00000&0.00003&-0.00003\\ [0.5ex]
      &&&&&&&&&1.00000&0.00052\\ [0.5ex]
      &&&&&&&&&&1.00000\\ [0.5ex]
      \hline
    \end{tabular}
  \end{center}
  \label{tab:ancoefficients}
\end{table*}

The form factor, with sub-threshold poles removed, can be expanded as an order $N$ polynomial in $z$, where $z<1$ for the physical region. On the lattice the coefficients of $z^n$ in Eq.~\eqref{Eq:zexpansion} contain discretisation effects, which appear as powers of the squared lattice spacing for the HISQ action. Since we are fitting results for multiple values of the heavy quark mass here the coefficients will carry dependence on the heavy quark mass. We must also allow for dependence on the light quark (spectator and sea quark mass) and we do this using a chiral logarithm factor $\mathcal{L}$ as well as analytic terms. For each form factor and each power, $n$, of $z$ we take  
\begin{equation}\label{Eq:an}
\begin{split}
  &a_n^{0,+,T}=\\
  &\Big(\frac{M_D}{M_H}\Big)^{\zeta_n}\Big(1+\rho_n^{0,+,T}\log\Big(\frac{M_{H}}{M_{D}}\Big)\Big)\times(1+\mathcal{N}^{0,+,T}_n)\times\\
  &\sum^{N_{ijkl}-1}_{i,j,k,l=0}d_{ijkln}^{0,+,T}\Big(\frac{\Lambda_{\text{QCD}}}{M_{H}}\Big)^i\Big(\frac{am_h^{\text{val}}}{\pi}\Big)^{2j}\times\\&\hspace{6.0em}\Big(\frac{a\Lambda_{\text{QCD}}}{\pi}\Big)^{2k}(x_{\pi}-x_{\pi}^{\mathrm{phys}})^{l}
\end{split}
\end{equation}
and will discuss the different pieces of this expression below. Note that the coefficients for each power of $z$ are independent in our fit.

\subsubsection{Discretisation Effects}
\label{sec:disc}
Discretisation effects are accounted for in two ways in Eq.~\eqref{Eq:an}. We allow for discretisation effects that vary with the heavy quark mass through the terms in $am_h$ with power $2j$. The size of these terms will vary between results for different $m_h$ on a given ensemble. Discretisation effects that do not vary with heavy quark mass but instead are set by some other scale (for example associated with the $K$ mesons) are allowed for in the powers of $a\Lambda_{\text{QCD}}$. These terms will be the same for all heavy quark masses on a given ensemble. We take $\Lambda_{\text{QCD}}$ = 0.5 GeV.

We also consider the possibility of logarithmic cutoff effects~\cite{Husung:2019ytz} via the addition of an $(am_h)^2 \log(am_h)$ term, discussed in Section~\ref{sec:stability}.

\subsubsection{Dependence on heavy quark mass} 
\label{sec:heavyqmass}
We include several terms in Eq.~\eqref{Eq:an} to model the physical dependence of the form factors on heavy quark mass, using as a proxy for this the heavy-light meson mass, $M_H$. This dependence connects the form factors for $D \to K$ to those for $B\to K$ and we can use insights from Heavy Quark Effective Theory (HQET) to suggest a functional form for it. We take a power series in inverse powers of $M_H$ (with power $i$ and $\Lambda_{\mathrm{QCD}}$ as above) multiplying a prefactor $(M_D/M_H)^{\zeta_n}$, with fitted power $\zeta_n$, and a logarithmic term. 
The $(M_D/M_H)^{\zeta_n}$ term models behaviour predicted by Large Energy Effective Theory (LEET)~\cite{Charles:1998dr}. The LEET expectation is for all form factors for a specific heavy to light transition to exhibit common $\backsim M_H^{-3/2}$ behaviour in the region of $q^2=0$ (where the light meson energy is close to $M_H/2$). 
This behaviour was observed in lattice QCD results for the closely related $B_s\to\eta_s$ decay~\cite{Parrott:2020vbe}, with an $M_H$ power between $-1.5$ and $-1$ towards $q^2=0$. In that case the behaviour was modelled with a $\log(M_{D_s}/M_{H_s})$ term multiplied by a series in inverse powers of $M_H$. Here we allow for this behaviour explicitly. 

Because we have taken $t_0=0$, the form factors at $q^2=0$ are set by the $z^0$ terms in the $z$-expansion. We therefore take a prior $P[\zeta_0]=1.5(5)$ as a common prior for the $a_0$ coefficients but set $\zeta_{n\neq0}=0$ for the other $a_n$. $M_K/M_H$ corrections to LEET can be accounted for in the form factor dependent $(\Lambda_{\text{QCD}}/M_{H})^i$ terms in our fit, as $M_K\approx\Lambda_{\mathrm{QCD}}$. We find that including this term in our fit increases log(GBF), reduces uncertainty at $q^2=0$, particularly for $f_T$, and returns a posterior of $\zeta_0=1.43(12)$. Allowing a broader prior $P[\zeta_0]=1.0(1.0)$ returns a posterior consistent with 1.5 ($1.42(12)$) and does not change the form factor result. Allowing $\zeta_0$ to vary between form factors  simply increases the uncertainty on $f_T(0)$, whilst leaving the central values unchanged. These tests confirm that our fit is not overly constrained by $\zeta$ and is flexible with regard to $M_H$ dependence. They will be discussed further in Section~\ref{sec:stability}, along with a test allowing $\zeta^{0,+,T}_{n\neq0}\neq0$.

For both the $n=0$ and the $n\neq 0$ coefficients we include the logarithmic term in Eq.~\eqref{Eq:an}, with priors on $\rho_n$ of 0.0(1.0). This term is motivated by the matching of HQET to QCD, as in~\cite{McLean:2019qcx,Parrott:2020vbe}. For $n=0$ this effectively allows for different form factors to have different powers $\zeta$ as well as allowing for sub-leading $M_H$ dependence from LEET~\cite{Charles:1998dr}. For $n\neq 0$ this term allows for an adjustable pre-factor non-integer power of $M_H$ for different dependence on $M_H$ in different regions of the $q^2$ range. The heavy mass dependence of the continuum form factors will be discussed below in Section~\ref{sec:connecting}.  

\subsubsection{Dependence on spectator quark mass}
\label{sec:spectq}
The dependence of the form factors on spectator quark mass is also a physical effect which connects $B \to K$ form factors (with a light spectator quark) smoothly to those for $B_c \to D_s$~\cite{Cooper:2021ofu} (with a charm spectator quark). We will discuss this comparison in Section~\ref{sec:results}. Here we include spectator quark masses varying from the physical value of $m_l$ up to $m_s$ (the latter corresponding to $B_s \rightarrow \eta_s$ form factors) in our dataset and aim to describe them all with our functional dependence on the spectator quark mass. This region of spectator masses is amenable to chiral perturbation theory~\cite{Bijnens:2010jg} and we use this to fix the chiral logarithm term, $\mathcal{L}$ in Eq.~\eqref{Eq:an}. We also include analytic terms to be discussed below. 
$\mathcal{L}$ takes the form
\begin{equation}\label{Eq:L}
  \begin{split}
    \mathcal{L} &= 1 - \frac{9g^2}{8}x_{\pi}\Big(\log x_{\pi}+\delta_{FV}\Big)\\
    &- \Big(\frac{1}{2}+\frac{3g^2}{4}\Big)x_K\log x_K -\Big(\frac{1}{6}+\frac{g^2}{8}\Big)x_{\eta}\log x_{\eta},
  \end{split}
\end{equation}
where $x_{\mathcal{M}}=\frac{M^2_{\mathcal{M}}}{(4\pi f_{\pi})^2}$ and $g$ is the coupling between $H$, $H^*$ and the light mesons. The form of $\mathcal{L}$ is appropriate for the vector and scalar form factors and, as in~\cite{Bouchard:2014ypa}, we make use of the fact that $f_T$ and $f_+$ in HQET are the same up to $\mathcal{O}(1/M_H)$ terms to use the same $\mathcal{L}$ for the tensor form factor. Any corrections to this are easily absorbed by our HQET expansion. 
In fact $\mathcal{L}$ does not have a big impact on our fit and we find no appreciable difference to the fit if we set $\mathcal{L}=1$ (see Section~\ref{sec:stability}).

$x_{\pi}$ in Eq.~\eqref{Eq:L} is constructed from the meson mass for a pseudoscalar meson made from the spectator quarks. This corresponds to the $\pi$ meson for the $B \to K$ case (albeit with an unphysically heavy light quark on some ensembles) but an $\eta_s$ meson in the $B_s \rightarrow \eta_s$ case. Likewise $x_K$ corresponds to a `$K$' meson constructed from a strange quark and a spectator quark. The value of $M_{\eta}$ appearing in $x_{\eta}$ is given by $M^2_{\eta}=(M^2_{\pi}+2M^2_{\eta_s})/3$. Since not all of these meson masses are available in our calculation we use leading-order chiral perturbation theory to rescale meson masses in proportion to the masses of the quarks they contain. Taking the ratio of $4\pi f_{\pi}$ to $M_{\eta_s}^{\text{phys}}$ = 0.6885(22) GeV~\cite{Dowdall:2013rya}, we use a proxy for $x_{\pi}$ of the form
\begin{equation}\label{eq:xpi}
x_{\pi}\approx2\frac{m_{\mathrm{spectator}}}{5.63m_s},
\end{equation}
where the factor of 2 accounts for the definition of $f_{\pi}^2$ in~\cite{Bijnens:2010jg}. $x_K$ and $x_{\eta}$ are constructed in an analogous way. The finite volume correction, $\delta_{FV}$, adjusts the $\pi$ chiral logarithm (Eq. 47 of~\cite{Bernard:2001yj}), and we include an error of $(0.7\%)^2$ to account for higher order terms.
 
In order to capture the heavy mass dependence of $g$, we take
\begin{equation}\label{Eq:g}
  g(M_H) = g_{\infty}+C_1\frac{\Lambda_{\text{QCD}}}{M_H} + C_2\frac{\Lambda_{\text{QCD}}^2}{M_H^2},
\end{equation}
with $g_{\infty}=0.48(11)$~\cite{Abada:2003un}, $g(M_D)=0.570(6)$~\cite{Lees:2013uxa} and $g(M_B)=0.500(33)$, an average of the values in~\cite{Flynn:2015xna,Detmold:2011bp,Bernardoni:2014kla}. Priors $P[C_1]=0.5(1.0)$ and $P[C_2]=0.0(3.0)$ are broad and based on a trial fit to just the $g$ data points given above. Our final fit has a slightly tighter value for $g_{\infty}$ giving posterior $g_{\infty}=0.457(56)$ with coefficients $C_1=0.73(62)$ and $C_2=-1.2(1.7)$.  

As well as the chiral logarithm term $\mathcal{L}$ that is common to all terms in the $z$-expansion, we include analytic terms in the spectator quark mass that can vary for different form factors and with the power of $z$, $n$. These appear through powers of $(x_{\pi}-x_{\pi}^{\text{phys}})$ in Eq.~\eqref{Eq:an} (with power $l$). 
$x_{\pi}^{\text{phys}}$ is defined as for $x_{\pi}$ in Eq.~\eqref{eq:xpi} and using 
\begin{equation}\label{mlmsrat}
  \frac{m_s^{\text{phys}}}{m_l^{\text{phys}}}=27.18(10)
\end{equation}
from~\cite{Bazavov:2017lyh}. 

We will quote our final form factors at the physical value of $m_l$ i.e. at the average of the physical $u$ and $d$ quark masses. We will discuss tests of isospin-breaking effects in Section~\ref{sec:stability} below.

\subsubsection{Mistuning effects for other quark masses}
\label{sec:mistune}
We must also account for any possible mistuning of the strange daughter quark and for mistuning of the quark masses in the sea. These are wrapped up in the quark mass-mistuning term, $\mathcal{N}$, in Eq.~\eqref{Eq:an}. The mass of the strange daughter quark is always the valence $s$ quark mass, listed in Table~\ref{tab:ensembles}. We take
\begin{equation}
\begin{split}
  \mathcal{N}_n^{0,+,T}&={c_{s,n}^{\text{val},0,+,T}\delta_s^{\text{val}}+c_{s,n}^{\text{sea},0,+,T}\delta_s^{\text{sea}}+2c_{l,n}^{\text{sea},0,+,T}\delta_l}\\
  &+c_{c,n}^{\text{sea},0,+,T}\delta_c^{\text{sea}}
\end{split}
\end{equation}
For the $s$ and $l$ quarks we use:
\begin{equation}
\delta_q=\frac{m_q-m_q^{\text{tuned}}}{10m_s^{\text{tuned}}}
\end{equation}
 Dividing by $m_s^{\text{tuned}}$ here makes this a physical, scale-independent ratio and the factor of 10 matches this approximately to the usual expansion parameter in chiral perturbation theory. 
As discussed in Section~\ref{sec:sim}, our valence $s$ quark masses are all well-tuned using the physical value of the $\eta_s$ mass to derive $m_s^{\text{tuned}}$~\cite{Dowdall:2013rya, Chakraborty:2014aca}; this is less true for the sea $s$ quarks. We include uncertainties in $m_s^{\text{tuned}}$ by defining it from the $\eta_s$ masses corresponding to our valence $s$ quark masses through
\begin{equation}\label{mstuned}
  m_s^{\text{tuned}}=m_s^{\text{val}}\Bigg(\frac{M_{\eta_s}^{\text{phys}}}{M_{\eta_s}}\Bigg)^2.
\end{equation}
$m_l^{\text{tuned}}$ is then defined from Eq.~\eqref{mlmsrat}. For the sea charm quarks we define 
\begin{equation}
\delta^{\text{sea}}_c=\frac{m_c^{\text{sea}}-m_c^{\text{tuned}}}{m_c^{\text{tuned}}}
\end{equation}
with $m_c^{\text{tuned}}$ fixed from the $\eta_c$ meson mass~\cite{Hatton:2020qhk}. These values, on each ensemble, correspond well with the lowest heavy valence mass that we have used (see Table~\ref{tab:renorm}).

\subsubsection{Prior choices}
\label{sec:priors}
We need to set priors for the parameters that appear in the $a_n$ coefficients of Eq.~\eqref{Eq:an}. As noted in Section~\ref{sec:heavyqmass} we include a parameter for an inverse power of $M_H$ as a prefactor for $n=0$ only, and take the prior for $\zeta_0$ as 1.5(5). For $\rho_n$ and $d_{ijkln}$ we take values of $0.0(1.0)$ in all cases except for terms which are $\mathcal{O}(a^2)$. We know such terms are highly suppressed in the HISQ action because it is $a^2$-improved~\cite{Follana:2006rc}, so we take a reduced width prior of $0.0(0.3)$ for $d_{i10ln}$ and $d_{i01ln}$ terms. Using such priors, we test the fit with different choices of $N_{ijkl}\equiv(N_i,N_j,N_k,N_l)$, and we find that the combination preferred by log(GBF) is $N_{ijkl}=(3,2,2,3)$. Note that the sum over each index, $i$, runs from 0 to $N_i-1$ in Eq.~\eqref{Eq:an}. We show below in Section~\ref{sec:stability} that increasing all of the entries in $N_{ijkl}$ by 1 makes almost no difference to the final results.

We also conduct an Empirical Bayes study in order to confirm that the priors listed above are of the right size. We do this using the facility built into lsqfit~\cite{peter_lepage_2020_3707868}. It works by varying a factor $w$ which multiplies all prior widths (or a subset of them) in order to find the $w$ choice which maximises log(GBF). In our case, we perform two such studies, on the whole set of $0.0(1.0)$ and $0.0(0.3)$ priors respectively. We find that our priors are conservative in both cases, with priors of $0.00(55)$ and $0.000(72)$ giving the optimal log(GBF). Taking these priors results in a log(GBF) increase of $\approx2$, which is not considered to be very significant, so we opt for our original, more conservative priors. The effect of doubling and halving the priors will be shown in Figure~\ref{fig:extrapstab}, discussed in Section~\ref{sec:stability}. 

The prior for the daughter strange quark mistuning parameter in $\mathcal{N}$, $c_{s,n}^{\text{val}}$, is taken as $0.0(1.0)$ for each $n$ and each form factor. This size is based on the variation seen between $B\to K$ and $B\to \pi$ form factors~\cite{Du:2015tda}. We expect smaller effects from sea quark mass mistuning and so take the $c_{s,n}^{\mathrm{sea}}$ and $c_{l,n}^{\mathrm{sea}}$ parameters to have priors of 0.0(0.5) and the $c_{c,n}^{\mathrm{sea}}$ parameters to have prior 0.0(1).

The choice of $t_0=0$ and the use of $x_{\pi}-x_{\pi}^{\mathrm{phys}}$ (which takes value 0 at the physical point) in Eq.~\eqref{Eq:an} makes it easy to apply the constraint that $f_+(0)=f_0(0)$ at the physical point for all heavy masses (Eq.~\eqref{eq:f0=fp}). We achieve this by setting $\rho_0^+=\rho_0^0$ and $d^+_{i0000}=d^0_{i0000}$. We take $N=3$ in Eq.~\eqref{Eq:zexpansion} so that the maximum power of $z$ corresponding to a fit parameter in the $z$-expansion is $z^2$ . We show below in Section~\ref{sec:stability} that increasing $N$ by 1 makes no appreciable difference to the final results.

\subsubsection{Tests of the fits}
\label{sec:stability}

We perform a variety of tests of the stability of our fits and these are summarised in Figure~\ref{fig:extrapstab}. This shows how the final $B\to K$ form factors at each end of the $q^2$ range (0 and $q^2_{\text{max}}$) vary as we change fit choices. Figure~\ref{fig:extrapstab} demonstrates that our preferred fit result is stable against reasonable variations and simultaneously optimises log(GBF) and $\chi^2$/d.o.f.. The only variations with a larger log(GBF) than our preferred fit are those which set $\mathcal{L}=1$ (test 10) and which halve the prior widths on $\zeta_0$ and all $d$ and $\rho$ parameters. However, $\mathcal{L}$ is theoretically motivated and we prefer to keep more conservative priors. Our final fit has an acceptable $\chi^2$/d.o.f. (0.3) when prior noise is included. No SVD cut is used in the fit.   

Tests 1--5 address variations of the power $\zeta_0$ of $M_H$ in the pre-factor term for the heavy-quark expansion in Eq.~\eqref{Eq:an}. Tests 1 and 2 change the prior for $\zeta_0$, whilst test 3 allows $\zeta_0$ to vary between form factors. Test 4 examines the effect of introducing $\zeta$ away from $q^2=0$. We take the usual correlated prior $\mathcal{P}[\zeta_0]=1.5(5)$, but allow uncorrelated priors for each of the form factors for $n\neq0$: $\mathcal{P}[\zeta^0_{n\neq0}]=1.5(5)$ and $\mathcal{P}[\zeta^{+,T}_{n\neq0}]=0.5(5)$. This allows approximately for the expected scaling at $q^2_{\mathrm{max}}$ from HQET~\cite{Hill:2005ju}, allowing for the single power of $M_H$ from the pole term. This is discussed in more detail below (Section~\ref{sec:connecting}). The scaling is not perfectly accounted for, as we are working in $z$ space, but we find that the output of the fit agrees very well with our preferred result, and indeed has smaller uncertainties, smaller $\chi^2/\mathrm{d.o.f}$ and larger log(GBF). We do not wish to constrain our fit so tightly, however, so we take the more conservative approach of only using $\zeta_0$. Test 5 drops this $\zeta$ term entirely.

Test 6 adds additional $z^3$ terms to the $z$-expansion and test 7 adds additional discretisation, heavy quark expansion and $m_l$ terms to each $a_n$. These do not change the fit output in any appreciable way. Tests 8 and 9 double and halve, respectively, the prior widths on $\zeta_0$ and all $d$ and $\rho$ priors. Again these make little difference, but we note that the log(GBF) grows for the case of smaller widths, indicating that our choice is conservative, as discussed in Section~\ref{sec:priors}. Test 10 drops the chiral logarithm term, $\mathcal{L}$ and we see little difference in this case as noted in Section~\ref{sec:spectq}. The analytic terms included in the $a_n$ are then capable of  modelling the dependence that we see for the range of spectator quark masses that we have.

In test 12, we allow for logarithmic terms $(am_h)^2\log(am_h)$ in the heavy mass~\cite{Husung:2019ytz}. We do this by including a term $(1+\omega^{0,+,T}_n\log(am_h))$ in Equation~\eqref{Eq:an} when $j=1$ with prior $\mathcal{P}[\omega^{0,+,T}_n] = 0(1)$. We find that the posteriors returned are consistent with zero, and the final form factors are not changed significantly.

With test 13 we show that removing the constraint of Eq.~\eqref{eq:f0=fp} also has little effect beyond a slightly larger uncertainty for $f_+$ at $q^2=0$. 

The tests from 14 upwards miss out various sets of data from the fit and some of these have a sizeable impact on the uncertainties. Dropping the results with the highest heavy quark mass from each ensemble (test 18), unsurprisingly increases the uncertainties considerably at $q^2_{\text{max}}$ since these results are the ones closest to the $b$ quark (and therefore closest to the physical $q^2_{\text{max}}$ for $B \to K$. This is also reflected in the contribution to the error budget from the HQET part of the expansion of the $a_n$. This will be discussed in Section~\ref{sec:results}. 

Dropping all the results from our finest lattice, set 8, also has a significant effect on uncertainties (test 15) because this set allows us to get closest to the $b$ mass. 
The gluon field ensembles on set 8 show only a slow variation of topological charge in Monte Carlo time. This could introduce a bias on this ensemble if the quantities we are studying are sensitive to topological charge. A study was made of this effect for decay constants in~\cite{Bernard:2017npd} and it was found that the impact of `topology freezing' was 1\% for $f_K/f_{\pi}$ on set 8 and 1\% for $f_D$. To allow for these effects, we therefore include an additional (correlated) uncertainty of $1\%$ on all set 8 results in our final fit (this is already incorporated in test 0 of Figure~\ref{fig:extrapstab}). We do this via a factor with prior $1.00(1)$, which returns a fit posterior of $0.993(5)$, showing that our set 8 results are consistent with those on our other sets. 

Test 14 drops the data with physical $m_l$ from the fit; in that case the fit uses the results with $m_l=m_s/5$ and $m_l=m_s$ to arrive at the physical light quark mass. This gives very similar central values but somewhat larger uncertainties. Test 16 instead drops the $m_l=m_s$ results; this has less impact on the uncertainties but shifts the central values at $q^2=0$ by about 1$\sigma$. 

\begin{figure}
\includegraphics[width=0.48\textwidth]{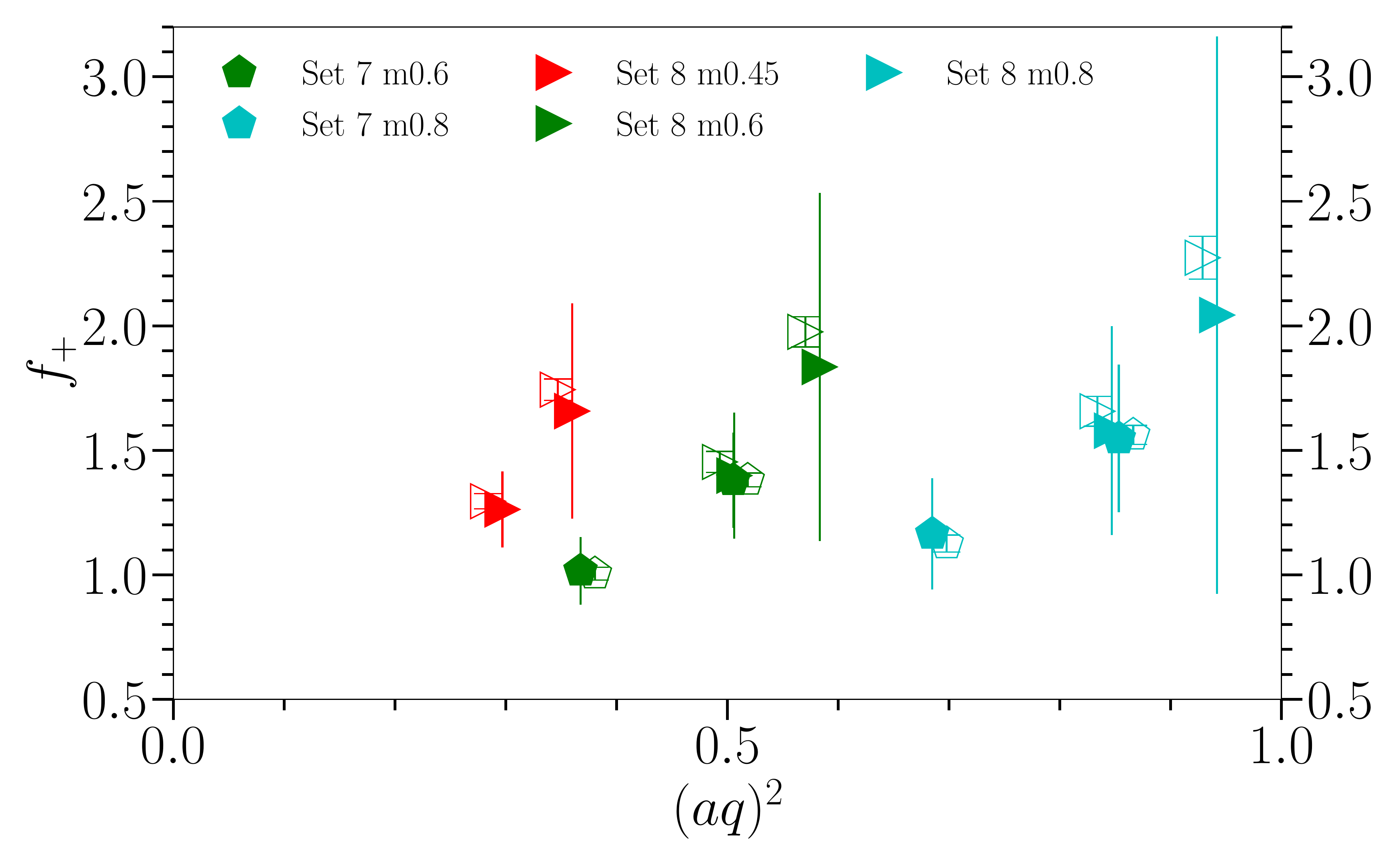}
\includegraphics[width=0.48\textwidth]{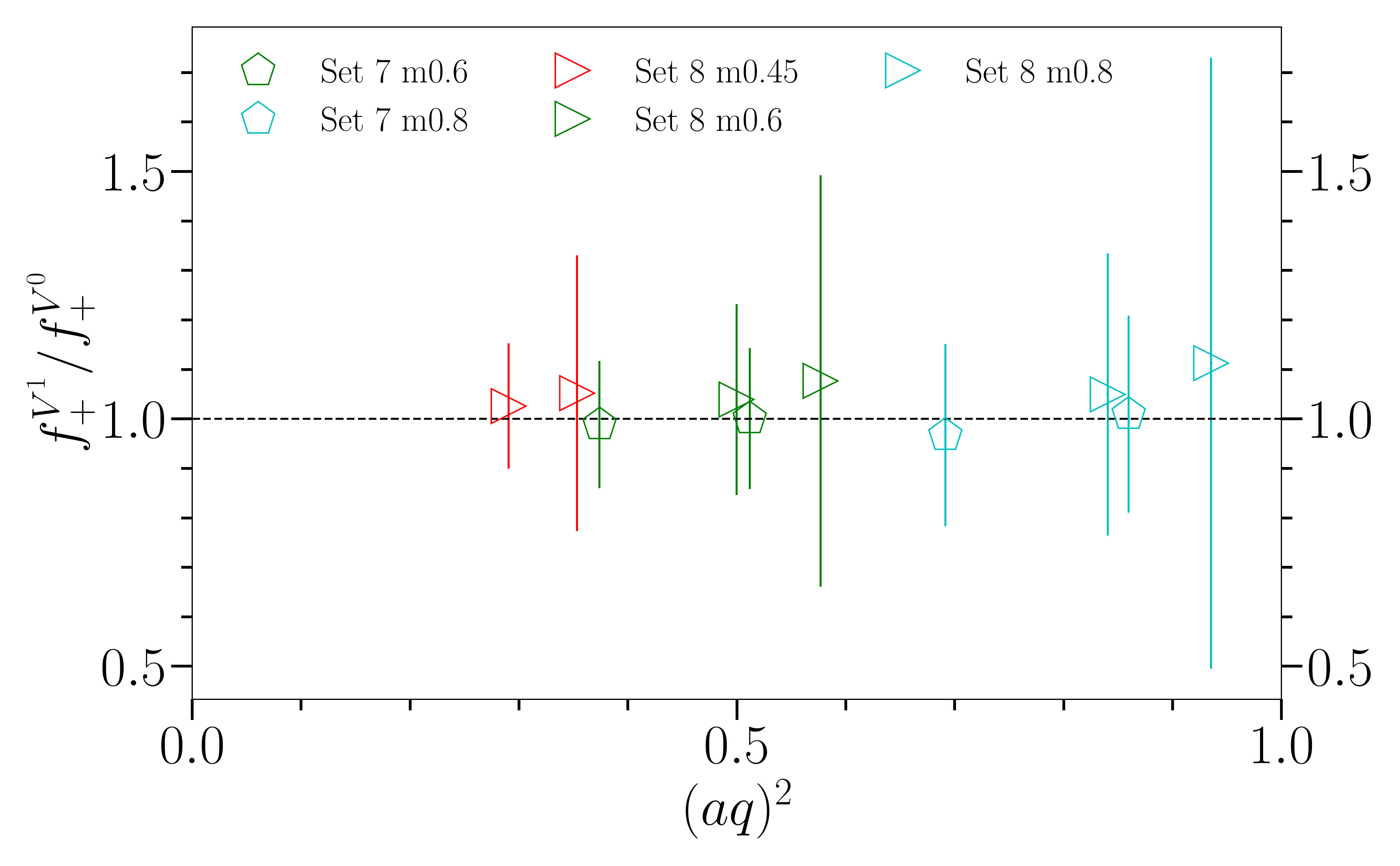}
\caption{Upper plot: A comparison of values and their statistical errors for the vector form factor derived from matrix elements for the spatial and temporal vector currents on ensembles where both are available. The filled symbols are the temporal vector results and the open symbols the spatial vector results. We have offset spatial vector results slightly on the $q^2$-axis for clarity. 
Lower plot: The ratio of the $f_+$ values for the spatial and temporal vector cases. We see no evidence of any differences between them (within our uncertainties) that would indicate discretisation effects.}
\label{fig:fpV0V1diff}
\end{figure}

Test 17 looks at the inclusion of results from the spatial vector current as well as the temporal vector current. As expected from the discussion in Section~\ref{sec:formfactors}, the use of the spatial vector current  improves the vector form factor at large values of $q^2$. Dropping these results, as in test 17, increases uncertainties on the vector form factor value at $q^2_{\text{max}}$ by a factor of 2.5. Figure~\ref{fig:fpV0V1diff} (top plot) shows the results for the form factor from spatial and temporal vector currents on sets 7 and 8 where we have both correlation functions (see Table~\ref{tab:fitresults4} in Appendix~\ref{sec:corr-results}). The plot shows the good agreement between the two sets of results and the considerably smaller uncertainties for the spatial vector current case, in agreement with what was seen in~\cite{Cooper:2021ofu}. 

As discussed in Section~\ref{sec:formfactors}, there is the possibility for $f_+^{V^0}$ and $f_+^{V^1}$ to differ by $q^2$-dependent discretisation effects. To address this explicitly we plot the ratio $f^{V^1}_+/f_+^{V^0}$ against $(aq)^2$ in the lower plot of Figure~\ref{fig:fpV0V1diff}, taking into account the correlations between the two values from the fits. We see no evidence of discretisation effects at the $\sim$10\% level nor any $q^2$ trend in the results. We include terms in the fit to account for such effects for each ensemble and heavy mass,
\begin{equation}
  f_+^{V^1}(q^2) = \left(1+\mathcal{C}^{a,m_h}\times(aq)^2\right)f_+^{V^0}(q^2),
\end{equation}
where the priors for all $\mathcal{C}$s are $0.0(1)$. We find that our fits do not constrain these coefficients and including such terms makes no difference at all to the results of our fit, in keeping with Figure~\ref{fig:fpV0V1diff}. We include these terms in our final fit (test 0 of Fig.~\ref{fig:extrapstab}) nevertheless. 

As noted above, we will quote our final form factors here at the physical value of $m_l$ i.e. at the average of the physical $u$ and $d$ quark masses and in pure QCD (i.e. neglecting QED effects). The physical processes correspond either to a charged $B$ meson decay with a $u$ spectator quark, or a neutral $B$ meson decay with a $d$ spectator quark, however. 
As a test of isospin-breaking effects we can monitor the change in our results as we change the physical ratio of $m_s/m_l$ (Eq.~\eqref{mlmsrat}) so that it matches that of $m_s/m_u$ or $m_s/m_d$. To do this we take $m_d/m_u\approx 2$~\cite{pdg}. We also switch to using the correct physical $B$ and $K$ masses, as opposed to using the average of the charged and neutral cases. We find that our form factors change by at most 0.5\%, or 0.2$\sigma$.  
Note that this test is in fact an overestimate of strong isospin-breaking effects because it also changes the sea $l$ masses to match either $u$ or $d$ which is not correct; the average of the light sea quark masses should remain $m_l$. No uncertainty is included in the form factors presented here to allow for QED effects or the isospin breaking effect discussed - both of these uncertainties will be addressed in the accompanying phenomenology paper.

\begin{figure*}
  
  \includegraphics[width=0.9\textwidth]{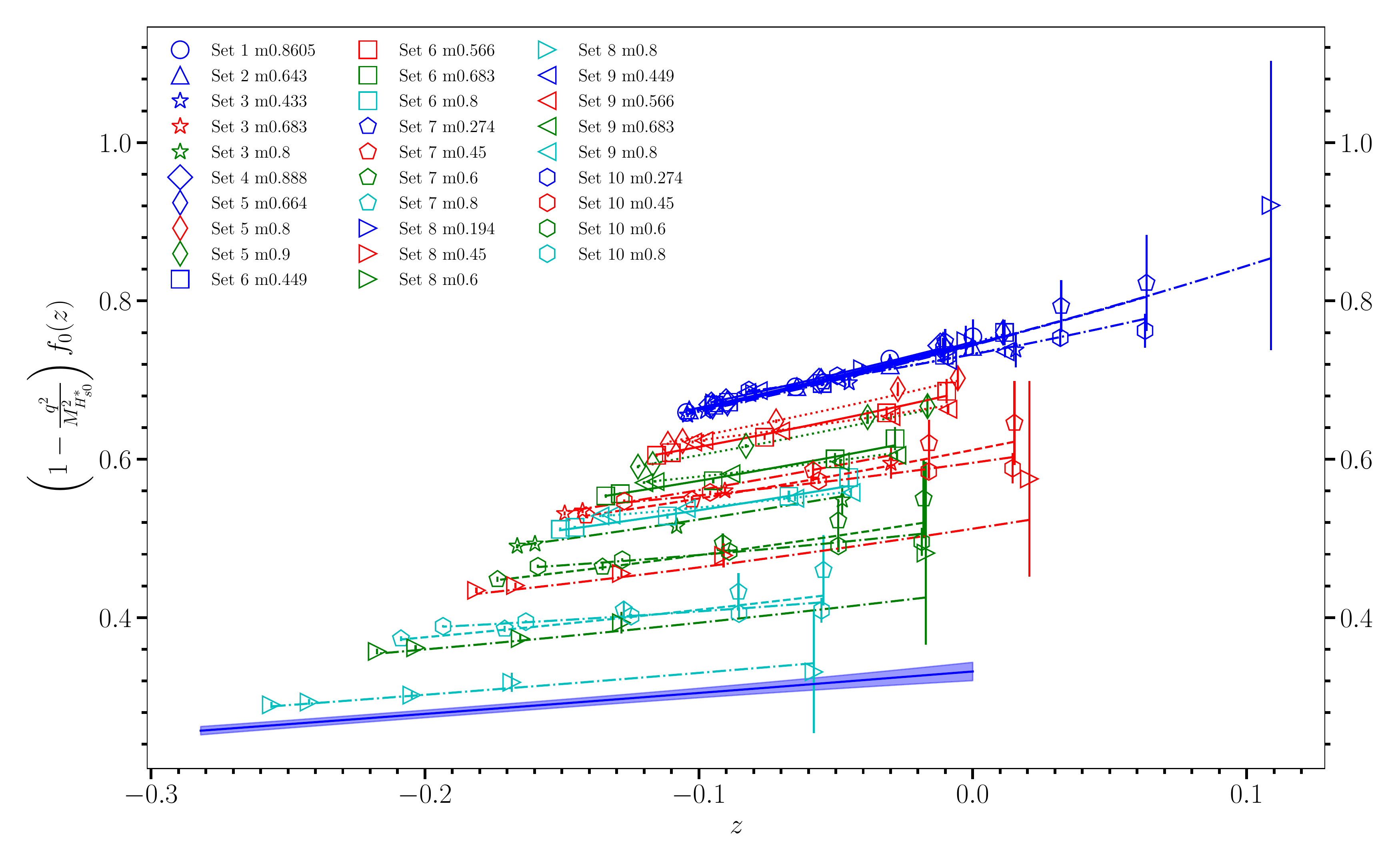}
  \caption{$\Big(1-\frac{q^2}{M^2_{H^*_{s0}}}\Big)f_0(z)$ data points and final result at the physical point (blue band). Data points are labelled by heavy quark mass, where e.g. m0.8 indicates $am_h=0.8$ on that ensemble. Lines between data points of a given heavy mass are the result of the fit evaluated on this ensemble and mass with all lattice artefacts present. Sets 9 and 10 are the $H_s\to\eta_s$ data from sets 1 and 2 in~\cite{Parrott:2020vbe}, which were fitted simultaneously with sets 6 and 7 respectively.}
  \label{Fig:f0nopoleinz}
  \includegraphics[width=0.9\textwidth]{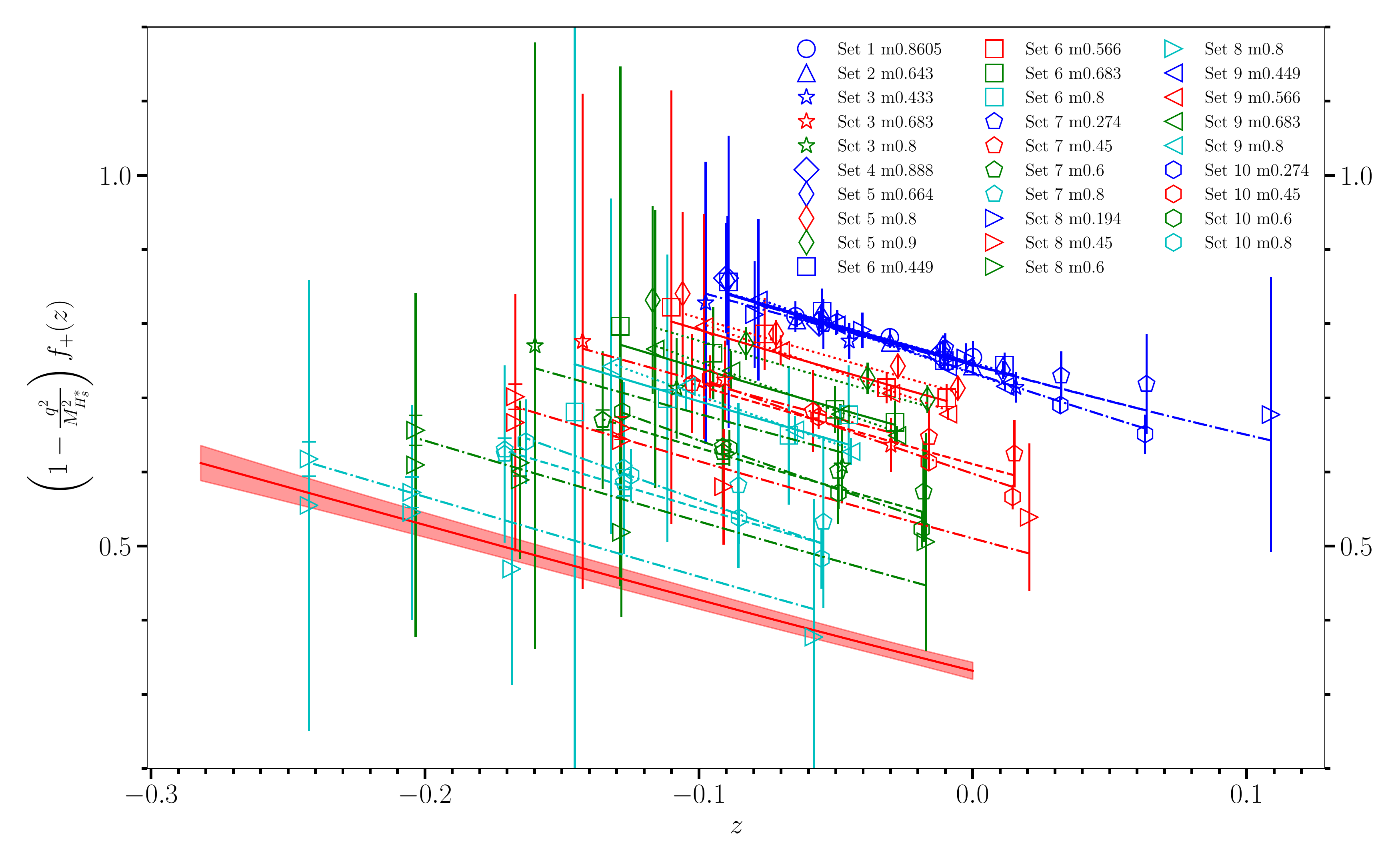}
  \caption{$\Big(1-\frac{q^2}{M^2_{H^*_s}}\Big)f_+(z)$ data points and final result at the physical point (red band). Data points are labelled by heavy quark mass, where e.g. m0.8 indicates $am_h=0.8$ on that ensemble. Lines between data points of a given heavy mass are the result of the fit evaluated on this ensemble and mass with all lattice artefacts present. Sets 9 and 10 are the $H_s\to\eta_s$ data from sets 1 and 2 in~\cite{Parrott:2020vbe}, which were fitted simultaneously with sets 6 and 7 respectively. At large $|z|$ (large $q^2$), data obtained from both temporal and spatial components of $V^{\mu}$ are shown, the latter with end caps specifying the associated uncertainty. As discussed in Section~\ref{sec:latticecalc}, errors for $f_+$ at large $q^2$ are significantly smaller when obtained from spatial vector components.}
  \label{Fig:fplusnopoleinz}
\end{figure*}
\begin{figure*}
  \includegraphics[width=0.9\textwidth]{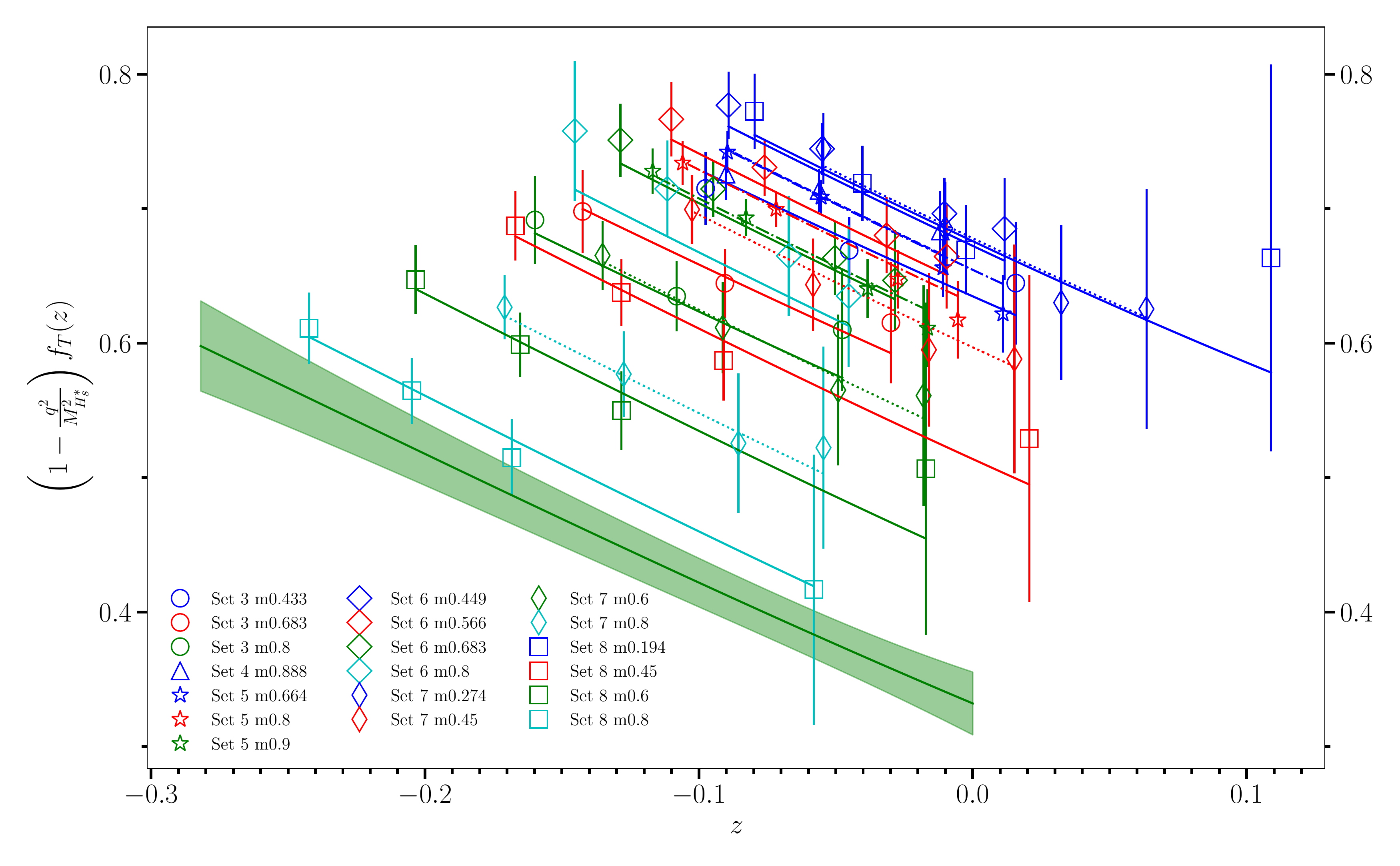}
  \caption{$\Big(1-\frac{q^2}{M^2_{H^*_s}}\Big)f_T(z)$ data points and final result at the physical point (green band). Data points are labelled by heavy quark mass, where e.g. m0.8 indicates $am_h=0.8$ on that ensemble. Lines between data points of a given heavy mass are the result of the fit evaluated on this ensemble and mass with all lattice artefacts present.}
  \label{Fig:fTnopoleinz}
\end{figure*}
\section{Results}\label{sec:results}
\subsection{Evaluating form factors at the physical point}\label{sec:evalphys}
When it comes to evaluating form factors at the physical point and in the continuum limit, we simply need to take physical inputs for values in Equation~\eqref{Eq:zexpansion}. Taking the valence and sea quark masses to their tuned values sets $\mathcal{N}_n^{0,+,T}=0$, and sending the lattice spacing $a\to 0$ means that, for any chosen $M_H$ (in GeV),
\begin{equation}\label{Eq:an_cont}
  \begin{split}
    &a_n^{0,+,T,(\mathrm{cont.})}(M_H)=\\
    &\Big(\frac{M^{\mathrm{phys}}_D}{M_H}\Big)^{\zeta_n}\Big(1+\rho_n^{0,+,T}\log\Big(\frac{M_{H}}{M^{\mathrm{phys}}_{D}}\Big)\Big)\times\\
    &\sum^{N_{i}-1}_{i=0}d_{i000n}^{0,+,T}\Big(\frac{\Lambda_{\text{QCD}}}{M_{H}}\Big)^i,
  \end{split}
\end{equation}
where $\Lambda_{\mathrm{QCD}}=0.5~\mathrm{GeV}$ as usual. These are the values for $a_n$ which are given in Tables~\ref{tab:ancoefficients} and~\ref{tab:DKan} for $M_H=M^{\mathrm{phys}}_B$ and $M_H=M^{\mathrm{phys}}_D$ respectively. As already discussed, our results are for $m_u=m_d=m_l$ so we use the average of the charged and neutral $B$, $K$ and $D$ masses from~\cite{pdg} when required. These masses can be used in Equation~\eqref{eq:zofq} to obtain $z$ from any given $q^2$ and $M_H$. Finally, $\mathcal{L}^{\mathrm{cont.}}(M_H)$ (Equation~\eqref{Eq:L}) is evaluated using $x_{\pi}^{\mathrm{phys}}$, $x_{K}^{\mathrm{phys}}$, $x_{\eta}^{\mathrm{phys}}$, $\delta_{FV}=0$ and evaluating $g$ (Equation~\eqref{Eq:g}) at $M_H$. The resulting $\mathcal{L}(M_H)$ values at $M_H=M^{\mathrm{phys}}_B$ and $M_H=M^{\mathrm{phys}}_D$ are also given in Tables~\ref{tab:ancoefficients} and~\ref{tab:DKan}.

Putting all of this together,
\begin{equation}\label{Eq:zexpansion_cont}
  \begin{split}
    f^{\mathrm{cont.}}_0(q^2,M_H)&=\frac{\mathcal{L}^{\mathrm{cont.}}(M_H)}{1-\frac{q^2}{M^2_{H_{s0}^{*}}}}\sum_{n=0}^{N-1}a_n^{0,(\mathrm{cont.})}(M_H)z^n\\
    f^{\mathrm{cont.}}_+(q^2,M_H)&=\frac{\mathcal{L}^{\mathrm{cont.}}(M_H)}{1-\frac{q^2}{M^2_{H_{s}^{*}}}}\sum_{n=0}^{N-1}a_n^{+,(\mathrm{cont.})}(M_H)\times\\
    &\Big(z^n-\frac{n}{N}(-1)^{n-N}z^N\Big)\\
    f^{\mathrm{cont.}}_T(q^2,M_H)&=\frac{\mathcal{L}^{\mathrm{cont.}}(M_H)}{1-\frac{q^2}{M^2_{H_{s}^{*}}}}\sum_{n=0}^{N-1}a_n^{T,(\mathrm{cont.})}(M_H)\times\\
    &\Big(z^n-\frac{n}{N}(-1)^{n-N}z^N\Big),
  \end{split}
\end{equation}
where the two pole masses are evaluated using $M_{H^*_{s0}}=M_H+\Delta$ and Equation~\eqref{eq:vectorpolemass} as usual, working in GeV and not lattice units. These pole masses are also given in Tables~\ref{tab:ancoefficients} and~\ref{tab:DKan}. We have used the superscript `cont.' here to emphasise that these expressions are valid in the continuum (with tuned quark masses) only, but we drop this superscript in the results tables and numerical results which follow, noting that all results are presented in this limit. For details on loading our results from the supplied python script, see Appendix~\ref{sec:reconstruct}.
\subsection{$B\to K$ form factor results}
\begin{figure}

\includegraphics[width=0.48\textwidth]{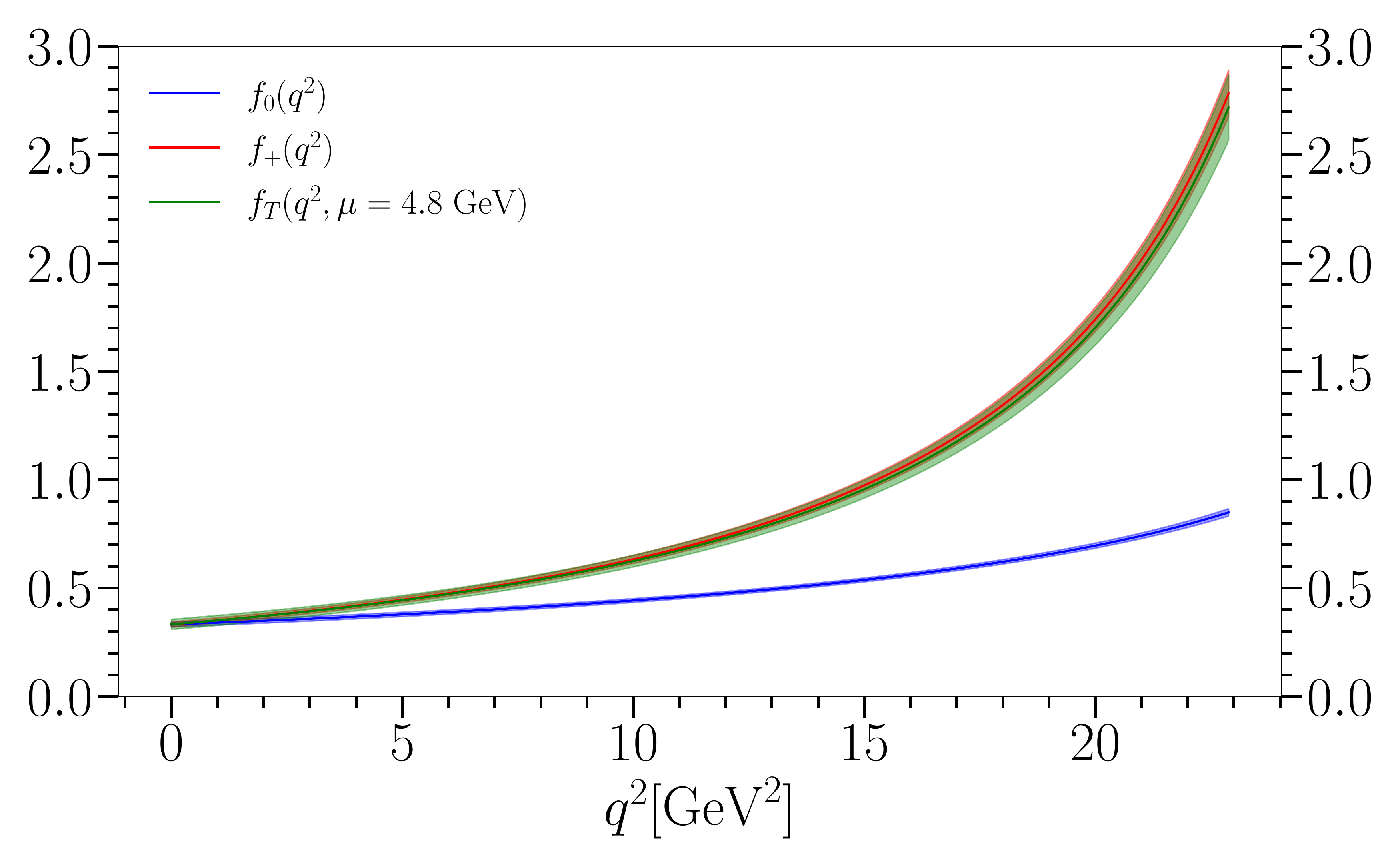}
\caption{Final $B\to K$ form factor results at the physical point across the full $q^2$ range.}
\label{fig:f0fpfTinqsq}
\end{figure}
\begin{figure*}
  
  \includegraphics[width=0.96\textwidth]{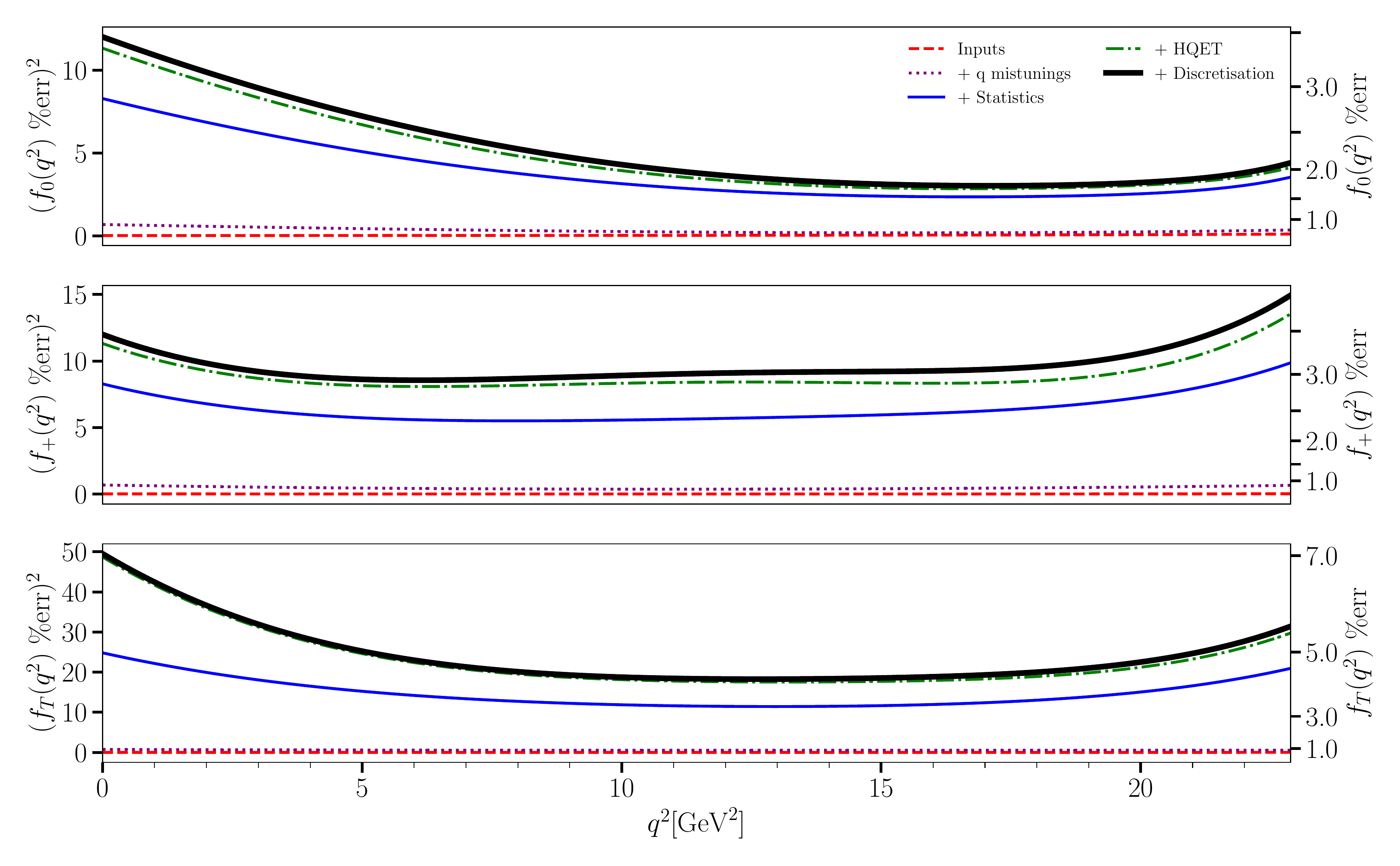}
  \caption{The contributions to the total percentage error (black line) of $B\to K$ form factors $f_0(q^2)$ (top) and $f_+(q^2)$ (middle) and $f_T(q^2)$ (bottom) from different sources, shown as an accumulating error. The red dashed line (`inputs') includes values for parameters, such as masses, taken from the PDG~\cite{pdg} and used in the fit as described above. The purple dotted line (`$q$ mistunings') adds, negligibly, to the inputs the error contribution from the quark mistunings associated with $c$ fit parameters and errors from the light quark chiral extrapolation, whilst the solid blue line (`statistics') further adds the error from our correlator fits. The green dot-dash line (`HQET') includes the contribution from the expansion in the heavy quark mass, and, finally, the thick black line (`Discretisation'), the total error on the form factor, also includes the discretisation errors. In the case of the tensor form factor, the difference here is so small as to obscure the HQET line. The percentage variance adds linearly and the scale for this is given on the left hand axis. The percentage standard deviation, the square root of this, can be read from the scale on the right-hand side.}
  \label{Fig:f0fpfTerr}
\end{figure*}
Figures~\ref{Fig:f0nopoleinz},~\ref{Fig:fplusnopoleinz} and~\ref{Fig:fTnopoleinz} show our lattice results and fit functions in $z$-space. The points plotted correspond to $(1-q^2/M^2)f$ where $(1-q^2/M^2)$ is the pole factor on the right-hand side of Eq.~\eqref{Eq:zexpansion} for each form factor. The figures show results on each ensemble for each value of $am_h$, joined by the line from the fit corresponding to those parameters. The final result in the continuum, at the $B$ mass and physical quark masses is shown by the solid band. We see that the lattice results lie on approximately linear curves in all cases. This is particularly clear for the scalar form factor case in Figure~\ref{Fig:f0nopoleinz}.  This makes for a benign $z$ expansion and justifies our choice of $N=3$, as is also confirmed by the log(GBF) value. Dark blue data points correspond to the charm quark mass on each ensemble. We can see here that, at this mass, discretisation effects are small with very good agreement between data on different ensembles, particularly in the scalar and vector cases. Otherwise we can see data points arranged according to mass, moving towards the $b$ mass, which is close to the $am_h=0.8$ value on our finest ensemble, set 8. This is shown from the proximity of set 8 data to the physical band in the plots. We see that the twist choices on our finest ensemble also give good coverage of the full $z$ range (shown by the physical band curves) at the physical point. 

We present our final scalar, vector and tensor form factors evaluated at the physical $B$ mass, physical quark masses, and in the continuum limit, across the full range of physical $q^2$ values in Figure~\ref{fig:f0fpfTinqsq}. The similarity of $f_+$ and $f_T(\mu=4.8~\mathrm{GeV})$ is very obvious, an assumption that was often used to estimate $f_T$ from $f_+$ before reliable $f_T$ calculations existed. 

\begin{figure}

\includegraphics[width=0.48\textwidth]{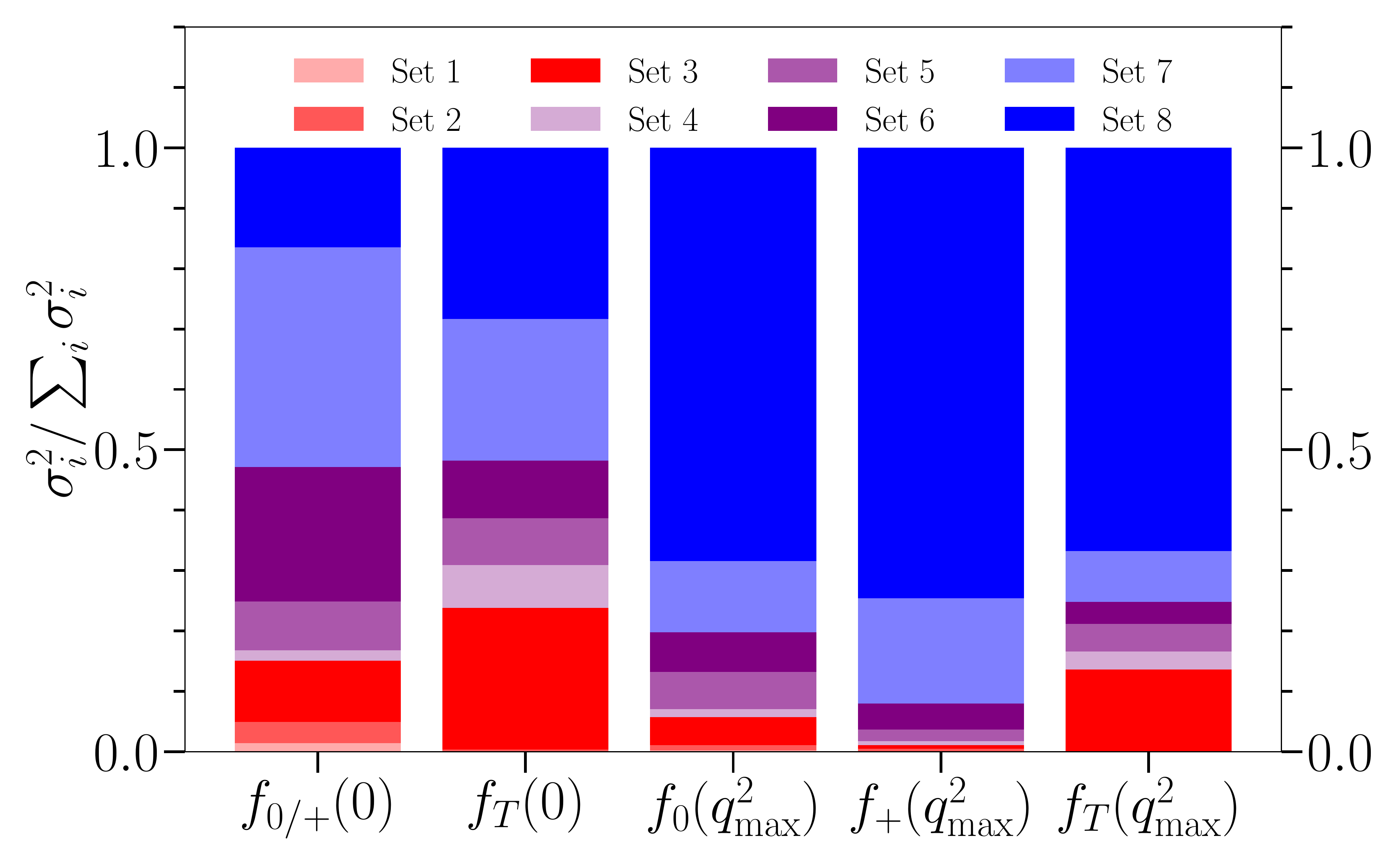}
\caption{Breakdown of the contributions to the statistical uncertainty of the $B\to K$ form factors at their extremes from data on each ensemble. Uncertainty from each ensemble $\sigma_i$ is added in quadrature, normalised by the total uncertainty squared $\sum_i\sigma_i^2$. Sets 6 and 7 include contributions from $H_s\to\eta_s$ data.}
\label{fig:BKenserr}
\end{figure}

A breakdown of the percentage error contributions to each form factor across the $q^2$ range is given in Figure~\ref{Fig:f0fpfTerr}. The largest contribution in all cases is from statistics, followed by the expansion in the heavy mass. All other errors, from quark mistuning (including the chiral logs $\mathcal{L}$ and analytic chiral terms), discretisation effects and input masses are small. As noted above, the contribution of the heavy mass expansion to the error is also apparent in Figure~\ref{fig:extrapstab}.

Further error analysis is displayed in Figure~\ref{fig:BKenserr}, which gives a breakdown of the contributions of each of the ensembles listed in Table~\ref{tab:ensembles} to the statistical uncertainty of each form factor at its extremal values of $q^2$. The contributions are normalised to a total variance of 1 in each case, and we note that sets 6 and 7 include contributions from the additional $H_s\to \eta_s$ data on those ensembles. We see that sets 7 and 8 make the largest contributions to the uncertainties of all form factors across the $q^2$ range, with set 1 making the smallest contribution in all cases. This error could be reduced with better statistics on the superfine and ultrafine ensembles (sets 7 and 8), perhaps also including an additional heavier mass at $am_h=0.9$ on set 8, or with a further, even finer ensemble, at the bottom mass. This would be a numerically expensive, but straightforward, exercise to reduce uncertainty in future.
\subsubsection{Results at $q^2_{\mathrm{max}}$}
\begin{figure}
  \begin{center}
    \includegraphics[width=0.48\textwidth]{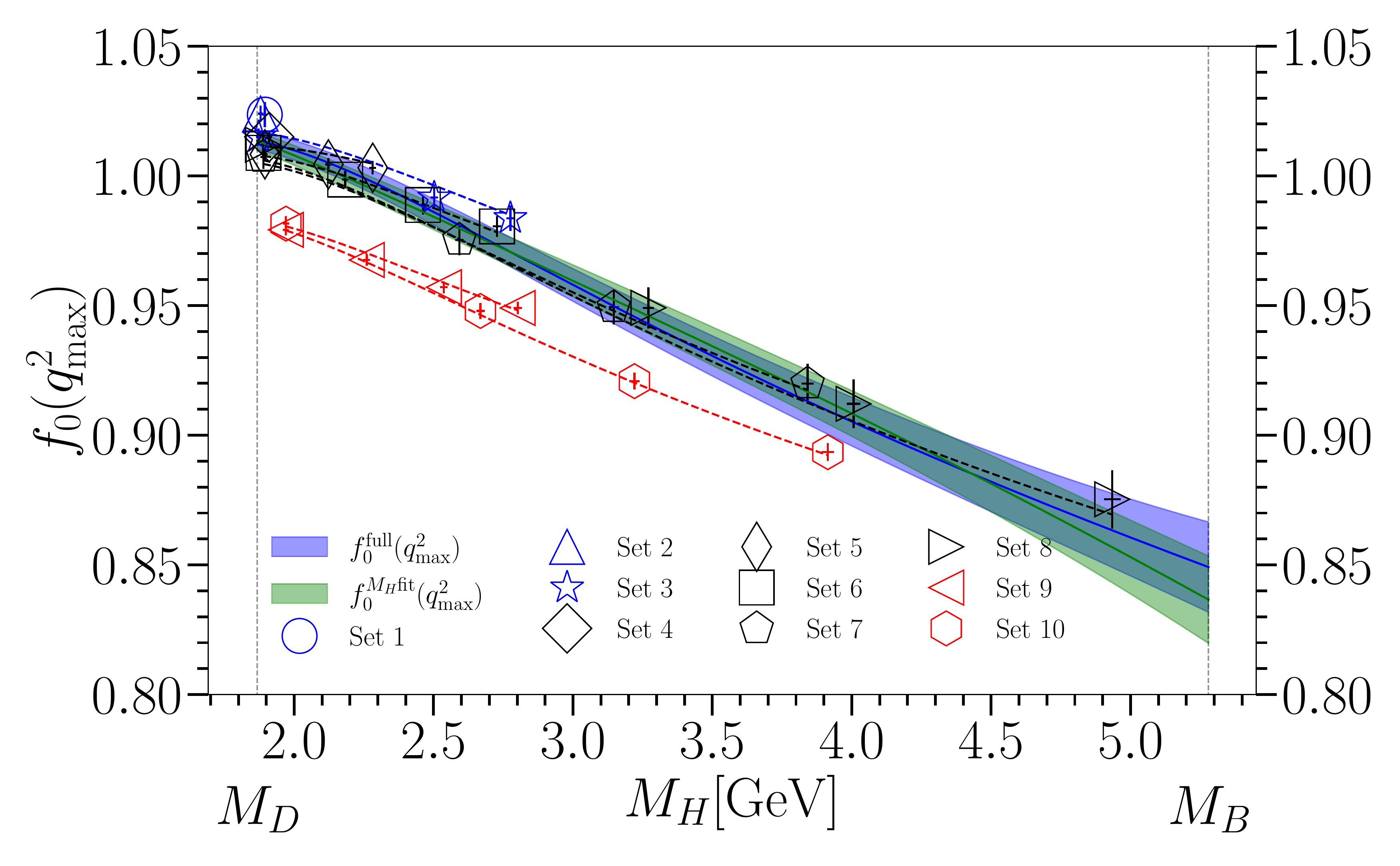}
    \caption{The $f_0(q^2_{\mathrm{max}})$ data points on each ensemble, plotted against $M_H$. Points in blue have physical $m_l$ values, black have $m_l=m_s/5$ and red have $m_l=m_s$. The blue band indicates the continuum result from our full fit (i.e. Equation~\eqref{Eq:zexpansion_cont}). The green band indicates the continuum results of a fit of just the $f_0(q^2_{\mathrm{max}})$ data to Equation~\eqref{eq:qsqmaxfit}. Dashed lines between data points indicate the full fit evaluated at that lattice spacing.}
    \label{fig:f0qsqmax}
  \end{center}
\end{figure} 

In order to test the ability of our fit to handle $M_H$ dependence independently of $q^2$ dependence we perform a simpler fit in $M_H$ at a fixed $q^2$ point for comparison to our full fit. For this we use our values for $f_0$ at $q^2_{\mathrm{max}}$ (only) and fit them to the functional form:

\begin{equation}\label{eq:qsqmaxfit}
  \begin{split}
    f^{M_H\mathrm{fit}}_0(q^2_{\mathrm{max}})=& \frac{\mathcal{L}}{1-\frac{q^2_{\mathrm{max}}}{M^2_{H_{s0}^{*}}}}\Big(1+\rho_0^{0}\log\Big(\frac{M_{H}}{M_{D}}\Big)\Big)\times(1+\mathcal{N}^{0}_0)\times\\
    &\sum^{N_{ijkl}-1}_{i,j,k,l=0}d_{ijkl0}^{0}\Big(\frac{\Lambda_{\text{QCD}}}{M_{H}}\Big)^i\Big(\frac{am_h^{\text{val}}}{\pi}\Big)^{2j}\times\\&\hspace{6.0em}\Big(\frac{a\Lambda_{\text{QCD}}}{\pi}\Big)^{2k}(x_{\pi}-x_{\pi}^{\mathrm{phys}})^{l},
  \end{split}
\end{equation}
taking the same choices for $N_{ijkl}$ and priors as for our full fit using Equations~\eqref{Eq:zexpansion} and~\eqref{Eq:an}. Figure~\ref{fig:f0qsqmax} shows the $f_0(q^2_{\mathrm{max}})$ data on each ensemble, as well as the result of our standard `full' fit to all data and the fit of the $f_0(q^2_{\mathrm{max}})$ alone (Equation~\eqref{eq:qsqmaxfit}).

Since Figure~\ref{fig:f0qsqmax} is a plot of results and a fit that depend only on $M_H$ it is easier to see here that we have good coverage of $M_H$ values from $M_D$ to $M_B$. The dashed lines connecting results at a fixed lattice spacing make clear how the discretisation effects behave, peeling away from the continuum curve for larger $am_h$ values. Our range of $am_h$ values (see Section~\ref{sec:sim}) allows our fit to track the discretisation effects. The smaller fit $f^{M_H\mathrm{fit}}_0(q^2_{\mathrm{max}})$ agrees well with our `full' fit result, indicating that we do indeed have good control of both $q^2$ and $M_H$ dependence in our fit. We repeat this test with versions of Equation~\eqref{eq:qsqmaxfit} which include $(M_D/M_H)^{\zeta_0}$ terms, where we trial $\mathcal{P}[\zeta_0]=1.5(0.5)$ and $\mathcal{P}[\zeta_0]=-0.5(0.5)$. In both cases, the fit agrees within one $\sigma$ with our full fit and the fit of Equation~\eqref{eq:qsqmaxfit}. Indeed, we find that the fit output hardly changes, when we fix the power $\zeta_0$ in Equation~\eqref{eq:qsqmaxfit} to be exactly $1.5$, showing that the output is determined by the lattice results and is not constrained by the presence or absence of the initial power term.

\subsection{Connecting $B\to K$ and $D\to K$ form factors}\label{sec:connecting}
\begin{figure*}
 \includegraphics[width=0.9\textwidth]{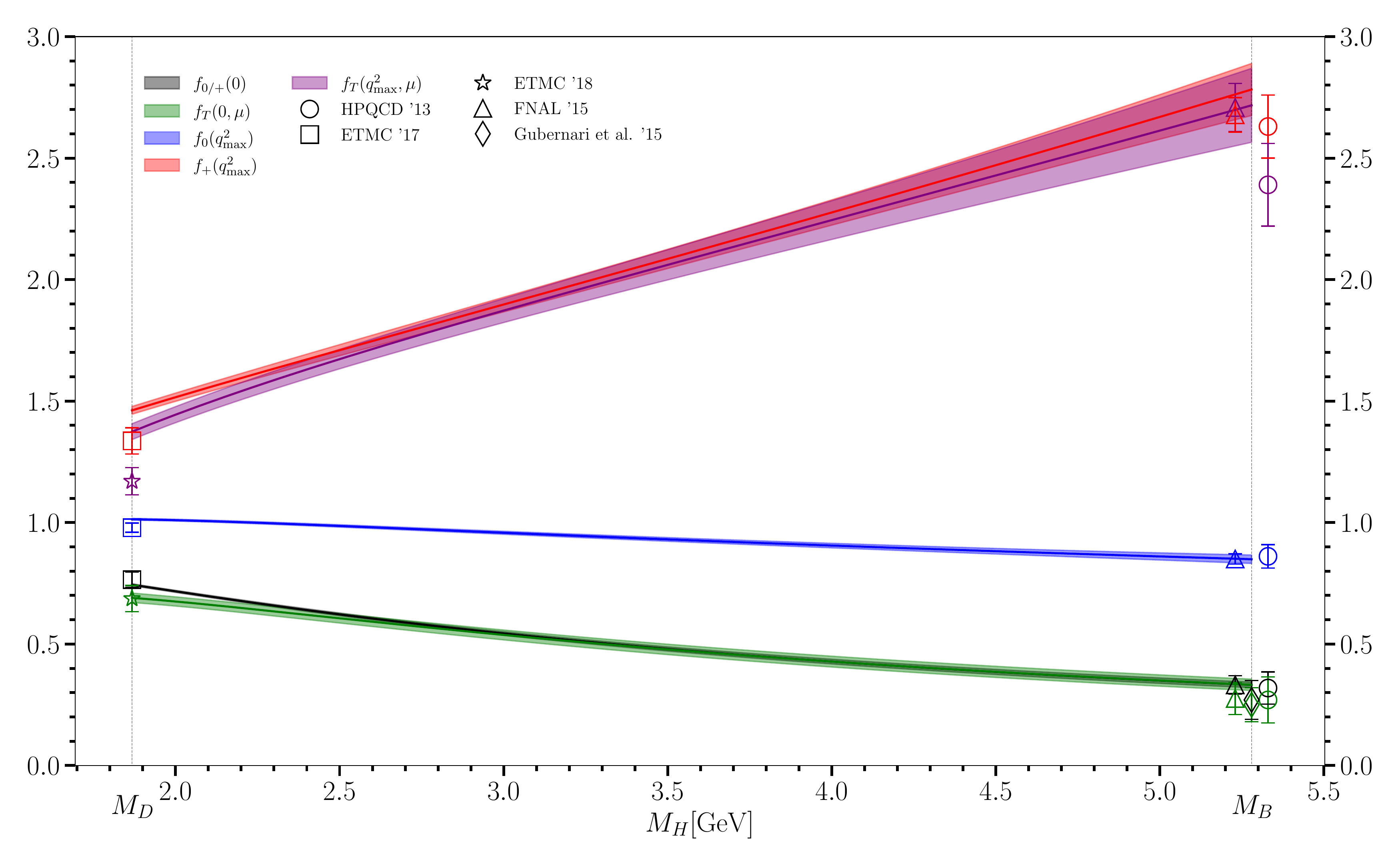}
  \caption{The form factors at $q^2_{\text{max}}$ and $q^2=0$ evaluated across the range of physical heavy masses from the $D$ to the $B$. Other lattice studies~\cite{Riggio:2017zwh,Lubicz:2018rfs,Bouchard:2013pna,Bailey:2015dka} of both $D\to K$ and $B\to K$ are shown for comparison. We also include some $B\to K$ results at $q^2=0$ from Gubernari et al.~\cite{Gubernari:2018wyi}, a calculation using light cone sum rules. We do not include HPQCD's $D\to K$ results that share data with our calculation here~\cite{Chakraborty:2021qav}; see text for a discussion of that comparison. At the $B$ end, data points are offset from $M_B$ for clarity. Note that we have run $Z_T$ to scale $\mu$ in this plot, where $\mu$ is defined linearly between $2~\mathrm{GeV}$ and $m_b=4.8~\mathrm{GeV}$, according to Equation~\eqref{eq:murunning}. The full running to $2~\mathrm{GeV}$ from $m_b$ results in a factor of 1.0773(17), applied to $f^{D\to K}_T$.}
  \label{fig:f0fpfTinmh}
\end{figure*}
Our heavy-HISQ approach allows us to study in detail the behaviour of the form factors at fixed $q^2$ with a change in heavy quark mass from the $c$ to the $b$. Figure~\ref{fig:f0fpfTinmh} illustrates this smooth variation with a plot of the continuum form factors, at extremal $q^2$ values, plotted against heavy mass $M_H$ from $M^{\text{phys}}_D$ to $M^{\text{phys}}_B$. This allows us to compare with previous calculations, both for $B\to K$ and $D\to K$, which we will discuss below. Firstly, however, we take a moment to address the running normalisation of $f_T(q^2,\mu)$.

In our calculation of the tensor form factor, we used $Z_T(\mu=4.8~\mathrm{GeV})$, calculated in~\cite{Hatton:2020vzp}. The scale $\mu$ is taken to be approximately equal to $m^{\mathrm{pole}}_b$. Whilst this is appropriate for the $B\to K$ results, we use a smaller scale, $\mu=2~\mathrm{GeV}$, for $D\to K$ to compare to previous results. In order to produce results at arbitrary $M_H$, we use a linear interpolation of $\mu$ between these two values,
\begin{equation}
  \mu(M_H)[\mathrm{GeV}]=2+\frac{2.8}{M^{\mathrm{phys}}_B-M^{\mathrm{phys}}_D}(M_H-M^{\mathrm{phys}}_D).
  \label{eq:murunning}
\end{equation}
Following~\cite{Hatton:2020vzp}, we then run from $\mu(M_B)=4.8~\mathrm{GeV}$ to our desired $\mu$ scale. The maximal extent of this running is down to $2~\mathrm{GeV}$ (i.e. for $M_H=M_D$), and this results in a factor of 1.0773(17) multiplying $f_T(q^2,4.8~\mathrm{GeV})$.

Returning to Figure~\ref{fig:f0fpfTinmh} and focusing on the $B\to K$ end of the results, we see very good agreement with previous work in general, adding confidence in the heavy-HISQ method. We find improvements in precision across the form factors, particularly at $q^2=0$, which is the important region for comparison to experiment in this case. At $q^2_{\mathrm{max}}$ our precision is not as high as that achieved in~\cite{Bailey:2015dka}. Our results have not been optimised for the $q^2 _{\mathrm{max}}$ region, however, so improvement there is readily possible. We also have the advantage that our renormalisation constants are more accurately calculated, which can otherwise be a source of systematic uncertainty. In~\cite{Bailey:2015dka}, one-loop perturbation theory is used to determine the current renormalisation factors and estimates made of the impact of missing $\alpha_s^2$ terms in these factors. Table~\ref{tab:ffres} provides numerical values for our $B\to K$ (as well as $D\to K$, see Section~\ref{sec:DKres}) form factors at the $q^2$ extremes shown in Figure~\ref{fig:f0fpfTinmh}.
\begin{table}
  \caption{Form factor results at the $q^2$ extremes. As described in the text, the $f^{D\to K}_0$ and $f_+^{D\to K}$ share data with the results in~\cite{Chakraborty:2021qav} (included for comparison) so should not be viewed as an independent calculation. }
  \begin{center} 
    \begin{tabular}{c c c}
      \hline
      &$q^2=0$&$q^2=q^2_{\text{max}}$\\ [0.5ex]
      \hline
      &\multicolumn{2}{c}{This work} \\ [0.5ex]
      \hline 
      $f^{B\to K}_0(q^2)$&0.332(12)&0.849(17)\\ [0.5ex]
      $f^{B\to K}_+(q^2)$&0.332(12)&2.78(11)\\ [0.5ex]
      $f^{B\to K}_T(q^2,\mu=4.8~\mathrm{GeV})$&0.332(24)&2.72(15)\\ [0.5ex]
      \hline
      $f^{D\to K}_0(q^2)$&0.7441(40)&1.0136(36)\\ [0.5ex]
      $f^{D\to K}_+(q^2)$&0.7441(40)&1.462(16)\\ [0.5ex]
      $f^{D\to K}_T(q^2,\mu=2~\mathrm{GeV})$&0.690(20)&1.374(33)\\ [0.5ex]
      \hline
      &\multicolumn{2}{c}{c.f. $D\to K$~\cite{Chakraborty:2021qav}}\\ [0.5ex]
      \hline
      $f^{D\to K}_0(q^2)$&0.7380(44)&1.0158(41)\\ [0.5ex]
      $f^{D\to K}_+(q^2)$&0.7380(44)&1.465(20)\\ [0.5ex]
    \end{tabular}
  \end{center}
  \label{tab:ffres}
\end{table}

\begin{figure}
\includegraphics[width=0.48\textwidth]{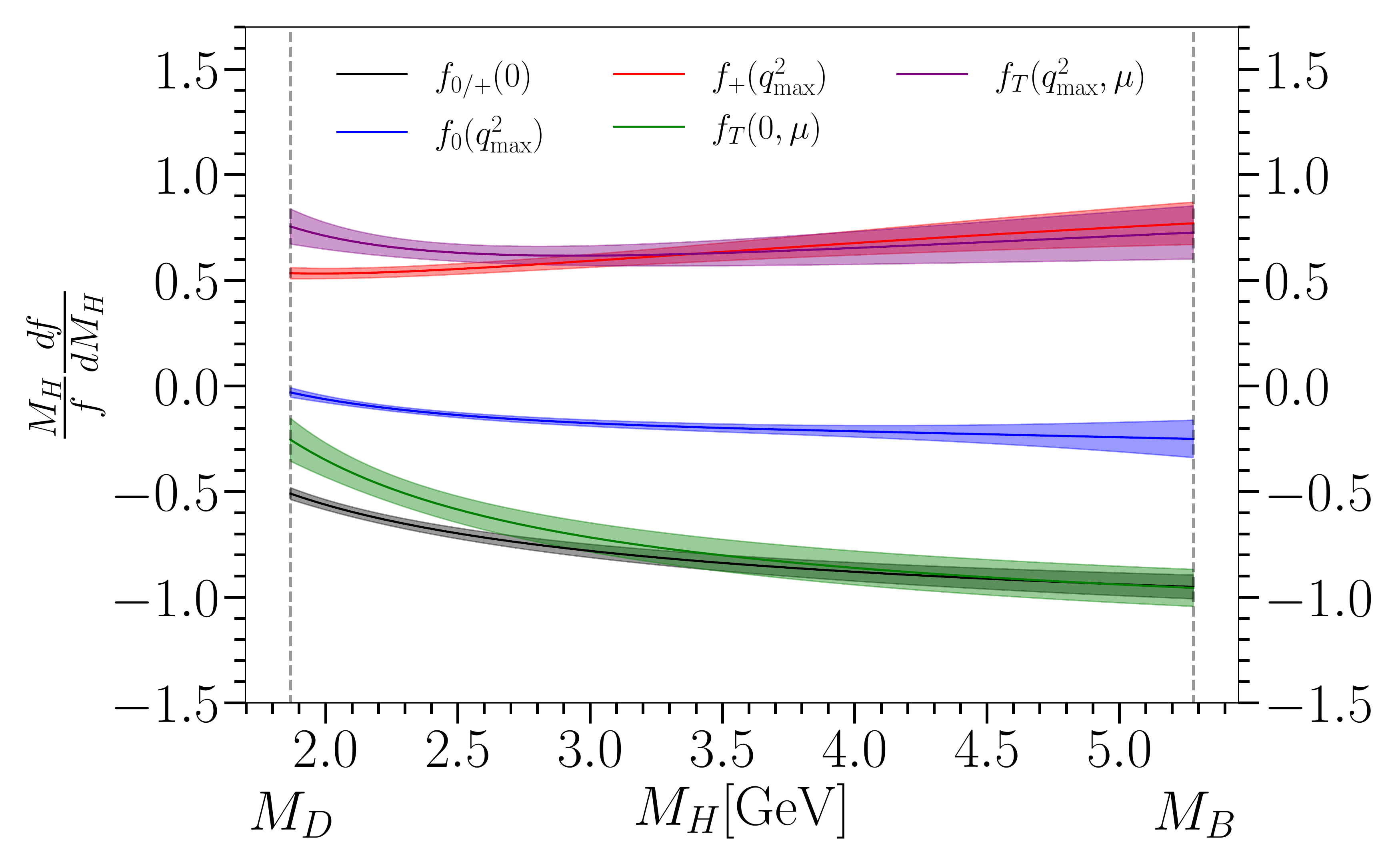}
\caption{The leading power of $M_H$ dependence in the form factors at $q^2=0$ and $q^2_{\mathrm{max}}$. The scale associated with the tensor form factor, $\mu$, is varied using equation~\eqref{eq:murunning}.}
\label{fig:qsqmaxderiv}
\end{figure}

Figure~\ref{fig:f0fpfTinmh} shows that the form factors at $q^2=0$ as well as $f_0(q^2_{\mathrm{max}})$ fall slowly as $M_H$ is increased. In contrast $f_+$ and $f_T$ at $q^2_{\mathrm{max}}$ increase. We can isolate the effective leading power of $M_H$ dependence by determining $X^{\mathrm{eff}}=M_H/f\times df/dM_H$, which returns $X$ for $f\propto M_H^X$. Our results for the form factors at $q^2=0$ and $q^2_{\mathrm{max}}$ are plotted in Figure~\ref{fig:qsqmaxderiv}. We see that at $q^2_{\mathrm{max}}$ for $M_H\to M_B$, the dependence is $X^{\mathrm{eff}}\approx-0.3$ for $f_0$, and $X^{\mathrm{eff}}\approx +0.7$ for $f_{+/T}$. This is roughly consistent with the values of $-0.5$ and $+0.5$ predicted in the infinite mass limit by HQET~\cite{Hill:2005ju}. It is clear that our fit is flexible enough to allow for $M_H$ dependence to vary with $q^2$. This flexibility arises from the fact that we are fitting in $z$-space with independent coefficients for different powers of $z$ (see Equation~\eqref{Eq:an}). Our lattice QCD results then have sufficient coverage of $z/q^2$ and $M_H$ space (in the range of masses from $M_D$ up to $0.9 M_B$) to give a robust fit result at $M_B$ across the full kinematic range. The effective powers of $M_H$ that we obtain at the two ends of the range, $q^2=0$ and $q^2_{\mathrm{max}}$, when $M_H=M_B$ provide a test of HQET. Note that Figure~\ref{fig:qsqmaxderiv} is essentially unchanged under the different fit possibilities that we test in Figure~\ref{fig:extrapstab} for the $(M_D/M_H)^{\zeta}$ terms in our fit. Thus the effective powers of $M_H$ that we obtain at $M_B$ are not dependent on the details of the fit form that we use, including how much input from HQET we impose.

A similar plot is presented in Appendix~\ref{sec:ffMHdep} (Figure~\ref{fig:ffMhdepderiv}), for fixed $q^2=M_D^2$.

\subsubsection{HQET tests of $B\to K$ results}
\begin{figure}
\includegraphics[width=0.48\textwidth]{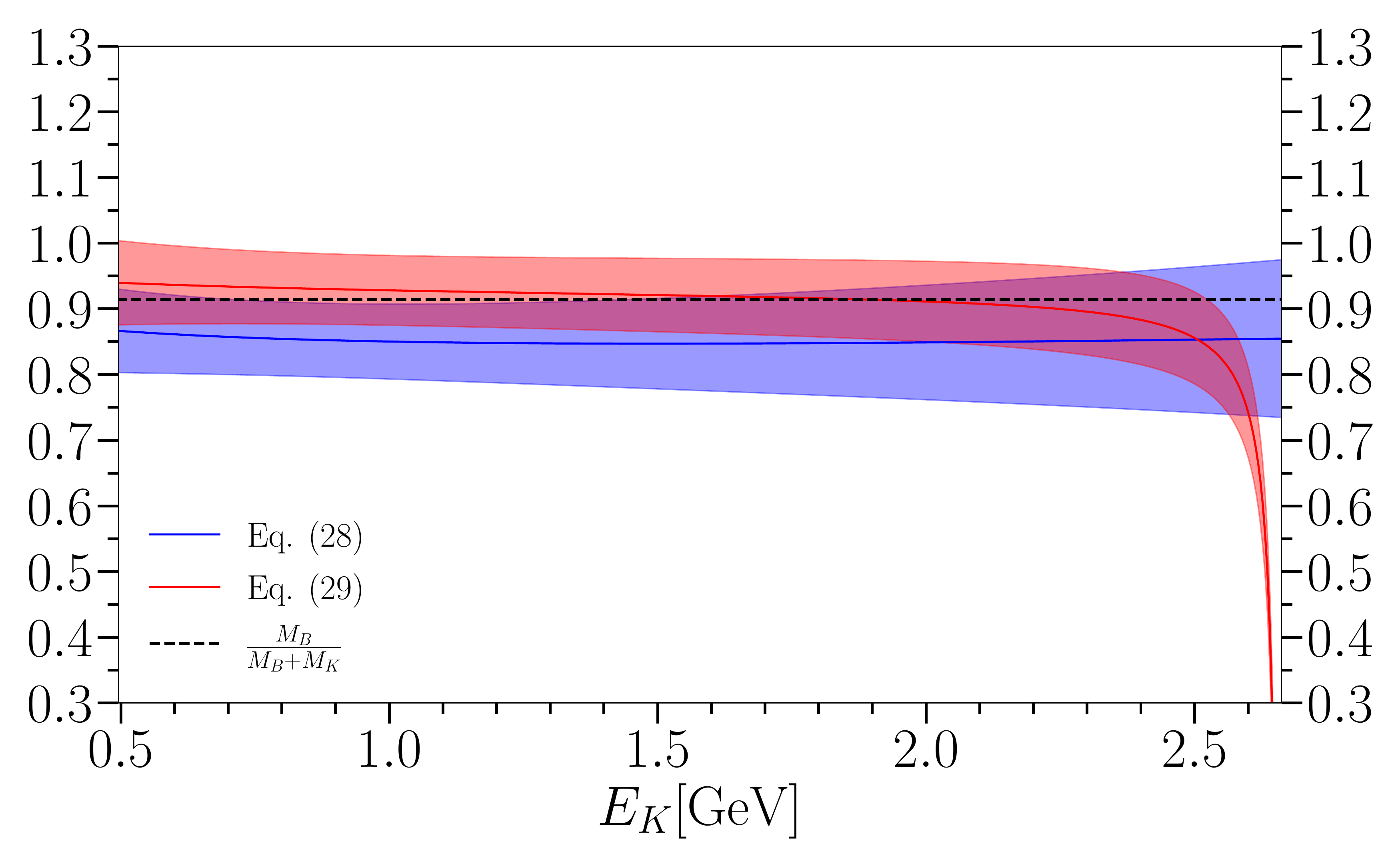}
\caption{Combinations of $B\to K$ form factors, $f_0$, $f_+$ and $f_T(\mu=4.8~\mathrm{GeV})$ in Equations~\eqref{eq:Hill19} and~\eqref{eq:Hill20} compared with expectations $\frac{M_B}{M_B+M_K}$ from HQET~\cite{Hill:2005ju} (dashed line). Uncertainties on the HQET expectations are not included. }
\label{fig:Hill1920}
\end{figure}
Returning to our $B\to K$ form factors, HQET expectations~(\cite{Hill:2005ju}, Equations (19) and (20)) give relations,
\begin{equation}\label{eq:Hill19}
  \frac{M_B}{M_B+M_K}=(f_+(E_K)-f_0(E_K))\frac{M_B^2}{q^2f_T(E_K)},
\end{equation}
\begin{equation}\label{eq:Hill20}
   \frac{M_B}{M_B+M_K}=\Big(\Big(1-\frac{E_K}{M_B}\Big)f_+(E_K)-\frac{f_0(E_K)}{2}\Big)\frac{M_B^2}{q^2f_T(E_K)}.
\end{equation}
Both Equations~\eqref{eq:Hill19} and~\eqref{eq:Hill20} are expected to be valid at small recoil (i.e. for $E_K\to M_K\approx0.5~\mathrm{GeV}$), whilst only Equation~\eqref{eq:Hill19} (\cite{Hill:2005ju} Equation (19)) should be valid for large recoil. Figure~\ref{fig:Hill1920} plots the form factor combinations (using $f_T(\mu=4.8~\mathrm{GeV})$) as a function of $E_K$. It shows the expected constant value of the form factor combination of Equation~\eqref{eq:Hill19} across the full $q^2$ range. It also shows the failure of Equation~\eqref{eq:Hill20} at large recoil (large $E_K$).

\subsection{Connecting $B\to K$ to other form factors}
\begin{figure}
  \begin{center}
    \includegraphics[width=0.48\textwidth]{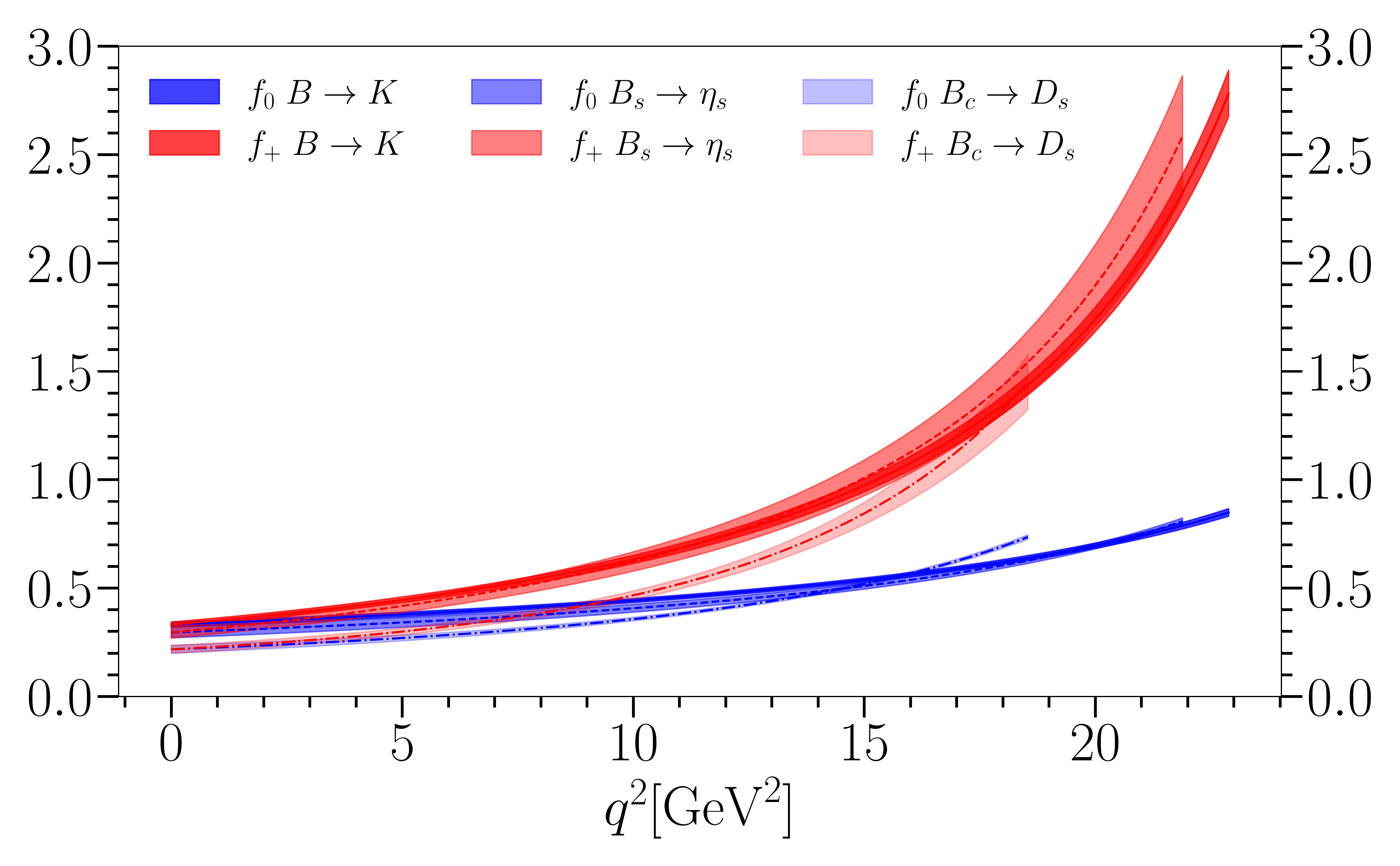}
    \includegraphics[width=0.48\textwidth]{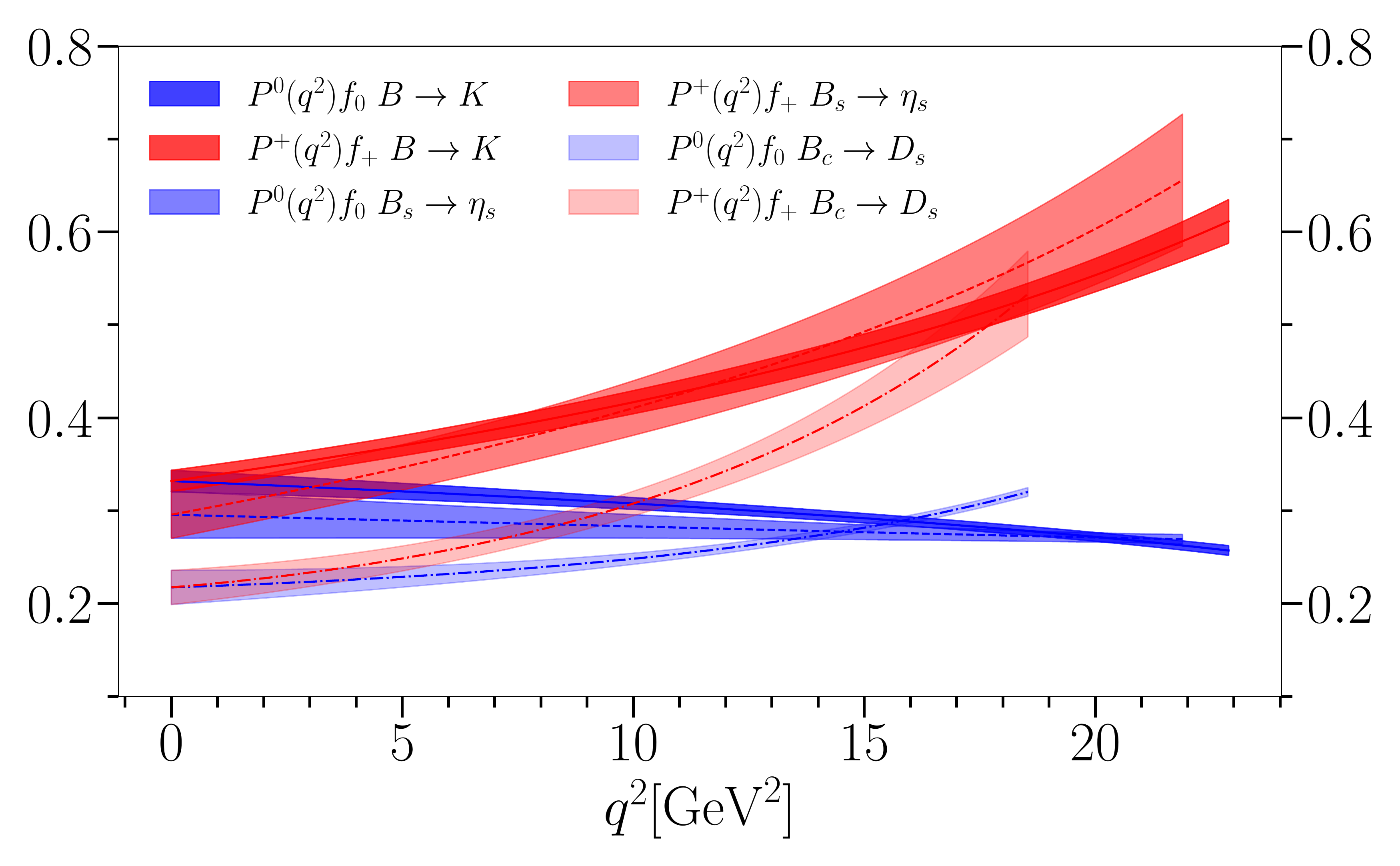}
    \caption{Comparison of our $B\to K$ scalar and vector form factors with those of $B_s\to\eta_s$~\cite{Parrott:2020vbe} and $B_c\to D_s$~\cite{Cooper:2021ofu} to show the impact of changing the spectator quark mass. In the lower pane, we have multiplied the form factors by their common pole factors to reduce the $y$-axis range and highlight the variation between the form factors. We take $P^0(q^2)=1-\frac{q^2}{M^2_{B^*_{s0}}}$, $P^+(q^2)=1-\frac{q^2}{M^2_{B^*_{s}}}$, using the central values of the masses in Table~\ref{tab:ancoefficients}.}
    \label{fig:f0fpBsetas}
  \end{center}
\end{figure} 
\begin{figure}
\includegraphics[width=0.48\textwidth]{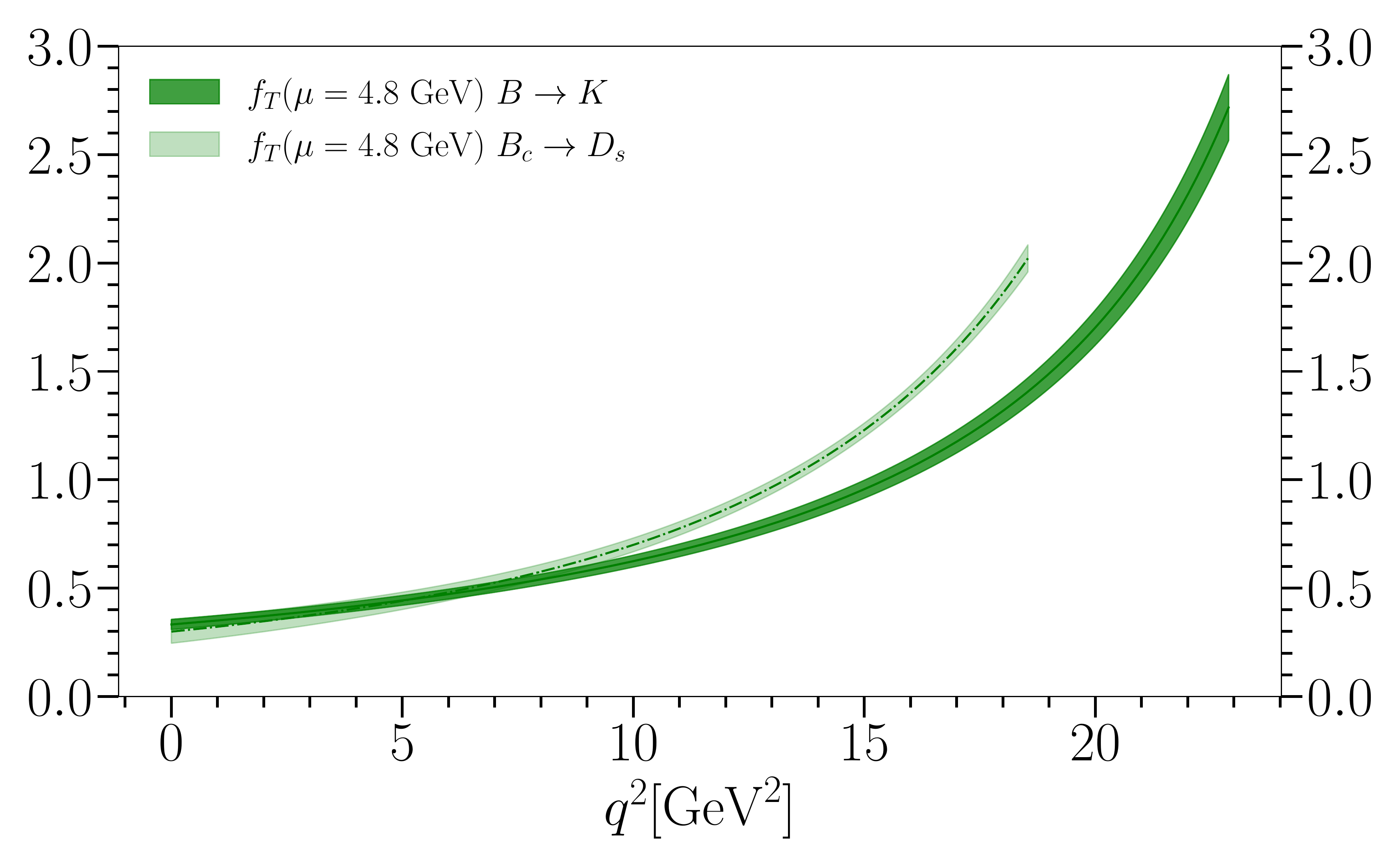}
\caption{Comparison of our $B\to K$ tensor form factor (at $\mu=4.8~\mathrm{GeV}$) with those of $B_c\to D_s$~\cite{Cooper:2021ofu} to show the impact of changing the spectator quark mass.}
\label{fig:fTBcDs}
\end{figure}
Figures~\ref{fig:f0fpBsetas} and~\ref{fig:fTBcDs} show the effect of changing spectator quark mass on the $b\to s$ pseudoscalar meson to pseudoscalar meson form factors. We compare our $B\to K$ results to the $B_s\to\eta_s$ results in~\cite{Parrott:2020vbe}, which differ only in the light spectator quark becoming a strange quark, and $B_c\to D_s$ results~\cite{Cooper:2021ofu}, where the spectator is a heavy (charm) quark. We see very mild spectator quark dependence for the light/strange quarks, at most a deviation of $\approx1\sigma$, which is roughly consistent with the modest effect of setting $\mathcal{L}=1$ in Figure~\ref{fig:extrapstab}. The transition to a heavy $c$ spectator leads to a much larger change. The heavier spectator gives a smaller form factor at $q^2=0$ that rises more steeply to $q^2_{\mathrm{max}}$, which has a smaller value. The behaviour of $f_T$ shown in Figure~\ref{fig:fTBcDs} is similar, but with a smaller shift at $q^2=0$.

We can also conduct a test of our chiral extrapolation by comparison with the $B_s\to\eta_s$ in~\cite{Parrott:2020vbe}. By setting $m_l/m_s = 1$, $M_{B}^{\mathrm{phys}}\to M_{B_s}^{\mathrm{phys}}$, $M_{D}^{\mathrm{phys}}\to M_{D_s}^{\mathrm{phys}}$ and $M_{K}^{\mathrm{phys}}\to M_{\eta_s}^{\mathrm{phys}}$ in our evaluation of Equation~\eqref{Eq:zexpansion_cont}, we can obtain results for $B_s\to\eta_s$. These are not completely independent of the results in~\cite{Parrott:2020vbe} as they include shared data on two sets (see Section~\ref{sec:sim}). However, the correlator and $z$-expansion fits used here and in that work are very different, and we do not include data on set 8, (set 3 in~\cite{Parrott:2020vbe}), nor the continuum $f_0(q^2_{\mathrm{max}})$ data point that was added there. This makes comparison of our results a strong test of our fit, particularly the chiral perturbation theory element.
\begin{table}
  \caption{A comparison of form factor results for $B_s\to\eta_s$ at the $q^2$ extremes, obtained here and in earlier work. As described in the text, the $f_0$ and $f_+$ obtained here share data with the results in~\cite{Parrott:2020vbe} (included for comparison) so should not be viewed as an independent calculation. }
  \begin{center} 
    \begin{tabular}{c c c}
      \hline
      &$q^2=0$&$q^2=q^2_{\text{max}}$\\ [0.5ex]
      \hline
      &\multicolumn{2}{c}{This work} \\ [0.5ex]
      \hline 
      $f^{B_s\to \eta_s}_0(q^2)$&0.3191(85)&0.819(17)\\ [0.5ex]
      $f^{B_s\to \eta_s}_+(q^2)$&0.3191(85)&2.45(19)\\ [0.5ex]
      $f^{B_s\to \eta_s}_T(q^2,\mu=4.8~\mathrm{GeV})$&0.370(78)&2.32(56)\\ [0.5ex]
      \hline
      &\multicolumn{2}{c}{c.f. $B_s\to \eta_s$~\cite{Parrott:2020vbe}}\\ [0.5ex]
      \hline
      $f^{B_s\to \eta_s}_0(q^2)$&0.296(25)&0.808(15)\\ [0.5ex]
      $f^{B_s\to \eta_s}_+(q^2)$&0.296(25)&2.58(28)\\ [0.5ex]
      \hline
      &\multicolumn{2}{c}{c.f. $B_s\to \eta_s$~\cite{Colquhoun:2015mfa}}\\ [0.5ex]
      \hline
      $f^{B_s\to \eta_s}_0(q^2)$&-&0.811(17)\\ [0.5ex]
    \end{tabular}
  \end{center}
  \label{tab:Bsetasffres}
\end{table}
Table~\ref{tab:Bsetasffres} gives the results of our form factors evaluated at $B_s\to\eta_s$, at extremal $q^2$ values. We see that they are in good agreement with the results in~\cite{Parrott:2020vbe}, supporting our extrapolation in the spectator mass. Additionally, we note that we agree very well with the $f_0(q^2_{\mathrm{max}})$ value given in~\cite{Colquhoun:2015mfa}. This point was included in the analysis in~\cite{Parrott:2020vbe}, and here we demonstrate that we are able to obtain a very similar result independently of this point.

\subsection{$D\to K$ form factors}\label{sec:DKres}
\begin{figure}

\includegraphics[width=0.48\textwidth]{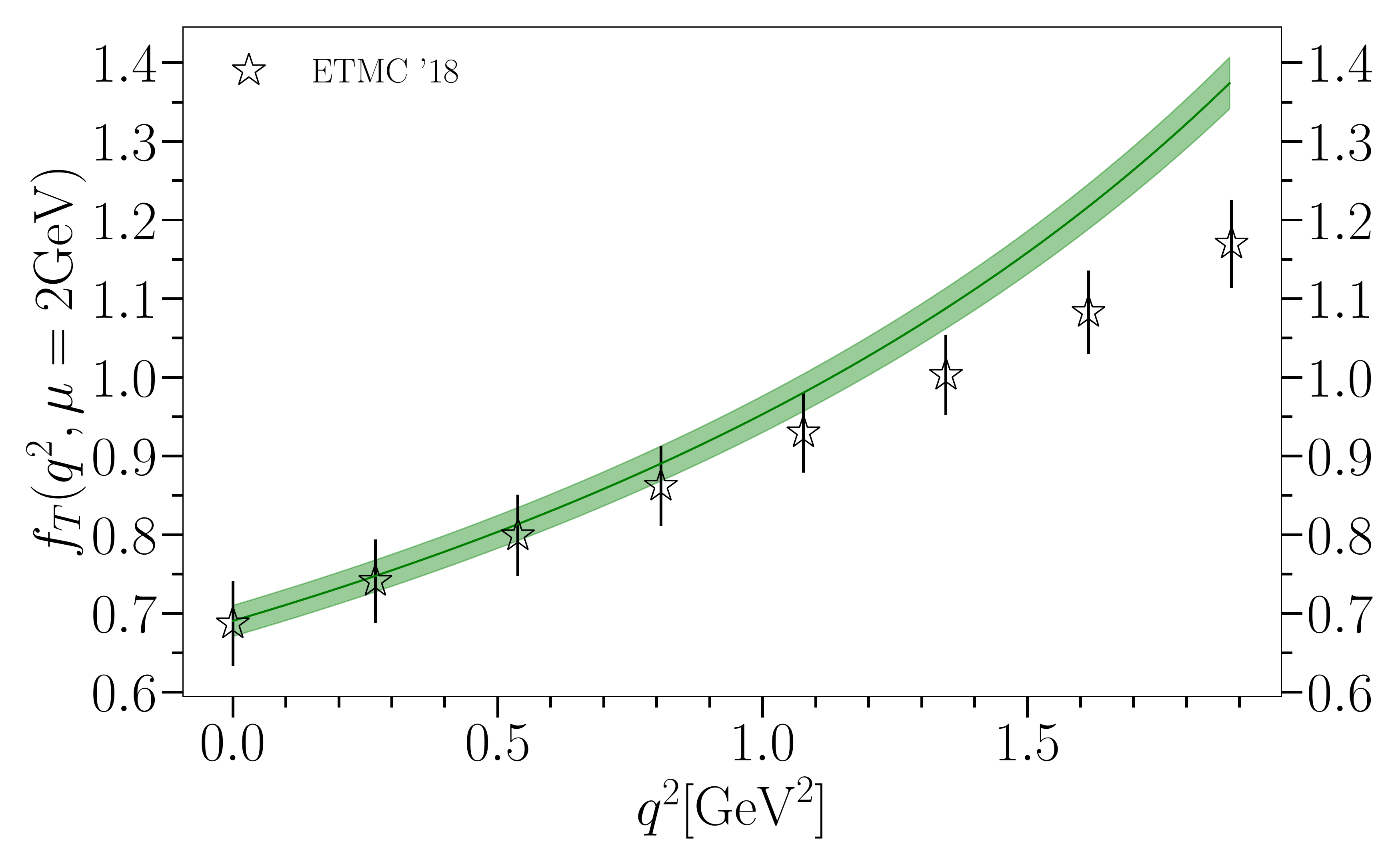}
\caption{The green band gives our $D\to K$ tensor form factor at $\mu=2~\mathrm{GeV}$, across the physical $q^2$ range. Results from~\cite{Lubicz:2018rfs} are included for comparison.}
\label{fig:DKfT}
\end{figure}
By evaluating our form factors at the $D$ mass, we are able to calculate scalar, vector and tensor form factors for the $D\to K$ decay. The scalar and vector form factors are in agreement with those in~\cite{Chakraborty:2021qav} (see Table~\ref{tab:ffres} for a comparison). Whilst these two calculations share a considerable amount of raw correlator data, this data (except for sets 1 and 2) is subject to very different and much larger correlator fits here, involving more masses as well as tensor three-point data, and the accompanying non-Goldstone kaons. As well as these different correlator fits, our heavy-HISQ method uses a very different modified $z$ expansion from that in~\cite{Chakraborty:2021qav}, in order to include heavy mass dependence. That we see agreement here, with a maximal difference of $1\sigma$ (assuming correlations are modest) indicates that our uncertainties are of an appropriate size.  

Our fit coefficients for the $D \to K$ form factors, along with their correlations are given in Table~\ref{tab:DKan}. Our form factors can be reconstructed from these values, or the python script described in Appendix~\ref{sec:reconstruct} can be used. As discussed above, we run the scale of $Z_T$ down to 2 GeV for the $D\to K$ results, as used in~\cite{Lubicz:2018rfs}. Following~\cite{Hatton:2020vzp} this involves multiplying by 1.0773(17), a factor which is included in the $a_n^T$ values in Table~\ref{tab:DKan} and in our results quoted in Table~\ref{tab:ffres}. Note that this is the same value used at the $M_D$ end of results in Figure~\ref{fig:f0fpfTinmh}.

Returning to the $D\to K$ end of the results in Figure~\ref{fig:f0fpfTinmh}, we again see good agreement with previous work, with the exception of $f_T(q^2_{\mathrm{max}},2~\mathrm{GeV})$ and $f_+(q^2_{\mathrm{max}})$ where we are in tension with ETMC~\cite{Riggio:2017zwh,Lubicz:2018rfs}. This was found previously for $f_+(q^2_{\mathrm{max}})$ in~\cite{Chakraborty:2021qav}. 

Our tensor form factor is compared to that from~\cite{Lubicz:2018rfs} in Figure~\ref{fig:DKfT}. We see that the uncertainty is reduced by roughly a factor of two across the $q^2$ range in our results. Good agreement is seen with~\cite{Lubicz:2018rfs} at low $q^2$. Additionally, we report the ratio $f^{D\to K}_T(0,\mu=2~\mathrm{GeV})/f^{D\to K}_+(0)=0.928(27)$, which agrees with the $0.898(50)$ given in~\cite{Lubicz:2018rfs}. However, our tensor form factor has a steeper slope in $q^2$ and at $q^2_{\mathrm{max}}$ there is disagreement at a level of $3.1\sigma$. 

\begin{figure}

\includegraphics[width=0.48\textwidth]{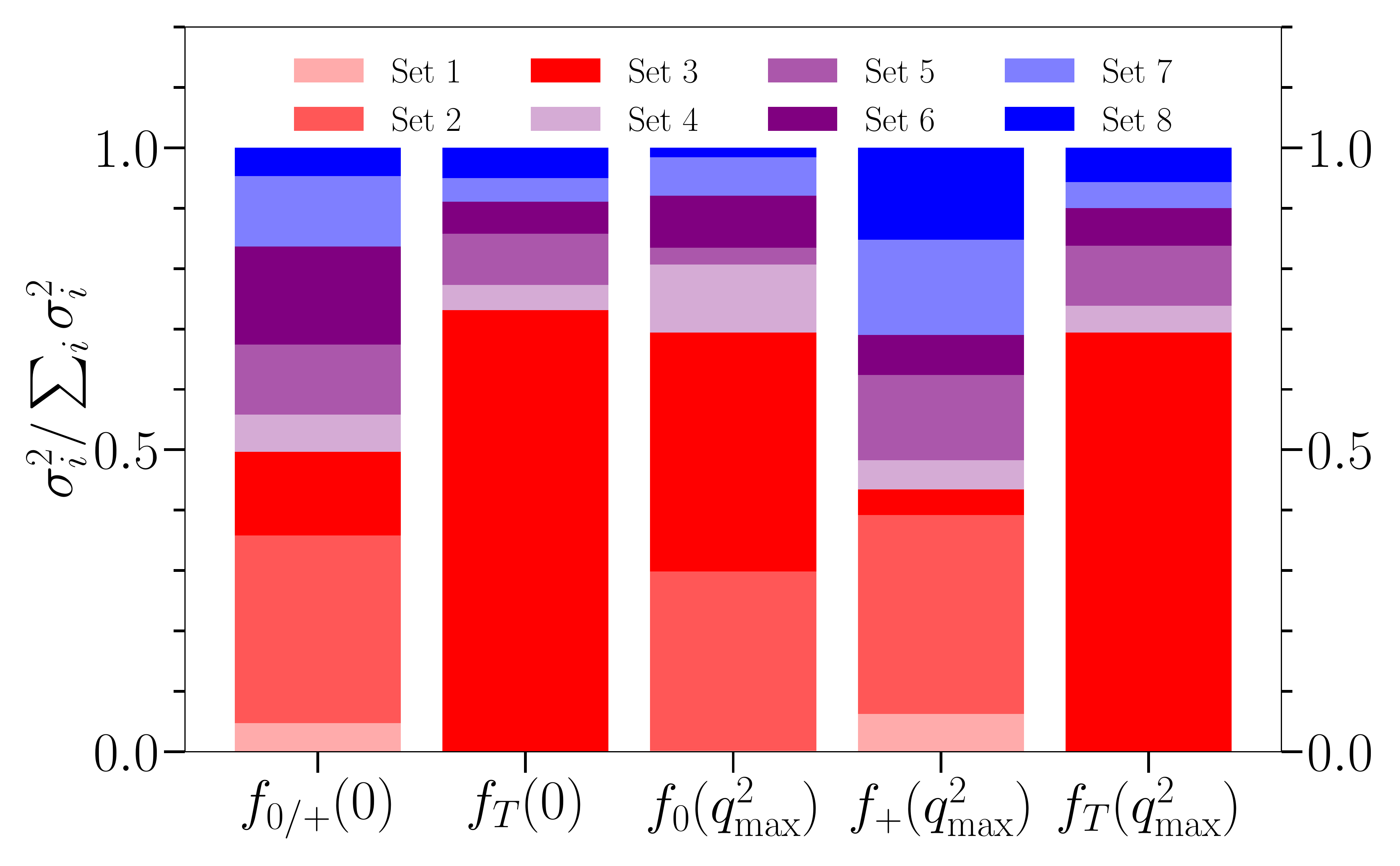}
\caption{Breakdown of the contributions to the statistical uncertainty of the $D\to K$ form factors at their extremes from data on each ensemble. Uncertainty from each ensemble $\sigma_i$ is added in quadrature, normalised by the total uncertainty squared $\sum_i\sigma_i^2$. Sets 6 and 7 include contributions from $H_s\to\eta_s$ data.}
\label{fig:DKenserr}
\end{figure}

Figure~\ref{fig:DKenserr} gives the breakdown of statistical uncertainty from each ensemble for the $D\to K$ form factors. It is clear from Figure~\ref{fig:DKenserr} that, unlike in the $B\to K$ case above (Figure~\ref{fig:BKenserr}), the errors on all $D\to K$ form factors across the $q^2$ range are dominated by the coarser lattices, specifically the physical sets 2 and 3, whilst again set 1 makes the smallest contribution in all cases except $f_+(q^2_{\mathrm{max}})$. This makes sense, as the physical charm mass is easily accessed on all ensembles, so the heavy quark extrapolation does not play much of a role here. The extrapolation to physical light quark mass is relatively more important, so sets 1, 2 and 3 play a bigger role. Sets 1 and 2 do not contain any tensor data, hence, set 3 is especially dominant in the case of $f_T$. 
\begin{table}
  \caption{$D\to K$ values of fit coefficients $a_n^{T}$, the pole mass in GeV, and the $\mathcal{L}$ term with correlation matrix below (see Equation~\eqref{Eq:zexpansion_cont} for fit form). The pole mass and $\mathcal{L}$ are very slightly correlated due to the way the fit function is constructed. These correlations are too small to have any meaningful effect on the form factor values, but we include them for completeness. For details on reconstructing our results, see Appendix~\ref{sec:reconstruct}. $a_n^T$ values include a factor of 1.0773(17) from running $Z_T$ to $\mu=2~\mathrm{GeV}$.}
  \begin{center} 
    \begin{tabular}{c c c c c }
      \hline
      $a_0^T$&$a_1^T$&$a_2^T$& $M^{\text{phys}}_{D^*_s}$ & $\mathcal{L}$\\ [0.5ex]
      0.522(15)&-0.74(13)&0.38(84)&2.11220(40)&1.3234(24)\\ [1ex]
      \hline 

      1.00000&0.34687&0.03704&-0.00005&-0.06075\\ [0.5ex]
      &1.00000&0.61200&-0.01018&0.01148\\ [0.5ex]
      &&1.00000&0.00069&-0.00046\\ [0.5ex]
      &&&1.00000&0.00003\\ [0.5ex]
      &&&&1.00000\\ [0.5ex]
      \hline
    \end{tabular}
  \end{center}
  \label{tab:DKan}
\end{table}

\subsection{Connecting $D\to K$ to other form factors}
\begin{figure}

\includegraphics[width=0.48\textwidth]{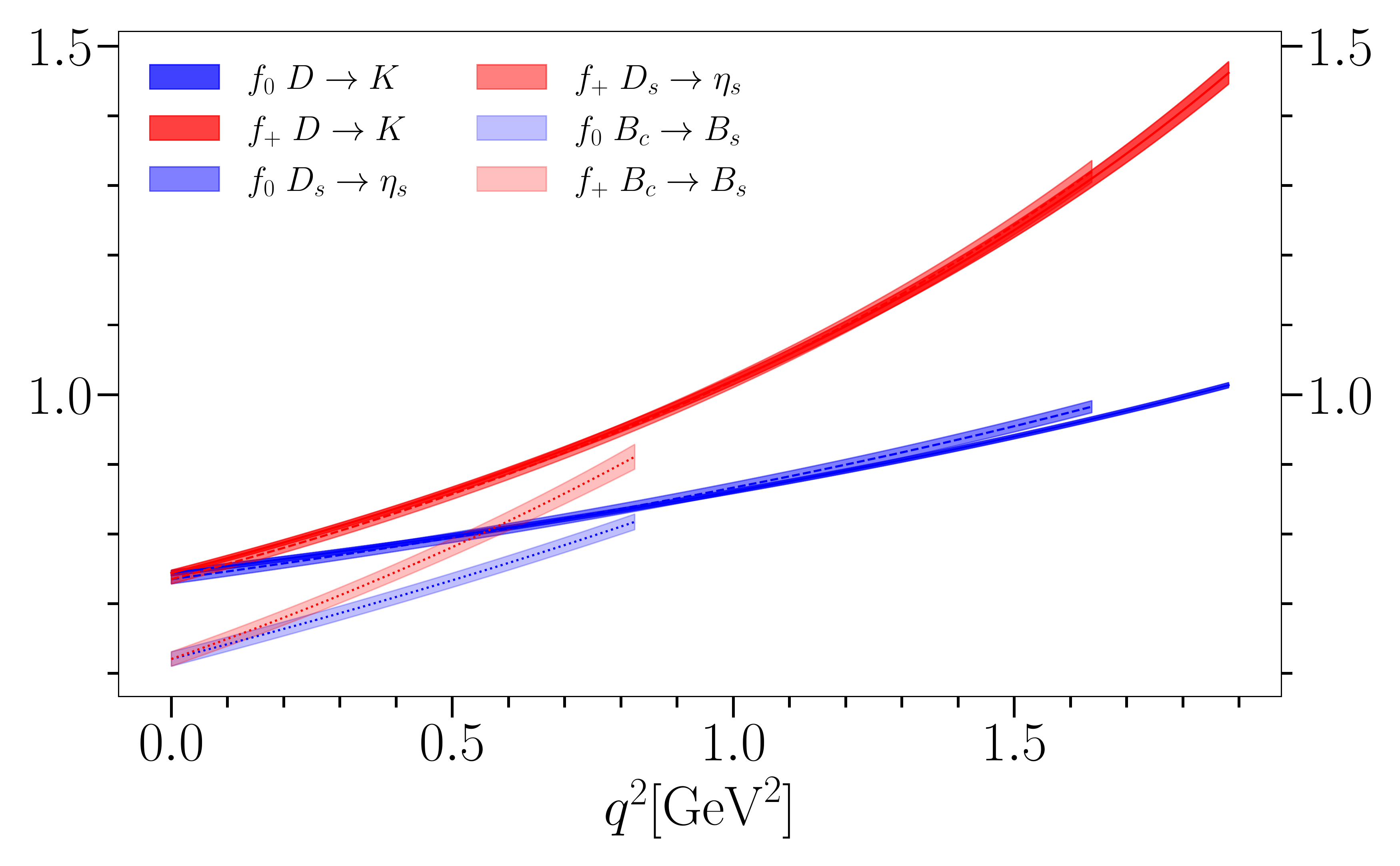}
\caption{Comparison of our $D\to K$ scalar and vector form factors with those for $D_s\to\eta_s$~\cite{Parrott:2020vbe} and $B_c\to B_s$~\cite{Cooper:2020wnj} to show the effect of changing spectator quark mass.}
\label{fig:f0fpDsetas}
\end{figure}
As with $b\to s$ decays above, we now have sufficient information from accurate lattice QCD calculations to test the impact on the pseudoscalar-to-pseudoscalar meson form factors of changing the quark mass for the spectator quark that accompanies the $c\to s$ decay. Figure~\ref{fig:f0fpDsetas} shows this effect. We compare our $D\to K$ results to the $D_s\to\eta_s$ results from~\cite{Parrott:2020vbe} and, as for $b \to s$, see a very gentle dependence when we change the spectator from light to strange. This agrees with the conclusions of~\cite{Koponen:2013tua} but is much more compelling here because of the high precision of both sets of form factors. The biggest deviation is for $f_0$, at the maximum $q^2$ for $D_s\to\eta_s$, where $D_s\to\eta_s$ is larger than $D\to K$ by $\approx2\sigma$ (or about 2\%).

We can also compare with $B_c\to B_s$ form factors~\cite{Cooper:2020wnj}, where the spectator quark is a $b$ quark. This is a very big change in spectator quark mass (roughly a factor of 1000) and unsurprisingly we see a much bigger change. The form factors for $B_c \to B_s$ have smaller values at $q^2=0$ (but only by $\sim$20\%) and rise much more steeply with $q^2$ than is the case with lighter spectator quarks. This trend is exactly the same, but magnified by the larger quark mass change, as that seen for the $b\to s$ case as we change from a light to a charm spectator (compare Figure~\ref{fig:f0fpBsetas}). 

\section{Conclusions}\label{sec:conclusions}
We have performed the first $N_f=2+1+1$ lattice QCD calculation of the scalar, vector and tensor form factors for semileptonic $B\to K$ decay. As well as including charm quarks in the sea and physical $u/d$ quarks, our calculation improves on earlier work in several ways.  We use the relativistic HISQ action for all valence quarks (as well as the sea quarks), extending further the use of HPQCD's heavy-HISQ technique. In contrast to earlier approaches, this method allows us to normalise the weak currents accurately. The scalar and vector currents are completely nonperturbatively normalised here, and the tensor current uses an $\alpha_s^3$-accurate matching from the lattice to $\overline{\text{MS}}$ via a symmetric momentum-subtraction scheme in which nonperturbative artefacts are fitted and removed~\cite{Hatton:2020vzp}. The heavy-HISQ approach combines results from multiple heavy quark masses with multiple values of the lattice spacing and multiple momenta for the daughter meson. The range of possible (physical) heavy meson masses grows on finer lattices as does the range of daughter meson momentum. Because the daughter meson momentum needed to reach $q^2=0$ is linear in the heavy meson mass, this means that we can cover the full $q^2$ range from $q^2_{\text{max}}$ down to $q^2=0$ in our lattice QCD calculation. This is also in contrast to earlier approaches that were restricted to a $q^2$ region close to $q^2_{\text{max}}$. Our form factors can be reconstructed using the results in Table~\ref{tab:ancoefficients}, or by using the code provided, and referring to Appendix~\ref{sec:reconstruct}.

Our form factors are compared at the extremes of $q^2$ to earlier values in Figure~\ref{fig:f0fpfTinmh}. This shows that our uncertainties are a factor of 3 smaller at $q^2=0$, and comparable to previous results at $q^2_{\text{max}}$. For $B\to K$, our uncertainties on $f_0$ and $f_+$ are now below $4\%$ across the whole physical $q^2$ range and for $f_T$ the uncertainty is below  $7\%$ across the same range (see Figure~\ref{Fig:f0fpfTerr}).Our calculational strategy is optimised for $q^2=0$ by the use of the temporal vector current. Results using the spatial vector current are more accurate at large $q^2$ (because of kinematic factors) and we show that in Figure~\ref{fig:fpV0V1diff}. Uncertainties at large $q^2$ in our results could straightforwardly be reduced by calculating more correlation functions with the spatial vector current. The important kinematic region for phenomenology is that of small $q^2$, however, so we have concentrated on that here. Our statistical uncertainties are dominated by our two finest (and most computationally costly) ensembles (see Figure~\ref{fig:BKenserr}), and we have demonstrated that our overall uncertainties are dominated by these statistics. They could then be straightforwardly reduced with more computing resources in future.

Because the heavy-HISQ approach requires multiple values of the heavy quark mass, a map of the form factors as a function of heavy meson mass is obtained connecting those for $D$ to those for $B$. The form factors are smooth functions of heavy meson mass in QCD and this is illustrated most clearly by Figure~\ref{fig:f0fpfTinmh}. We can also test expectations from HQET (see Figures~\ref{fig:qsqmaxderiv} and~\ref{fig:Hill1920} ). 

Our results here for the vector and scalar form factors for $D \to K$ are not independent of, and agree with, those given in a recent HPQCD publication~\cite{Chakraborty:2021qav} based on $D \to K$ correlators only. Here we give in addition the tensor form factor. For $f_T(q^2,\mu=2~\mathrm{GeV})$ we have an uncertainty below $3\%$ across the full $q^2$ range, roughly halving the uncertainty given in earlier calculations~\cite{Lubicz:2018rfs}. Our results for $f_T(q^2,\mu=2~\mathrm{GeV})$  for $D\to K$ are significantly higher than those of~\cite{Lubicz:2018rfs} at large $q^2$ values. Our form factors can be reconstructed using the results in Table~\ref{tab:DKan}, or from the code provided (see Appendix~\ref{sec:reconstruct}).

The smooth connection between $B\to K$ and $D\to K$ form factors obtained in the heavy-HISQ approach is a useful one, because at least the $D\to K$ vector form factor can be compared to accurate experimental results for the semileptonic $D\to K$ decay process. In~\cite{Chakraborty:2021qav} it was shown that the shape of the $D\to K$ vector form factor obtained from lattice QCD using HISQ agrees well with that inferred from the experimental differential decay rate. Since we do not expect new physics in the tree-level $D\to K$ decay, this is a stringent test of (lattice) QCD. It also provides a firm basis for the $B\to K$ form factors that we obtain here as an extension to heavier mass of the $D\to K$ results. 

As well as being a smooth function of heavy (parent) quark mass, form factors in QCD are also a smooth function of spectator quark mass. With accurate form factors covering the full $q^2$ range now available for a range of processes using the HISQ formalism for all quarks, we can make comparisons that show the impact of changing the spectator quark mass. The conclusion is that very large changes in mass are needed to achieve sizeable effects (see Figures~\ref{fig:f0fpBsetas} and ~\ref{fig:f0fpDsetas}); very little is seen on substituting a strange quark for a light one. Increasing the spectator quark mass by a larger factor (substituting a charm or bottom quark for a light one) makes the trend clearer, pushing the form factor downwards at $q^2=0$ and compressing the $q^2$ range. More comparisons of this kind will become possible also for pseudoscalar to vector meson decay channels as further sets of form factors become available from lattice QCD. This will yield a more complete picture of form factor behaviour with implications for our understanding of  meson internal structure.

In an accompanying paper we will lay out in detail the phenomenological implications of the improved form factors for $B\to K$ that we have calculated here. 

\section{Acknowledgements}
We are grateful to the MILC collaboration for the use of  their  configurations  and  their  code, which we use to generate quark propagators and construct correlators. We would also like to thank  L. Cooper, J. Harrison, D. Hatton and G. P. Lepage for useful discussions and B. Chakraborty, J. Koponen and A. T. Lytle for generating propagators/correlators in previous projects that we could make use of here. Computing was done on the Cambridge Service for Data Driven Discovery (CSD3) supercomputer, part of which is operated by the University of Cambridge Research Computing Service on behalf of the UK Science and Technology Facilities Council (STFC) DiRAC HPC Facility. The DiRAC component of CSD3 was funded by BEIS via STFC capital grants and is operated by STFC operations grants. We are grateful to the CSD3 support staff for assistance. Funding for this work came from STFC.

\begin{appendix}
\section{Reconstructing our results}\label{sec:reconstruct}
Tables~\ref{tab:ancoefficients} and~\ref{tab:DKan} should allow the reader to reconstruct our form factors using details given in Section~\ref{sec:evalphys} and Equation~\eqref{Eq:zexpansion_cont}. However, to make this easier, we attach an ancillary python script and text file, which will reproduce our fully correlated $B\to K$ and $D\to K$ form factors at any $q^2$ value chosen. The reader should only need a python installation with the packages gvar~\cite{peter_lepage_2020_3715065} and numpy to run this script.

The python script \textit{make\_BK\_DK\_ffs.py} loads data and correlations from \textit{BtoKandDtoKformfacs.txt} and contains functions \textit{make\_fX\_Y(qsq)}, where `X' can be `0', `p' or `T' for $f_0$, $f_+$ and $f_T$ respectively and `Y' values of `B' or `D' give the $B\to K$ or $D\to K$ form factors.
Running `python3 \textit{make\_BK\_DK\_ffs.py}' with \textit{BtoKandDtoKformfacs.txt} in the same directory (and numpy and gvar installed) should produce as terminal output a number of tests. These give the form factors obtained by evaluating the functions at various $q^2$ values, and compare them with saved results (which are given in the form `c.f. \textit{value}'). These numbers should agree, and you may also wish to compare the relevant ones with Table~\ref{tab:ffres} as a sanity check. After this, using \textit{make\_BK\_DK\_ffs.py} as a module and calling the functions \textit{make\_fX\_Y(qsq)} from another python script will report the form factors for any chosen float or gvar $q^2$ value. Our form factors are only valid over the physical $q^2$ range. Values of $q^2$ outside of the range will result in a warning but will still work, providing they do not cause $z$ to be imaginary.

\section{Additional comments on form factor $M_H$ dependence}\label{sec:ffMHdep}
As well as the heavy mass dependence shown in Figure~\ref{fig:f0fpfTinmh}, we also provide information on the heavy mass dependence of our form factors at fixed $q^2=M_D^2$. This dependence for $B\to \pi$ form factors, which are related to ours by SU(3) flavour symmetry, is of interest in QCD factorisation studies of $B\to D\pi$ (e.g.~\cite{Beneke:2000ry}), where the heavy mass dependence of $f^{B\to\pi}(q^2=M_D^2)$ is expected to be $f^{B\to\pi}\propto M_H^{-3/2}$. We are able to test this behaviour explicitly by varying $M_H$ to reveal the $M_H$ dependence of $f^{B\to K}(q^2=M_D^2)$, relevant for a similar analysis to~\cite{Beneke:2000ry}, for the $B\to KD$ decay.    
\begin{figure}
  \includegraphics[width=0.48\textwidth]{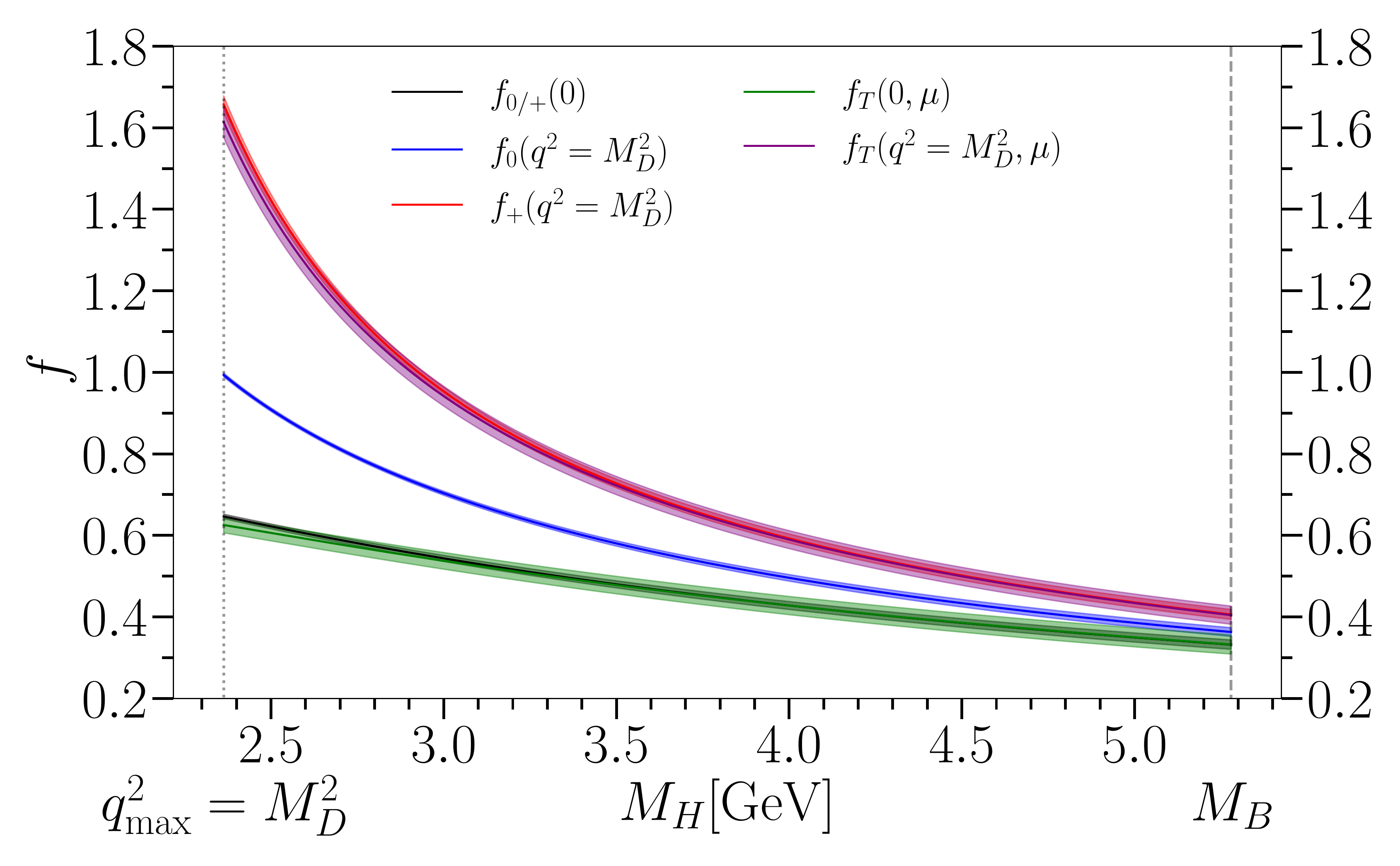}
\caption{The $M_H$ dependence of our form factors at $q^2=0$ and $q^2=M_D^2$.}
\label{fig:ffMhdep}
\end{figure}
\begin{figure}
 \includegraphics[width=0.48\textwidth]{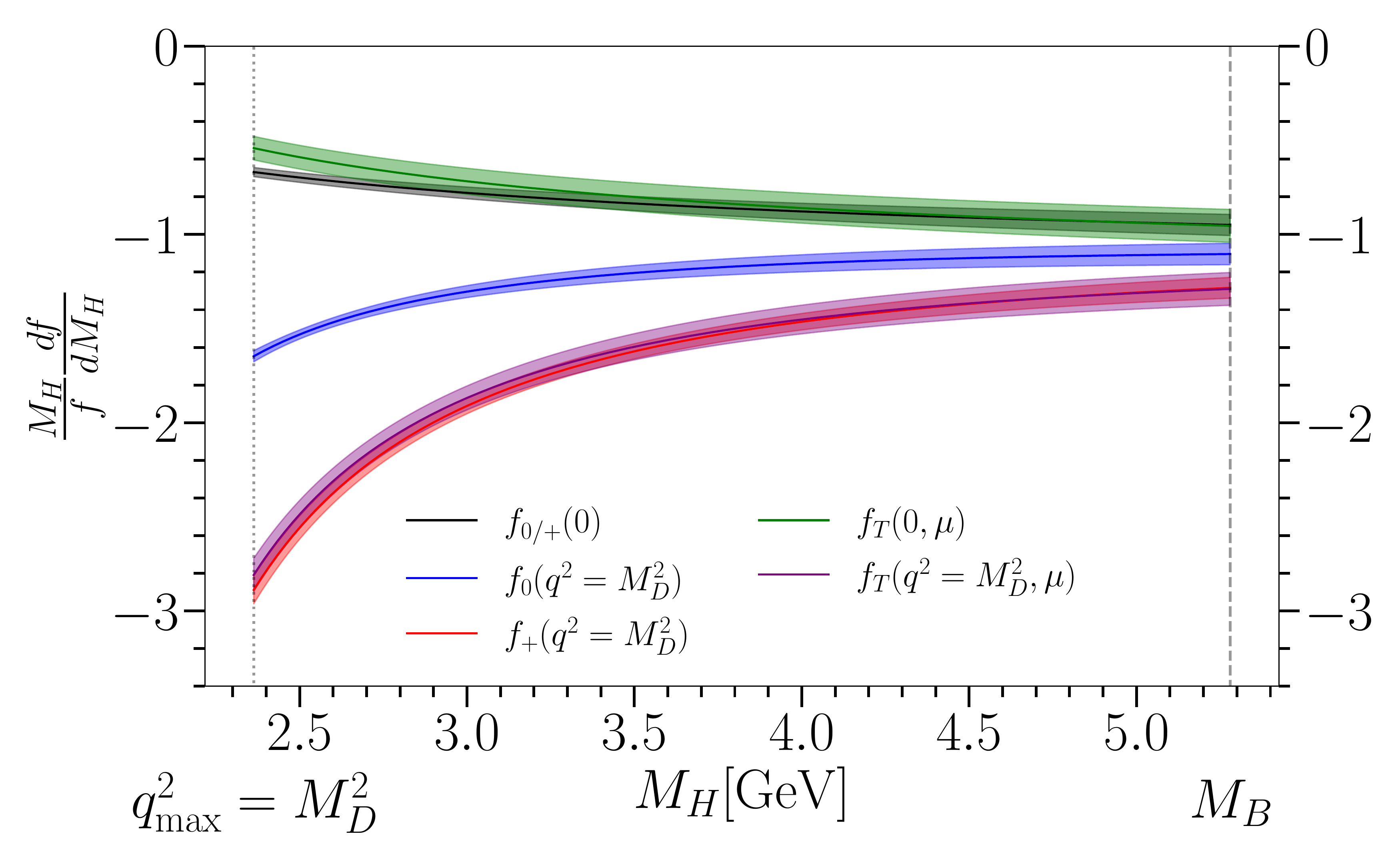}
\caption{The dominant power of the $M_H$ dependence of our form factors at $q^2=0$ and $q^2=M_D^2$, isolated using $(M_H/f)\times df/dM_H$.}
\label{fig:ffMhdepderiv}
\end{figure}

Figure~\ref{fig:ffMhdep} shows the heavy mass dependence of the form factors at $q^2=0$ and $q^2=M_D^2$. As above, the scale $\mu$ is given by Equation~\eqref{eq:murunning}. The lower bound on the mass range accessible to us is set at the point where $q^2_{\mathrm{max}}=M_D^2$. We see that towards $M_H=M_B$, the heavy mass dependence of the form factors at the two different $q^2$ is similar, but that this diverges quite rapidly at lower values of $M_H$, particularly for the vector and tensor form factors.

The dominant power in the $M_H$ dependence is isolated in Figure~\ref{fig:ffMhdepderiv}, by plotting $(M_H/f)\times df/dM_H$ (i.e. returning $X$ if $f\propto M_H^X$). The $M_H$ dependence of the form factors evaluated at $q^2=0$ is relatively unchanged (i.e. nearly constant $X$) for the plotted range of $M_H$, whilst the form factors evaluated at $q^2=M_D^2$ show a more variable exponential dependence.

\section{Correlator fit results}
\label{sec:corr-results}
Table~\ref{tab:fitresults1} gives the results of our correlator fits for the Goldstone kaon, $K$, for each twist (momentum) on each ensemble listed in Table~\ref{tab:ensembles}. The non-Goldstone kaon ($\hat{K}$) differs from the Goldstone by discretisation effects which we account for in our fit and does not feature directly in our analysis. Likewise in the last two columns of Table~\ref{tab:fitresults1} we present for each energy the raw fit result  $E^{\mathrm{fit}}_K$, as well at the theoretical value, $E_K^{\mathrm{theory}}=\sqrt{M_K^2+|\vec{p}|^2}$, which again differs only by discretisation effects. In practice, we only need to consider $E_K$ in our analysis when we calculate $q^2$. In this case, we use the theoretical value as it is more precise, and we account for discretisation effects elsewhere in our fit. 
Tables~\ref{tab:fitresults2},~\ref{tab:fitresults3} and~\ref{tab:fitresults4} contain numerical results from our two- and three-point correlator fits, across all 8 gluon ensembles listed in Table~\ref{tab:ensembles}. For each heavy mass and at each $q^2$ value, we provide the Goldstone heavy mass $M_H$, as well as the matrix elements and the form factor values obtained from these via Equations~\eqref{Eq:sca},~\eqref{Eq:vec} and~\eqref{Eq:ten}. As before with the kaon above, the non-Goldstone heavy meson mass does not feature directly in our analysis. 
\begin{table*}
  \caption{Goldstone kaon energies from fits to correlators on all gluon ensembles. The first column, $\theta$, is the twist value applied, which is converted to $|a\vec{p}|$ (shown in column 2) using $|a\vec{p}_K|=\theta(\sqrt{3}\pi)/N_x$. We then provide the theoretical $E_k^{\mathrm{theory}}=\sqrt{M_K^2+|\vec{p}|^2}$, which is the value used in our analysis. The fit result, $E^{\mathrm{fit}}_K$, which differs from this only by discretisation effects, is given in the final column.}
  \begin{center} 
    \begin{tabular}{ c | c c c c}
      \hline
      \hline
      Set& $\theta$ & $|a\vec{p}|$ &$E^{\mathrm{theory}}_K$&$E^{\mathrm{fit}}_K$\\ [0.5ex]
      \hline 
      \hline 
      \multirow{5}{*}{1}&0&0.0000&0.37886(17)&0.37886(17)\\ [1ex]
      &2.013&0.3423&0.51059(13)&0.50984(54)\\ [1ex]
      &3.05&0.5186&0.64227(10)&0.6411(17)\\ [1ex]
      &3.969&0.6749&0.773970(85)&0.7702(53)\\ [1ex]
      \hline
      \hline
      \multirow{5}{*}{2}&0&0.0000&0.303983(49)&0.303983(49)\\ [1ex]
      &2.405&0.2726&0.408334(36)&0.40820(25)\\ [1ex]
      &3.641&0.4128&0.512611(29)&0.51204(97)\\ [1ex]
      &4.735&0.5368&0.616870(24)&0.6148(23)\\ [1ex]
      \hline
      \hline 
      \multirow{5}{*}{3}&0&0.0000&0.218672(66)&0.218672(66)\\ [1ex]
      &0.8563&0.0728&0.230473(62)&0.230482(89)\\ [1ex]
      &2.998&0.2549&0.335841(43)&0.33554(88)\\ [1ex]
      &5.140&0.4370&0.488669(29)&0.4884(30)\\ [1ex]
      \hline
      \hline 
      \multirow{5}{*}{4}&0&0.0000&0.41621(18)&0.41621(18)\\ [1ex]
      &0.3665&0.1246&0.43447(17)&0.43443(19)\\ [1ex]
      &1.097&0.3731&0.55894(13)&0.55735(72)\\ [1ex]
      &1.828&0.6217&0.748141(99)&0.7451(33)\\ [1ex]
      \hline
      \hline 
      \multirow{7}{*}{5}&0&0.0000&0.33311(11)&0.33311(11)\\ [1ex]
      &0.441&0.1000&0.34780(11)&0.34790(12)\\ [1ex]
      &1.323&0.3000&0.448262(83)&0.44816(35)\\ [1ex]
      &2.205&0.4999&0.600744(62)&0.6001(19)\\ [1ex]
      &2.646&0.5999&0.686193(54)&0.6850(36)\\ [1ex]
      \hline
      \hline 
      \multirow{7}{*}{6}&0&0.0000&0.24238(11)&0.24238(11)\\ [1ex]
      &0.4281&0.0728&0.25308(10)&0.25306(12)\\ [1ex]
      &1.282&0.2180&0.325993(79)&0.32565(34)\\ [1ex]
      &2.141&0.3641&0.437369(59)&0.43642(86)\\ [1ex]
      &2.570&0.4370&0.499729(52)&0.4986(14)\\ [1ex]
      \hline
      \hline 
      \multirow{7}{*}{7}&0&0.0000&0.160189(88)&0.160189(88)\\ [1ex]
      &1.261&0.1430&0.214698(66)&0.21405(30)\\ [1ex]
      &2.108&0.2390&0.287691(49)&0.28712(73)\\ [1ex]
      &2.946&0.3340&0.370397(38)&0.3693(14)\\ [1ex]
      &3.624&0.4108&0.440952(32)&0.4390(25)\\ [1ex]
      \hline
      \hline 
      \multirow{7}{*}{8}&0&0.0000&0.118509(63)&0.118509(63)\\ [1ex]
      &0.706&0.0600&0.132843(57)&0.132953(97)\\ [1ex]
      &1.529&0.1300&0.175909(43)&0.17617(22)\\ [1ex]
      &2.235&0.1900&0.223949(34)&0.22405(39)\\ [1ex]
      &4.705&0.4000&0.417213(18)&0.4177(31)\\ [1ex]
      
      \hline
      \hline
    \end{tabular}
  \end{center}
  \label{tab:fitresults1}
\end{table*}
\begin{table*}
  \caption{Results from fits to correlators on sets 1, 2 and 3, all of which have approximately physical light quark masses. For each heavy quark mass there are four values for the $K$ momentum, giving four different values for $q^2$. For each of these values we give the current matrix elements (the matrix elements for the vector and tensor are given before their respective normalisations $Z_V$ and $Z_T$ have been applied). The final three columns give the values for $f_0(q^2)$, $f_+(q^2)$ (in this case only data from $V^0$ is available) and $f_T(q^2,\mu=4.8~\mathrm{GeV})$, determined using Equations~\eqref{Eq:vec},~\eqref{Eq:sca} and~\eqref{Eq:ten}. No tensor data was calculated on sets 1 or 2.}
  \begin{center} 
    \begin{tabular}{ c | c c c c c c c}
      \hline
      \hline
      Set   &&&&&&&\\[0.5ex]
      $am_h$&$(aq)^2$&$\bra{K}S\ket{H}$ & $\bra{K}V^0\ket{\hat{H}}$ & $\bra{\hat{K}}T^{10}\ket{\hat{H}}$ &$f_0(q^2)$&$f^{V^0}_+(q^2)$&$f_T(q^2)$\\ [0.5ex]
      $aM_H$ &&&&&&&\\[0.5ex]
      \hline 
      \hline 
      \multirow{1.6666666666666667}{*}{1}&1.1443(10)&2.524(13)&1.792(16)&-&1.0236(49)&-&-\\ [1ex]
      \multirow{1.6666666666666667}{*}{0.8605}&0.76263(88)&2.236(12)&1.605(14)&-&0.9066(46)&1.133(29)&-\\ [1ex]
      \multirow{1.6666666666666667}{*}{1.44857(46)}&0.38113(75)&2.033(18)&1.480(21)&-&0.8243(72)&0.912(14)&-\\ [1ex]
      &-0.00042(62)&1.861(54)&1.425(59)&-&0.755(22)&0.755(22)&-\\ [1ex]

      \hline
      \hline 
      \multirow{1.6666666666666667}{*}{2}&0.72338(50)&2.1519(74)&1.4643(83)&-&1.0240(31)&-&-\\ [1ex]
      \multirow{1.6666666666666667}{*}{0.643}&0.48244(44)&1.9015(60)&1.3104(68)&-&0.9049(26)&1.123(13)&-\\ [1ex]
      \multirow{1.6666666666666667}{*}{1.15450(30)}&0.24166(38)&1.713(10)&1.193(11)&-&0.8154(49)&0.9029(90)&-\\ [1ex]
      &0.00092(32)&1.561(21)&1.093(22)&-&0.7428(98)&0.7430(98)&-\\ [1ex]

      \hline
      \hline 
      \multirow{1.6666666666666667}{*}{3}&0.37848(37)&1.6554(59)&1.0665(83)&-&1.0149(33)&-&-\\ [1ex]
      \multirow{1.6666666666666667}{*}{0.433}&0.35880(36)&1.6209(59)&1.0463(81)&0.0796(30)&0.9937(34)&1.39(32)&1.199(46)\\ [1ex]
      \multirow{1.6666666666666667}{*}{0.83388(30)}&0.18307(30)&1.371(12)&0.900(15)&0.1959(72)&0.8405(71)&0.978(31)&0.843(31)\\ [1ex]
      &-0.07181(21)&1.128(34)&0.776(42)&0.238(17)&0.692(21)&0.661(19)&0.597(42)\\ [1ex]
      \hline 
      \multirow{1.6666666666666667}{*}{3}&0.80193(77)&1.8293(87)&1.307(13)&-&0.9916(43)&-&-\\ [1ex]
      \multirow{1.6666666666666667}{*}{0.683}&0.77563(76)&1.7896(85)&1.279(13)&0.1051(46)&0.9701(43)&1.67(72)&1.501(66)\\ [1ex]
      \multirow{1.6666666666666667}{*}{1.11418(43)}&0.54084(67)&1.504(14)&1.080(19)&0.2522(99)&0.8155(78)&1.154(85)&1.028(41)\\ [1ex]
      &0.20028(54)&1.242(41)&0.930(54)&0.300(22)&0.673(22)&0.738(43)&0.714(52)\\ [1ex]
      \hline 
      \multirow{1.6666666666666667}{*}{3}&1.03474(98)&1.905(10)&1.407(16)&-&0.9837(49)&-&-\\ [1ex]
      \multirow{1.6666666666666667}{*}{0.8}&1.00557(97)&1.863(10)&1.376(15)&0.1164(55)&0.9620(48)&1.82(97)&1.634(77)\\ [1ex]
      \multirow{1.6666666666666667}{*}{1.23589(48)}&0.74512(87)&1.562(16)&1.154(22)&0.276(11)&0.8067(81)&1.25(12)&1.109(46)\\ [1ex]
      &0.36736(72)&1.293(44)&1.002(59)&0.330(25)&0.667(23)&0.771(65)&0.773(58)\\ [1ex]

      \hline
      \hline
    \end{tabular}
  \end{center}
  \label{tab:fitresults2}
\end{table*}
\begin{table*}
  \caption{Results from fits to correlators on sets 4, 5 and 6, all of which have $m_l=m_s/5$. For each heavy quark mass there are several values for the $K$ momentum, giving different values for $q^2$. For each of these values we give the current matrix elements (the matrix elements for the vector and tensor are given before their respective normalisations $Z_V$ and $Z_T$ have been applied). The final three columns give the values for $f_0(q^2)$, $f_+(q^2)$ (in this case only data from $V^0$ is available) and $f_T(q^2,\mu=4.8~\mathrm{GeV})$ determined using Equations~\eqref{Eq:vec},~\eqref{Eq:sca} and~\eqref{Eq:ten}.}
  \begin{center} 
    \begin{tabular}{ c | c c c c c c c}
      \hline
      \hline
      Set   &&&&&&&\\[0.5ex]
      $am_h$&$(aq)^2$&$\bra{K}S\ket{H}$ & $\bra{K}V^0\ket{\hat{H}}$ & $\bra{\hat{K}}T^{10}\ket{\hat{H}}$ &$f_0(q^2)$&$f^{V^0}_+(q^2)$&$f_T(q^2)$\\ [0.5ex]
      $aM_H$ &&&&&&&\\[0.5ex]
      \hline
      \hline 
      \multirow{1.6666666666666667}{*}{4}&1.16024(75)&2.5539(63)&1.8680(99)&-&1.0150(22)&-&-\\ [1ex]
      \multirow{1.6666666666666667}{*}{0.888}&1.10569(74)&2.5089(58)&1.8358(95)&0.1413(37)&0.9972(20)&1.41(12)&1.192(32)\\ [1ex]
      \multirow{1.6666666666666667}{*}{1.49335(38)}&0.73394(65)&2.243(10)&1.660(13)&0.3420(75)&0.8913(40)&1.080(20)&0.964(21)\\ [1ex]
      &0.16885(52)&1.969(34)&1.530(39)&0.430(18)&0.782(13)&0.807(16)&0.727(31)\\ [1ex]

      \hline
      \hline 
      \multirow{2.0}{*}{5}&0.73640(43)&2.1641(51)&1.5039(56)&-&1.0083(21)&-&-\\ [1ex]
      \multirow{2.0}{*}{0.664}&0.70142(42)&2.1254(49)&1.4775(54)&0.1124(23)&0.9903(20)&1.40(14)&1.214(26)\\ [1ex]
      \multirow{2.0}{*}{1.19125(26)}&0.46206(37)&1.9046(63)&1.3326(76)&0.2647(47)&0.8874(28)&1.088(18)&0.953(17)\\ [1ex]
      &0.09877(30)&1.673(20)&1.182(24)&0.322(11)&0.7797(94)&0.811(12)&0.694(23)\\ [1ex]
      &-0.10481(25)&1.557(28)&1.107(33)&0.327(15)&0.725(13)&0.697(12)&0.588(27)\\ [1ex]
      \hline 
      \multirow{2.0}{*}{5}&1.00245(58)&2.2493(62)&1.6260(68)&-&1.0044(24)&-&-\\ [1ex]
      \multirow{2.0}{*}{0.8}&0.96327(57)&2.2085(59)&1.5966(66)&0.1249(27)&0.9862(23)&1.51(20)&1.316(29)\\ [1ex]
      \multirow{2.0}{*}{1.33434(30)}&0.69516(51)&1.9761(71)&1.4356(87)&0.2925(55)&0.8824(31)&1.155(27)&1.028(20)\\ [1ex]
      &0.28823(43)&1.733(22)&1.269(26)&0.354(12)&0.7738(99)&0.856(19)&0.747(25)\\ [1ex]
      &0.06019(38)&1.610(31)&1.187(36)&0.361(17)&0.719(14)&0.733(15)&0.635(30)\\ [1ex]
      \hline 
      \multirow{2.0}{*}{5}&1.21276(70)&2.3094(69)&1.7106(79)&-&1.0031(27)&-&-\\ [1ex]
      \multirow{2.0}{*}{0.9}&1.17065(69)&2.2672(66)&1.6791(76)&0.1335(30)&0.9848(26)&1.59(24)&1.388(32)\\ [1ex]
      \multirow{2.0}{*}{1.43437(33)}&0.88243(62)&2.0265(77)&1.5073(96)&0.3118(60)&0.8803(32)&1.205(35)&1.081(21)\\ [1ex]
      &0.44500(53)&1.771(23)&1.329(28)&0.376(13)&0.769(10)&0.886(26)&0.782(27)\\ [1ex]
      &0.19987(47)&1.647(33)&1.242(39)&0.384(18)&0.715(14)&0.760(21)&0.665(31)\\ [1ex]

      \hline
      \hline 
      \multirow{2.0}{*}{6}&0.38668(32)&1.6875(52)&1.1190(74)&-&1.0089(29)&-&-\\ [1ex]
      \multirow{2.0}{*}{0.449}&0.36819(31)&1.6571(51)&1.0994(75)&0.0830(27)&0.9907(28)&1.40(32)&1.268(41)\\ [1ex]
      \multirow{2.0}{*}{0.86422(27)}&0.24216(27)&1.4757(74)&0.980(13)&0.1958(50)&0.8823(44)&1.096(41)&0.999(26)\\ [1ex]
      &0.04966(22)&1.279(14)&0.872(25)&0.2405(91)&0.7644(84)&0.791(11)&0.735(28)\\ [1ex]
      &-0.05813(19)&1.210(26)&0.864(46)&0.254(14)&0.723(16)&0.701(16)&0.646(36)\\ [1ex]
      \hline 
      \multirow{2.0}{*}{6}&0.57131(47)&1.7721(64)&1.2348(98)&-&0.9986(33)&-&-\\ [1ex]
      \multirow{2.0}{*}{0.566}&0.54996(46)&1.7395(62)&1.2124(99)&0.0949(34)&0.9802(32)&1.51(54)&1.407(51)\\ [1ex]
      \multirow{2.0}{*}{0.99823(32)}&0.40439(41)&1.5451(85)&1.077(16)&0.2218(63)&0.8707(47)&1.181(73)&1.098(32)\\ [1ex]
      &0.18203(35)&1.336(16)&0.951(30)&0.270(11)&0.7531(91)&0.840(25)&0.801(33)\\ [1ex]
      &0.05753(31)&1.268(28)&0.928(50)&0.282(16)&0.714(16)&0.735(19)&0.698(40)\\ [1ex]
      \hline 
      \multirow{2.0}{*}{6}&0.78019(63)&1.8513(75)&1.341(11)&-&0.9888(37)&-&-\\ [1ex]
      \multirow{2.0}{*}{0.683}&0.75611(62)&1.8165(72)&1.315(11)&0.1054(38)&0.9702(35)&1.62(71)&1.529(55)\\ [1ex]
      \multirow{2.0}{*}{1.12566(37)}&0.59195(57)&1.6103(94)&1.165(17)&0.2453(70)&0.8601(49)&1.26(10)&1.188(34)\\ [1ex]
      &0.34121(49)&1.392(18)&1.026(32)&0.297(12)&0.7434(94)&0.889(41)&0.861(36)\\ [1ex]
      &0.20081(45)&1.322(31)&0.999(54)&0.310(18)&0.706(17)&0.771(34)&0.748(43)\\ [1ex]
      \hline 
      \multirow{2.0}{*}{6}&1.00992(80)&1.9254(85)&1.385(42)&-&0.9805(40)&-&-\\ [1ex]
      \multirow{2.0}{*}{0.8}&0.98324(79)&1.8886(82)&1.361(40)&0.1186(82)&0.9618(38)&1.5(2.4)&1.69(12)\\ [1ex]
      \multirow{2.0}{*}{1.24733(40)}&0.80134(73)&1.672(10)&1.211(49)&0.273(14)&0.8515(50)&1.27(35)&1.299(65)\\ [1ex]
      &0.52349(64)&1.444(19)&1.065(76)&0.330(22)&0.7355(96)&0.92(13)&0.941(63)\\ [1ex]
      &0.36793(59)&1.372(33)&0.967(98)&0.337(28)&0.699(17)&0.853(84)&0.800(66)\\ [1ex]

      \hline
      \hline
    \end{tabular}
  \end{center}
  \label{tab:fitresults3}
\end{table*}
\begin{table*}
  \caption{Results from fits to correlators on sets 7 and 8, both of which have $m_l=m_s/5$. For each heavy quark mass there are five values for the $K$ momentum, giving five different values for $q^2$. For each of these values we give the the current matrix elements (the matrix elements for the vector and tensor are given before their respective normalisations $Z_V$ and $Z_T$ have been applied). The final four columns give the values for $f_0(q^2)$, $f_+(q^2)$ (from both $V^0$ and $V^1$ where present) and $f_T(q^2,\mu=4.8~\mathrm{GeV})$ determined using Equations~\eqref{Eq:vec},~\eqref{Eq:sca} and~\eqref{Eq:ten}.}
  \begin{center} 
    \begin{tabular}{ c | c c c c c c c c c}
      \hline
      \hline
      Set   &&&&&&&&&\\[0.5ex]
      $am_h$&$(aq)^2$ &$\bra{K}S\ket{H}$ & $\bra{K}V^0\ket{\hat{H}}$ &$\bra{\hat{K}}V^1\ket{H}$& $\bra{\hat{K}}T^{10}\ket{\hat{H}}$ &$f_0(q^2)$&$f^{V^0}_+(q^2)$&$f^{V^1}_+(q^2)$&$f_T(q^2)$\\ [0.5ex]
      $aM_H$ &&&&&&&&&\\[0.5ex]
      \hline
      \hline 
      \multirow{2.0}{*}{7}&0.16548(20)&1.1890(57)&0.7398(68)&-&-&1.0073(45)&-&-&-\\ [1ex]
      \multirow{2.0}{*}{0.274}&0.10367(17)&1.0435(72)&0.6559(85)&-&0.1241(44)&0.8841(60)&1.071(45)&-&0.997(35)\\ [1ex]
      \multirow{2.0}{*}{0.56698(25)}&0.02090(14)&0.923(20)&0.591(25)&-&0.1502(76)&0.782(17)&0.808(21)&-&0.722(37)\\ [1ex]
      &-0.072889(96)&0.816(34)&0.529(39)&-&0.156(14)&0.691(29)&0.620(28)&-&0.535(49)\\ [1ex]
      &-0.152895(62)&0.741(55)&0.507(60)&-&0.163(23)&0.628(46)&0.523(49)&-&0.455(65)\\ [1ex]
      \hline 
      \multirow{2.0}{*}{7}&0.38171(46)&1.3252(88)&0.922(10)&-&-&0.9753(60)&-&-&-\\ [1ex]
      \multirow{2.0}{*}{0.45}&0.29690(42)&1.1584(95)&0.810(11)&-&0.1605(59)&0.8525(68)&1.25(12)&-&1.213(45)\\ [1ex]
      \multirow{2.0}{*}{0.77802(37)}&0.18332(36)&1.021(24)&0.723(29)&-&0.193(10)&0.751(18)&0.923(75)&-&0.871(47)\\ [1ex]
      &0.05462(30)&0.902(43)&0.647(51)&-&0.200(19)&0.664(31)&0.702(43)&-&0.645(62)\\ [1ex]
      &-0.05516(25)&0.823(68)&0.616(76)&-&0.207(30)&0.606(50)&0.579(42)&-&0.545(79)\\ [1ex]
      \hline 
      \multirow{2.0}{*}{7}&0.61475(71)&1.426(11)&1.051(13)&-&-&0.9493(67)&-&-&-\\ [1ex]
      \multirow{2.0}{*}{0.6}&0.51181(66)&1.243(11)&0.920(13)&0.1917(48)&0.1870(72)&0.8277(74)&1.38(19)&1.381(28)&1.371(53)\\ [1ex]
      \multirow{2.0}{*}{0.94425(46)}&0.37396(60)&1.092(28)&0.815(33)&0.2278(92)&0.224(12)&0.727(19)&1.02(14)&1.005(26)&0.980(55)\\ [1ex]
      &0.21777(52)&0.965(50)&0.724(58)&-&0.231(23)&0.643(34)&0.770(92)&-&0.724(72)\\ [1ex]
      &0.08453(46)&0.891(81)&0.698(96)&-&0.240(35)&0.593(54)&0.627(74)&-&0.613(90)\\ [1ex]
      \hline 
      \multirow{2.0}{*}{7}&0.9854(11)&1.544(14)&1.199(16)&-&-&0.9199(77)&-&-&-\\ [1ex]
      \multirow{2.0}{*}{0.8}&0.8597(10)&1.344(14)&1.046(15)&0.2224(62)&0.2177(83)&0.8009(79)&1.55(30)&1.562(38)&1.554(59)\\ [1ex]
      \multirow{2.0}{*}{1.15285(54)}&0.69140(93)&1.179(32)&0.917(35)&0.263(11)&0.260(14)&0.703(19)&1.16(22)&1.126(34)&1.109(61)\\ [1ex]
      &0.50070(84)&1.040(57)&0.798(64)&-&0.264(26)&0.620(34)&0.89(17)&-&0.806(80)\\ [1ex]
      &0.33802(77)&0.969(94)&0.78(11)&-&0.275(39)&0.578(56)&0.70(15)&-&0.682(98)\\ [1ex]

      \hline
      \hline 
      \multirow{2.0}{*}{8}&0.09183(13)&0.9331(48)&0.5553(58)&-&-&1.0120(48)&-&-&-\\ [1ex]
      \multirow{2.0}{*}{0.194}&0.07974(13)&0.8903(44)&0.5328(57)&-&0.0616(22)&0.9657(45)&1.25(11)&-&1.192(43)\\ [1ex]
      \multirow{2.0}{*}{0.42154(22)}&0.04344(11)&0.7833(71)&0.4727(91)&-&0.0995(39)&0.8496(77)&0.979(29)&-&0.890(35)\\ [1ex]
      &0.002933(86)&0.699(18)&0.435(22)&-&0.1109(55)&0.758(19)&0.764(20)&-&0.678(34)\\ [1ex]
      &-0.1600035(19)&0.53(11)&0.337(93)&-&0.134(29)&0.58(12)&0.40(11)&-&0.389(84)\\ [1ex]
      \hline 
      \multirow{2.0}{*}{8}&0.37432(52)&1.137(10)&0.808(10)&-&-&0.9491(79)&-&-&-\\ [1ex]
      \multirow{2.0}{*}{0.45}&0.35338(51)&1.0813(94)&0.7705(94)&0.1036(30)&0.0973(36)&0.9026(72)&1.66(43)&1.743(43)&1.708(64)\\ [1ex]
      \multirow{2.0}{*}{0.73033(43)}&0.29048(47)&0.943(12)&0.675(13)&0.1660(55)&0.1546(60)&0.7868(97)&1.26(15)&1.295(30)&1.254(49)\\ [1ex]
      &0.22031(43)&0.842(27)&0.620(30)&-&0.1687(86)&0.703(23)&0.92(12)&-&0.936(48)\\ [1ex]
      &-0.06198(27)&0.63(14)&0.46(12)&-&0.182(42)&0.53(11)&0.488(90)&-&0.48(11)\\ [1ex]
      \hline 
      \multirow{2.0}{*}{8}&0.60243(79)&1.229(14)&0.922(13)&-&-&0.9121(95)&-&-&-\\ [1ex]
      \multirow{2.0}{*}{0.6}&0.57678(78)&1.168(12)&0.877(12)&0.1205(41)&0.1139(45)&0.8667(86)&1.83(70)&1.976(61)&1.949(77)\\ [1ex]
      \multirow{2.0}{*}{0.89467(51)}&0.49972(73)&1.016(14)&0.766(15)&0.1911(70)&0.1798(72)&0.754(10)&1.40(25)&1.453(42)&1.421(57)\\ [1ex]
      &0.41376(68)&0.911(32)&0.704(35)&-&0.195(10)&0.676(24)&1.00(22)&-&1.056(56)\\ [1ex]
      &0.06795(49)&0.70(17)&0.54(15)&-&0.214(52)&0.52(12)&0.55(16)&-&0.55(13)\\ [1ex]
      \hline 
      \multirow{2.0}{*}{8}&0.9670(12)&1.341(18)&1.055(17)&-&-&0.875(11)&-&-&-\\ [1ex]
      \multirow{2.0}{*}{0.8}&0.9354(12)&1.273(16)&1.003(16)&0.1411(55)&0.1344(58)&0.831(10)&2.0(1.1)&2.273(86)&2.250(98)\\ [1ex]
      \multirow{2.0}{*}{1.10185(60)}&0.8405(11)&1.105(18)&0.873(18)&0.2218(91)&0.2114(92)&0.722(11)&1.58(42)&1.657(60)&1.634(71)\\ [1ex]
      &0.7346(11)&0.991(37)&0.799(40)&-&0.228(13)&0.647(24)&1.10(37)&-&1.203(67)\\ [1ex]
      &0.30870(82)&0.65(15)&0.55(15)&-&0.218(53)&0.421(98)&0.50(25)&-&0.55(13)\\ [1ex]

      \hline
      \hline
    \end{tabular}
  \end{center}
  \label{tab:fitresults4}
\end{table*}
\end{appendix}

\bibliography{BKpaper}
\end{document}